# Diffusion model for fast $\alpha$-particle transport in Magneto-Inertial Fusion plasmas

A. L. Velikovich, F. García-Rubio, S. Davidson, C. L. Ellison,
J. Hamilton, and P. F. Schmit

Pacific Fusion, 6082 Stewart Ave., Fremont, California 94538, USA

## Abstract

A diffusion model is developed to describe fast $\alpha$-particle transport and confinement in dense magnetized DT fusion plasmas, tailored to Magneto-Inertial Fusion conditions. The model, derived from the Fokker-Planck kinetic equation using the Chapman-Enskog approximation, includes dynamic Coulomb friction from electrons and ions. This approach improves upon earlier heuristic model [Liberman and Velikovich, J. Plasma Phys. **31**, 369 (1984)], which, while valid in the ignition range near 5 keV, lacks accuracy at the higher temperatures ~20 keV required for high-gain Magnetized Liner Inertial Fusion (MagLIF). This paper provides explicit analytic formulas for all transport coefficients and characterizes the relationship between the $\alpha$-particle pressure gradient and the associated Ampère force in magnetized plasmas, which are both relevant for MagLIF at high temperatures.

## I. INTRODUCTION

We present the development of a diffusion model to describe the transport and confinement of fast α-particles in deuterium-tritium (DT) fusion plasmas, focusing on the parameter range characteristic of Magneto-Inertial Fusion (MIF). Fast α-particles are produced in the DT fusion reaction T(d,n)$^4$He at the local rate

$$\dot{n}_\alpha = n_D n_T \langle \sigma v \rangle_{DT}, \tag{1}$$

where $n_D$ and $n_T$ are deuterium and tritium ion number densities, respectively, and $\langle \sigma v \rangle_{DT}$ is the reactivity averaged over the energy distributions of fusing ions. When fusion conditions are met, a significant number of fast α-particles are produced. However, to sustain and propagate the fusion burn necessary for achieving ignition and high gain – and ultimately for the production of fusion energy – these particles must be successfully confined to deposit their energy within the DT fuel plasma.

The two primary approaches to controlled thermonuclear fusion – magnetic (MCF) and inertial (ICF) confinement – rely upon different mechanisms for this. In MCF, the α-particles and the nearly collisionless, low-$\beta$ fusion plasma are confined magnetically. In ICF, fast α-particles are confined by assembling a high fuel areal density, $\rho R$, which ensures sufficient α-energy deposition within the hot spot to initiate self-heating and ignite the surrounding fuel. The latter was demonstrated in the National Ignition Facility (NIF) shot N221204, which achieved target energy gain greater than unity [1] with a measured hot spot $(\rho R)_{\text{hs}}$ of approximately 0.44 g/cm$^2$. Although all approaches to controlled thermonuclear fusion aim to operate at similar DT plasma temperatures, ~10 to 20 keV, the characteristic densities of MCF and ICF plasmas differ by over ten orders of magnitude, ranging from the relatively large for MCF value of $n_e = 3\times10^{14}$

cm$^{-3}$ projected for the high-field SPARC tokamak [2] to $n_e \approx 1.5\times10^{25}$ cm$^{-3}$ measured in NIF shot N221204.

Research aimed at achieving controlled fusion in the vast intermediate density range between MCF and ICF has a long history. It began in the 1960s [3] with efforts to increase the magnetic field confining the DT plasma via magnetic flux compression [4, 5], reaching well into the megagauss (MG) range from the ~100 kG fields achievable for high-field tokamaks [2]. The corresponding increase in plasma density allowed for a reduction in confinement time and the development of more compact fusion systems. This research field became known as Magnetized Target Fusion (MTF). Its progress, following pioneering theoretical and experimental works [3, 6, 7], is documented in the monograph [8]. The MTF parameter space has been systematically characterized by Lindemuth in a series of publications over several decades [9, 10].

Magneto-Inertial Fusion refers to a high-density domain in this parameter space that is adjacent to the Inertial Confinement Fusion (ICF) range. It aims to achieve high fusion yields with fuel densities that are one to two orders of magnitude greater than that of solid DT ice (0.2 g/cm$^3$), but still less than the peak densities achieved in extreme ICF implosions, such as NIF shot N221204. MIF employs inertial confinement of the compressed DT plasma while using magnetization for thermal insulation and $\alpha$-particle trapping. The most prominent example of the MIF approach is Magnetized Liner Inertial Fusion (MagLIF), a concept proposed in Refs. [11, 12] and pursued on the Z machine at Sandia National Laboratories [13]. The scaling of MagLIF performance toward the high-gain regime at driving currents of ~60 MA is presented in Refs. [14–17]. Design variants of MagLIF utilizing Auto-magnetizing (AutoMAG) liners and Dynamic Screw Pinch (DSP) concepts are reported in Refs. [18–20] and [21–23], respectively.

Fast α-particles lose energy, heating the fusion plasma, through dynamic friction or drag from binary small-angle Coulomb collisions with plasma electrons and ions. In nearly collisionless magnetically confined fusion plasmas, this is just one of many processes affecting fast α-particle dynamics and confinement, and it is not the most significant one (see Sec. 2.1 of Ref. [24] and Refs. [25, 26]). This complexity makes their behavior in MCF plasmas more intricate than what diffusion models can represent; see Refs. [26-28]. In contrast, in dense inertially confined laser fusion plasmas, the dynamics of fast α-particles are primarily governed by Coulomb friction, allowing their transport models to be simplified using methods from radiation transport theory, such as forward-reverse [29] or single-ray [30] approximations.

Modeling the transport of fast α-particles in MIF presents a complexity that sits between MCF and ICF. The trajectories of the slowing-down α-particles in MIF are helical, not straight, due to their magnetization. This complexity prevents the use of the simplified models of Refs. [29, 30]. However, we can still assume that Coulomb friction governs the dynamics of fast α-particles, whose trajectories are affected by external fields, especially the applied magnetic field. This assumption allows us to utilize a diffusion model, provided that the parameters of the fusion plasma change gradually over time and vary smoothly in space, thereby justifying the diffusion approximation.

The simplest diffusion equation for fast α-particle energy density ($\mathcal{E}_\alpha$) transport across a magnetic field in a dense plasma at rest is [31]

$$\frac{\partial \mathcal{E}_\alpha}{\partial t} = -\nabla \cdot \mathbf{q}_\alpha - Q_\alpha + \dot{n}_\alpha E_{\alpha 0} \,. \qquad (2)$$

According to this model, the term $Q_\alpha$, which represents the collisional energy loss from the α-particles to heating the fusion plasma, equals $2\nu_{\alpha e}\mathcal{E}_\alpha$. The energy flux carried by fast α-particles is expressed as $\mathbf{q}_\alpha = -\kappa_\perp \nabla \mathcal{E}_\alpha$, where the perpendicular "thermal diffusivity" coefficient of the fast α-particles' energy density is given by

$$\kappa_\perp = \frac{u_{\alpha 0}^2 \tau_{e\alpha}}{2\left(9+\chi_\alpha^2\right)}. \tag{3}$$

The divergence term $-\nabla \cdot \mathbf{q}_\alpha$ in the right-hand side of Eq. (2) accounts for the non-locality of the α-heating and the effect of magnetization. Here, $u_{\alpha 0} = \sqrt{2E_{\alpha 0}/m_\alpha} = 1.3\times 10^9$ cm/s is the birth velocity of an α-particle in the center-of-mass reference frame of fusing ions, $E_{\alpha 0} = 3.518$ MeV is its birth energy (its numerical value is taken from the MCF sources [25, 32]), $\tau_{\alpha e}$ is the characteristic time of α-*e* collisions defined by Eq. (23) below, and $\chi_\alpha = \omega_\alpha \tau_{\alpha e}$ is the Hall parameter for the fast α-particles, where $\omega_\alpha = 2eB/\left(m_\alpha c\right)$ denotes their gyro frequency in magnetic field $B$. (Note that while a true Hall parameter for fast α-particles would include collisions with ions, the product $\omega_\alpha \tau_{\alpha e}$ is used here for convenience.)

Equation (2) has passed a verification test of Ref. [33], where Basko *et al*. compared its predictions to their Monte Carlo simulations. It had been used to model the fast α-particle transport in linear z-pinches [34] and high-gain MagLIF targets [17]. A similar diffusion model, constructed to match Basko's results [33] and incorporating the same physics, was used in the pioneering paper [12] on high-yield, ice-layered MagLIF; this was also the α-transport model utilized in the HYDRA simulations that benchmarked the FLASH code with α-transport package [17]. A counterpart diffusion equation for the α- particle flux derived from the model [31] was used in Ref. [35] to model the α-driven Nernst effect manifesting during thermonuclear burn propagation in magnetized plasma.

Note that the diffusion equation (2) and those used in [12, 35] only take into account the deceleration of the fast α-particles in their collisions with electrons, a simplification that is acceptable near ignition, at DT plasma temperatures about 5 keV. But it becomes invalid at temperatures ~20 keV characteristic of high-gain MagLIF [17], when the electron and ion contributions to dynamic friction become comparable; cf. Refs. [36, 37], and Eqs. (8), (9) in Ref. [38].

We report revisiting the fast α-particle diffusion model to extend its applicability for the whole range of DT plasma temperatures relevant for Magneto-Inertial Fusion, including MagLIF. Using the Fokker-Planck kinetic equation to calculate the transport coefficients under the Chapman-Enskog approximation and taking into account both the electron and ion contributions to dynamic friction, we completed the work started in Refs. [31] and [39]. We report new exact analytic formulas for α-particle transport coefficients and their simplified closed-form asymptotic expressions in the limiting cases of lower temperatures and strong magnetization. The new results confirm that earlier formulas provide a good approximation at lower temperatures but diverge from the more precise values calculated for higher temperatures. Additionally, they reveal effects related to fast α-particle transport that must be taken into account in high-temperature modeling, including the relation between the α-particle pressure gradient [40, 41] and the associated $\mathbf{j}_\alpha \times \mathbf{B}$ force specific to magnetized plasmas. We provide an elementary derivation to clarify why magnetization reduces the perpendicular α-particle diffusion coefficient by a factor $1+\chi_\alpha^2$, whereas the corresponding reduction factor for the energy density diffusion coefficient is $1+\left(\chi_\alpha/3\right)^2$, as in Eq. (3). We also explain why the dependence of the diffusion coefficients on the α-particle Hall parameter $\chi_\alpha$ is monotonic, as in Eq. (3), if only the electron friction is accounted for, but including ion friction makes it oscillatory.

To ensure a self-contained presentation, we derive these results from first principles, recovering numerous established formulas from the literature. We emphasize the continuity between this work and decades of prior research on fast-ion transport in plasma, while clarifying the physical significance of the assumptions and approximations employed at each stage of the derivation.

This article is organized as follows: Section II introduces the governing equations; Section III provides explicit formulas for transport coefficients and their low-temperature approximations; Section IV describes the α-transport coupling into MHD fluid equations; and Section V evaluates the relative magnitudes of the model terms and explores the relevant transport regimes by post-processing high-yield MagLIF simulation profiles from FLASH. Finally, Section VI concludes with a discussion of our findings and future directions.

## II. GOVERNING EQUATIONS

### A. Background plasma and fusion source terms

Since the DT fusion reaction does not change the number of electrons in the plasma, below we denote the electron number density by $n \equiv n_e$. The background ion component of the burning plasma, through which a fast α-particle moves, consists of at least three species: single-charged deuterium and tritium ions, and double-charged thermalized α-particles, the "helium ash". In this model, we neglect α-α scattering and approximate the total ion number density as:

$$n_i = n_D + n_T + n_\alpha + n_{\mathrm{ash}} \approx n_D + n_T + n_{\mathrm{ash}}, \tag{4}$$

where $n_\alpha$ stands for the number density of the fast (non-thermalized) α-particles. We will justify the approximation $n_\alpha \ll n$ in Section II.D. Following conventional α-particle transport theory, when background ions belong to different species, they are modeled as a "fictitious ion" [39] with a mean mass and charge. Let us specify these mean values. If $\phi$ is the burn-up fraction of the fuel – initially an equimolar DT mixture – and the thermalized α-particles are retained within the fuel, the total number of the ions has been reduced by the factor of $1-\phi/2$. Consequently, the fractions of D, T, and ash ions are $\xi_D = \xi_T = (1-\phi)/(2-\phi)$, and $\xi_{ash} = \phi/(2-\phi)$, respectively. The average ion mass is then

$$m_i = \frac{2(1-\phi)}{2-\phi}\bar{m}_{DT} + \frac{\phi}{2-\phi}m_\alpha . \tag{5}$$

where $\bar{m}_{DT} = (m_D + m_T)/2$. Below, we will use a dimensionless parameter

$$\mu = 1 - \frac{m_i}{m_\alpha} = \frac{2(1-\phi)}{2-\phi}\left(1 - \frac{\bar{m}_{DT}}{m_\alpha}\right). \tag{6}$$

The average ion charge and its square are

$$\langle Z_i \rangle = \frac{2}{2-\phi}, \quad \langle Z_i^2 \rangle = \frac{2(1+\phi)}{2-\phi}. \tag{7}$$

The quasineutrality condition relates the ion number density to the electron density $n$ as

$$n_i = \frac{n}{\langle Z_i \rangle}. \tag{8}$$

If the burn-up fraction is negligible, $\phi \ll 1$, then the value of $\mu = 0.372$ corresponds to $m_i = \bar{m}_{DT}$, and $\langle Z_i \rangle = \langle Z_i^2 \rangle = 1$. For example, the measured value of $\phi = 0.0433$ in NIF shot N221204 [1] reduces $\mu$ from the above value by 2.1%, while increasing $\langle Z_i \rangle$ and $\langle Z_i^2 \rangle$ to 1.022 and 1.066, respectively. To illustrate the dependence of our coefficients on the plasma ion composition, in some examples below we will consider a high burn-up case of $\phi = 0.5$, which reduces the above value of $\mu$ by a factor of 2/3 to $\mu = 0.248$ and increases $\langle Z_i \rangle$ and $\langle Z_i^2 \rangle$ to 4/3 and 2, respectively. Full burn-up of a DT plasma $\phi = 1$ corresponds to $\mu = 0$. Note that the presence of high-atomic-number liner impurities mixed into the fuel, as well as α-particle diffusion into the cold liner itself, are both accounted for by increasing $m_i$, $\langle Z_i \rangle$ and $\langle Z_i^2 \rangle$, which correspondingly reduces $\mu$ and can make it negative.

The normalized Maxwellian distribution of the deuterium and tritium ions at temperature $T_i$ can be expressed as a combination of statistically independent distributions of their center-of-mass velocities, $\mathbf{v}_{\rm cm}$, and relative velocities, $\mathbf{v}_{\rm rel}$:

$$f_{i,\rm cm}\left(\mathbf{v}_{\rm cm}\right) = \left(\frac{M}{2\pi T_i}\right)^{3/2} \exp\left(-\frac{M v_{\rm cm}^2}{2T_i}\right), \tag{9}$$

$$f_{i,\rm rel}\left(\mathbf{v}_{\rm rel}\right) = \left(\frac{m_r}{2\pi T_i}\right)^{3/2} \exp\left(-\frac{m_r v_{\rm rel}^2}{2T_i}\right), \tag{10}$$

where $M = m_D + m_T = 2\bar{m}_{DT}$ and $m_r = m_D m_T / \left(m_D + m_T\right) \approx \bar{m}_{DT} / 2$ are the total and reduced masses of the fusing ions. The isotropic velocity distribution of the fast α-particles created in the center-of-mass reference frame with the same birth energy $E_{\alpha 0}$ is

$$\Sigma_\alpha\left(u\right) = \frac{\dot{n}_\alpha}{\left(2\pi\right)^{3/2} u u_{\alpha 0}} \left(\frac{M}{T_i}\right)^{1/2} \exp\left[-\frac{M\left(u^2 + u_{\alpha 0}^2\right)}{2T_i}\right] \sinh\left(\frac{M u u_{\alpha 0}}{T_i}\right), \tag{11}$$

where $u$ is the α-particle velocity in the reference frame of the background plasma, and $\dot{n}_\alpha$ is the production rate given by (1); cf. Eq. (39) of Ref. [42]. The normalization of the source function (11) is

$$\int_0^\infty 4\pi u^2 \Sigma_\alpha(u)\,du = \dot{n}_\alpha \,. \tag{12}$$

The total kinetic energy generation rate of newborn α-particles is

$$\int_0^\infty 4\pi u^2 \Sigma_\alpha(u) \times \frac{m_\alpha u^2}{2} du = \dot{n}_\alpha \left( E_{\alpha 0} + \Delta E_+ \right), \tag{13}$$

where the thermal addition $\Delta E_+ = 3 m_\alpha T_i / (2M)$ to the birth energy $E_{\alpha 0}$ accounts for the center-of-mass motion of the fusing ions.

Conversely, fusion removes two ions from the plasma. The average ion thermal energy loss per reaction, weighted by the reactivity of the colliding ions, is:

$$\begin{aligned} \Delta E_- &= \frac{\left\langle \sigma v \left( \frac{1}{2} M v_{\text{cm}}^2 + \frac{1}{2} m_r v_{\text{rel}}^2 \right) \right\rangle}{\langle \sigma v \rangle_{DT}} = \frac{\langle \sigma v \rangle_{DT} \times \left\langle \frac{1}{2} M v_{\text{cm}}^2 \right\rangle + \left\langle \sigma v \times \frac{1}{2} m_r v_{\text{rel}}^2 \right\rangle}{\langle \sigma v \rangle_{DT}} \\ &= \left\langle \frac{1}{2} M v_{\text{cm}}^2 \right\rangle + \frac{\langle \sigma v E \rangle_{DT}}{\langle \sigma v \rangle_{DT}} = \frac{3}{2} T_i + \frac{\langle \sigma v E \rangle_{DT}}{\langle \sigma v \rangle_{DT}}, \end{aligned} \tag{14}$$

where $E = m_r v_{\text{rel}}^2 / 2$ is the energy available in the center-of-mass frame. The energy loss $\Delta E_-$ exceeds $3T_i$ (the average thermal energy of two ions) due to the Gamow shift. This shift reflects the fact that the loss term is weighted by fusion reactivity, which disproportionately removes the most energetic ions. Using the Bosch and Hale parametrization [43] for the DT fusion cross-section $\sigma(E)$, we find that in the 2 to 50 keV range, this loss is well-approximated by:

$$\Delta E_- = 18.5\sqrt{T_i} + 0.852 \times T_i - 12.4 \text{ keV}, \tag{15}$$

where $T_i$ is in keV. Figure 1 plots the characteristic loss and gain energies, $\Delta E_-$ and $\Delta E_+$, and the thermal energy of a background plasma ion, $3T_i/2$, all normalized with respect to the center-of-mass birth energy of an α-particle, $E_{\alpha 0}$. The loss term $\Delta E_-$, containing the Gamow shift, is clearly the largest, representing approximately 3% of $E_{\alpha 0}$ at $T_i = 25$ keV. The extra thermal energy received at birth by the fast α-particle, $\Delta E_+$, is close to the thermal energy returned by a thermalized α-particle to the background plasma, $3T_i / 2$; at the same ion temperature, their difference is only 0.3% of $E_{\alpha 0}$.

## B. Electron and ion stopping powers

We employ the same expressions for the stopping power as Ref. [44], utilizing a slightly modified notation. Their origin is the classic textbook [45] and earlier fundamental sources [46-49]. The stopping power of the free plasma electrons and ions is given by, respectively,

$$\begin{aligned}\left(\frac{dE_\alpha}{ds}\right)_e &= -\frac{8\pi e^4 m_\alpha n \ln\Lambda_{\alpha e}}{m_e E_\alpha} G_e\left(y_e\right),\\ \left(\frac{dE_\alpha}{ds}\right)_i &= -\frac{8\pi e^4 m_\alpha \left\langle Z_i^2\right\rangle n \ln\Lambda_{\alpha i}}{m_i E_\alpha \left\langle Z_i\right\rangle} G_i\left(y_i\right),\end{aligned} \tag{16}$$

where $E_\alpha$ is the α-particle kinetic energy, $s$ is the distance traveled along its deceleration trajectory, $\ln\Lambda_{\alpha e}$ and $\ln\Lambda_{\alpha i}$ are the Coulomb logarithms for α-*e* and α-*i* collisions given by Eqs. (19) and (20), respectively. The dimensionless variables are defined as

$$y_e = \frac{m_e E_\alpha}{m_\alpha T_e}, \quad y_i = \frac{m_i E_\alpha}{m_\alpha T_i}. \tag{17}$$

The dimensionless functions $G_{e,i}\left(y\right)$ are the Maxwell integrals [48, 49],

$$G_{e,i}\left(y\right) = \operatorname{erf}\left(\sqrt{y}\right) - 2\sqrt{\frac{y}{\pi}}\left(1 + \frac{m_{e,i}}{m_\alpha}\right)\exp\left(-y\right). \tag{18}$$

For electrons, we can neglect $m_e / m_\alpha$ on the right-hand side of Eq. (18), so in the limit

$y \ll 1$, the function $G_e(y)$ simplifies to $4/\left(3\sqrt{\pi}\right)\cdot y^{3/2}$. In the opposite limit $y \gg 1$, both functions $G_{e,i}(y)$ approach unity. They are plotted in Fig. 2(a).

For the α-*e* Coulomb logarithm, we use the high-temperature formula

$$\ln\Lambda_{\alpha e} = \ln\left[\frac{T_e}{e\hbar}\left(\frac{3m_e}{\pi n}\right)^{1/2}\right] = 25.3 - \frac{1}{2}\ln\left[n\left(\mathrm{cm}^{-3}\right)\right] + \ln\left[T_e\left(\mathrm{eV}\right)\right]. \quad (19)$$

This formula assumes that the relative velocity of the α-particle and an electron is the electron thermal velocity, which is significantly higher than $4e^2/\hbar = 8.75\cdot 10^8$ cm/s. This condition is satisfied in the fusion-relevant temperature range. Note that the right-hand side of (19) is independent of the mass or charge of the ions that the electrons collide with, and can be used to calculate $\ln\Lambda_{ei}$ as well; cf. Ref. [50], p. 215. Alternatively, Sivukhin's formula [49] is sometimes used instead of Braginskii's [50]; in the high-temperature (quantum) regime, the former differs from Eq. (19) simply by replacing the constant 25.3 with 24.8. The theory presented here is unaffected by this choice. This classical theory is applicable only if the Coulomb logarithm is sufficiently large – meaning that the electron density $n$ is not too high. For high-yield MagLIF conditions, substituting $n = 5\times 10^{23}$ cm$^{-3}$ and $T_e = 20$ keV into (19) yields a reasonably large value, $\ln\Lambda_{\alpha e} = 7.9$. While NIF shot N221204 reached a density of $1.5\times 10^{25}$ cm$^{-3}$ [1], laser fusion research often explores even higher densities in the $10^{26}$– $10^{27}$ cm$^{-3}$ range. In this high-density domain, the plasma is moderately to strongly coupled, requiring the introduction of corrections to the Coulomb logarithm [51, 52].

The α-*i* Coulomb logarithm is calculated using the classical formula for counterstreaming ions in the presence of warm electrons [49],

$$\ln \Lambda_{\alpha i} = \ln\left[\left(\frac{T_e}{4\pi e^2 n}\right)^{1/2} \times \frac{m_i m_\alpha \bar{u}^2}{2\langle Z_i\rangle(m_\alpha + m_i)e^2}\right] = \ln\left[\frac{1-\mu}{2e^3\langle Z_i\rangle(2-\mu)}\left(\frac{T_e}{\pi n}\right)^{1/2} E_\alpha\right]. \quad (20)$$

Since the fast α-particle velocity is much higher than the thermal velocity of the background ions, we identify the relative velocity of the α-particle with respect to the ion, $\bar{u}$, with the α-particle velocity; hence, $m_\alpha \bar{u}^2/2 = E_\alpha$. While Eq. (20) with variable $E_\alpha$ is directly applicable to Monte Carlo simulations, calculating transport coefficients requires Coulomb logarithms averaged over the α-particle energy distribution. We define the appropriate characteristic value for $E_\alpha$ as explained below.

In Fig. 2(a), colored solid lines represent the Maxwell integrals (18) vs. $y$, the dashed lines show their asymptotics, and the symbols illustrate our parameter range. The values of $y_e$ are shown for $T_e$ varying between 10 and 25 keV, at $E_\alpha = E_{\alpha 0}$. The $y_i$ symbols range between the thermalization regime ($E_\alpha / T_i = 4.5$), and the birth energy at $T_i = 20$ keV for the background ion mass $m_i = \bar{m}_{DT}$. These symbols fall onto the asymptotic lines corresponding to $y_e \ll 1$ and $y_i \gg 1$, respectively, consistent with the ordering

$$v_{Ti} \ll u \ll v_{Te}, \quad (21)$$

where the thermal velocities are defined as $v_{Te} = \sqrt{T_e / m_e}$ and $v_{Ti} = \sqrt{T_i / m_i}$.

Substituting the low-$y_e$ asymptotic expression (18) into (16) yields the well-known formula for α-particle deceleration in collisions with electrons:

$$\left(\frac{du}{dt}\right)_e = -\nu_{\alpha e} u, \quad (22)$$

where the *α-e* Coulomb collision frequency is given by

$$\nu_{\alpha e} = \frac{1}{\tau_{\alpha e}} = \frac{16e^4\sqrt{2\pi m_e} n \ln\Lambda_{\alpha e}}{3m_\alpha T_e^{3/2}}. \tag{23}$$

The characteristic time $\tau_{\alpha e}$ is of the same order as the electron-ion equilibration time, $\tau_{ei}^{\varepsilon}$, and much longer than the electron-electron and ion-ion equilibration times: $\tau_{\alpha e} \sim \tau_{ei}^{\varepsilon} \sim (m_i / m_e)^{1/2} \tau_{ii} \sim (m_i / m_e)\tau_{ee}$ (cf. §3 of Ref. [50]). This ordering justifies treating electrons and ions as distinct Maxwellian fluids at separate temperatures, $T_e$ and $T_i$, without requiring the background plasma to be isothermal.

As a fast α-particle moves through the electron fluid slowly compared to the electron thermal velocity, the Coulomb collision cross-section is determined by the electron temperature and varies as $T_e^{-2}$. Conversely, substituting $G_i(y_i) = 1$ into (16), we find the deceleration due to α−*i* collisions. To write this equation in a compact form, we introduce the characteristic velocity $u_c$, at which the ion friction equals the electron friction:

$$u_c = \left[\frac{3\sqrt{\pi}\langle Z_i^2\rangle \ln\Lambda_{\alpha i}}{4(1-\mu)\langle Z_i\rangle \ln\Lambda_{\alpha e}}\right]^{1/3} \left(\frac{m_\alpha}{m_e}\right)^{1/6} \left(\frac{T_e}{E_{\alpha 0}}\right)^{1/2} u_{\alpha 0}; \tag{24}$$

this reproduces Eq. (1.23) of Ref. [25] in our notation. The ratio $\varepsilon = u_c / u_{\alpha 0}$ is an important parameter in the theory of fast α-particle transport. The deceleration of a fast α-particle due to ion friction is

$$\left(\frac{du}{dt}\right)_i = -\nu_{\alpha e}\frac{u_c^3}{u^2}. \tag{25}$$

Equation (25) implies that the background ions are stationary scattering centers. The collision cross-section is determined by the α-particle energy only (proportional to $E_\alpha^{-2}$), resulting in a “hard stop” as a particle loses energy. Electron and ion friction dominate at $u > u_c$ and $u < u_c$, respectively. Consequently, the kinetic energy $E_{\alpha c} = m_\alpha u_c^2 / 2$ is the appropriate average value of $E_\alpha$ to substitute into the formula (20) for α-*i* Coulomb logarithm. By combining Eqs. (20) and (24), the ratio of Coulomb logarithms in (24) can be expressed self-consistently as:

$$\frac{\ln\Lambda_{\alpha i}}{\ln\Lambda_{\alpha e}} = -\frac{2}{3\ln\Lambda_{\alpha e}} W_{-1}\left( -\frac{2^{5/2}\pi^{1/4}(2-\mu)^{3/2}\langle Z_i\rangle^{5/2}\ln\Lambda_{\alpha e}}{(1-\mu)^{1/2}\langle Z_i^2\rangle}\left(\frac{m_e}{m_\alpha}\right)^{1/2}\frac{e^{9/2}n^{3/4}}{T_e^{9/4}}\right), \tag{26}$$

where $W_{-1}(z)$ is the Lambert W-function [53]. This ratio is a slowly varying function of the background plasma parameters, and in the high-gain MagLIF range it stays within 2% of its average value, $\ln\Lambda_{\alpha i} / \ln\Lambda_{\alpha e} = 1.64$. Substituting it into Eq. (24) yields the approximation:

$$\varepsilon \approx \left[\frac{\langle Z_i^2\rangle}{(1-\mu)\langle Z_i\rangle}\right]^{1/3}\left(\frac{T_e}{107\ \text{keV}}\right)^{1/2}. \tag{27}$$

Note that the classical formula (20) applies for relative velocities below the boundary value derived from Sivukhin's condition (5.14) [49], which for $Z_\alpha = 2$ gives $u_{bound} = 5\times10^8$ cm/s. Identifying this limit with $u_c$ corresponds to $\varepsilon \approx 0.4$, which translates into $T_e \approx 12.5$ keV in equimolar DT. At higher electron temperatures, a quantum-mechanical Coulomb logarithm, such as in Eq. (3a) of Ref. [36], is more appropriate. For simplicity, we retain the classical formula valid near ignition; adopting the quantum expression at higher temperatures would not alter the present theoretical framework, but only modify the functional dependence of $\varepsilon$ on plasma parameters, which remains determined self-consistently as described above.

The approximation $\ln\Lambda_{\alpha i} \approx \ln\Lambda_{\alpha e}$ sometimes used in the literature, including Refs. [31, 39], significantly underestimates the value of $\varepsilon$. The value of $\varepsilon$ obtained from Eq. (24) is typically less than unity but it is not exceedingly small, varying in the range between 0.25 and 0.5 as the electron temperature increases from 5 to 20 keV. The power of this parameter $\varepsilon^3$ is denoted by $\varepsilon$ and $\alpha$ in Refs. [31] and [39], respectively. Our present notation is preferable because the low-temperature expansion of the transport coefficients in Section III is in powers of small $\varepsilon$, not $\varepsilon^3$. (Equation (3) represents the lowest order of this expansion, which is independent of $\varepsilon$, and, therefore, not affected by the assumption that $\ln\Lambda_{\alpha i} \approx \ln\Lambda_{\alpha e}$.) Increase of the parameter $\varepsilon$ implies a reduced relative contribution of the electrons to the dynamic friction because the collision frequency (23) scales as $\varepsilon^{-3}$. Figure 2(b) illustrates that $\varepsilon$ is primarily a function of the electron temperature, proportional to $\sqrt{T_e}$, while its dependence on other plasma parameters is relatively weak.

### C. Fast α-particle deceleration

Combining Eqs. (22) and (25), we arrive at the equation of motion of a fast α-particle decelerating in a uniform plasma at rest with a constant magnetic field $\mathbf{B}$:

$$\frac{d\mathbf{u}}{dt} = -\nu_{\alpha e}\mathbf{u}\left(1+\frac{u_c^3}{u^3}\right)+\omega_\alpha \mathbf{u}\times\mathbf{b}\,, \tag{28}$$

where $\mathbf{b}=\mathbf{B}/B$ is the unit vector in the direction of the magnetic field and $u=|\mathbf{u}|$. The first term on the right-hand side of Eq. (28) describes α-particle deceleration and energy loss via dynamic friction, while the second term accounts for gyromotion. Because fast α-α collisions are neglected under the assumption $n_\alpha \ll n$ (justified below), collective transport is governed by test-particle dynamics; consequently, as demonstrated below, this single-particle equation of motion directly yields key transport properties – including characteristic deceleration scales, qualitative magnetic suppression features, and the equilibrium quantities (such as number and energy densities) – prior to solving the kinetic equation.

The deceleration time of the α-particle from its initial velocity $u = u_{\alpha 0}$ to a full stop, $u = 0$, is found from Eq. (28):

$$\tau_{\rm dec} = \tau_{\alpha e} \int_0^{u_{\alpha 0}} \frac{u^2 du}{u^3 + u_c^3} = \frac{1}{3} \tau_{\alpha e} \ln\left(1 + \varepsilon^{-3}\right). \tag{29}$$

In the limit $\varepsilon \to 0$, the deceleration time (29) diverges logarithmically because the α-particle slows down exponentially without ever coming to a complete rest. However, the deceleration path remains finite even in this low-temperature limit:

$$l_\alpha = \tau_{\alpha e} \int_0^{u_{\alpha 0}} \frac{u^3 du}{u^3 + u_c^3} = l_{\alpha e} \left(1 - \frac{\varepsilon}{6} L_+\right), \tag{30}$$

where $l_{\alpha e} = u_{\alpha 0} \tau_{\alpha e}$ is the deceleration path determined by the α-*e* collisions only, and $L_+$ is one of the positive-definite auxiliary functions defined as

$$L_\pm(\varepsilon) = 2\sqrt{3}\left[\arctan\left(\sqrt{3}\frac{1-\varepsilon}{1+\varepsilon}\right) + \frac{\pi}{3}\right] \pm \ln\left[\frac{(1+\varepsilon)^3}{1+\varepsilon^3}\right]. \tag{31}$$

Figure 3(a) presents $l_\alpha$ and $l_{\alpha e}$ vs. the electron temperature $T_e$ for $n = 10^{24}$ cm$^{-3}$ in a log-log scale. The temperature dependence of $l_{\alpha e}$ determined by Eq. (23) is close to a power law but with an index 1.35, less than 3/2, due to the Coulomb logarithm (19). At $T_e = 25$ keV, we find: $l_{\alpha e} = 4.2$ mm, $l_\alpha = 1.7$ mm, smaller by a factor of 2.5 determined by the expression in parentheses in the right-hand side of Eq. (30).

To evaluate the efficacy of magnetic confinement, we identify single-particle analogs of particle and energy fluxes: $\mathbf{u}$ and $\left(m_\alpha u^2 / 2\right)\mathbf{u}$, respectively. Integrating the velocity $\mathbf{u}$ over time, we obtain the α-particle displacement from its point of birth after it has lost its kinetic energy: $\Delta\mathbf{r}_p = \int \mathbf{u} dt$. Similarly, the energy displacement is defined by $\Delta\mathbf{r}_E = \left(1/E_{\alpha 0}\right)\int\left(m_\alpha u^2/2\right)\mathbf{u} dt$, indicating the average location where the energy has been dissipated. As this integrand scales as the α-particle velocity cubed, the energy displacement is smaller than the particle displacement.

We assume that the magnetic field $\mathbf{B}$ is in the *z*-direction, the α-particle is born at the origin, $x = y = 0$, with its velocity $u_{\alpha 0}$ in the *x*-direction, and integrate Eq. (28) until it reaches the end of its deceleration path: at $t \to \infty$ if only the electron friction is accounted for ($\varepsilon = 0$), or at $t = \tau_{\text{dec}}$ (29) for finite $\varepsilon$. In the former case, the integration is done analytically, resulting in the displacement of the α-particle,

$$\Delta x_p = \frac{1}{1+\chi_\alpha^2} l_{\alpha e}, \quad \Delta y_p = -\frac{\chi_\alpha}{1+\chi_\alpha^2} l_{\alpha e}. \tag{32}$$

and of its energy:

$$\Delta x_E = \frac{3}{9+\chi_\alpha^2} l_{\alpha e}, \quad \Delta y_E = -\frac{\chi_\alpha}{9+\chi_\alpha^2} l_{\alpha e}. \tag{33}$$

As the Hall parameter $\chi_\alpha$ is increased from 0 to infinity, the stopping point (32) moves from $\Delta x_p = l_{\alpha e}$, $\Delta y_p = 0$ (its position at $\chi_\alpha = 0$) back towards the origin along the semicircle with the diameter $l_{\alpha e}$ in the lower half-plane $(x, y)$. Similarly, the locus of points (33) is a semicircle whose diameter is $l_{\alpha e} / 3$. These two semicircles are shown in Fig. 3(b) as black lines labeled “Electrons only”, specified by subscripts *p* and *E*, respectively, with the arrows indicating the direction of increased $\chi_\alpha$.

When the ion friction is taken into account, the displacements are found numerically. The results of the numerical solution for $T_e = 25\,\text{keV}$ are shown in Fig. 3(b) as a red line, with $\Delta x_p$ and $\Delta y_p$ normalized with respect to $l_\alpha$ given by Eq. (30). In contrast with the semicircle (32), the red line is not monotonic. It approaches the origin in an oscillatory manner, featuring a sequence of narrowing loops caused by the "hard stop" of the α-particle – a result of ion friction – making the stopping position sensitive to the phase of the α-particle's gyration.

Figures 3(c) and 3(d) plot, in the log scale, the particle displacements, $\left|\Delta x_p\right|$ and $-\Delta y_p$, and the energy displacements, $\Delta x_E$ and $-\Delta y_E$, respectively, in units of $l_{\alpha e}$, vs. the Hall parameter $\chi_\alpha$. We plot the absolute value of $\Delta x_p$, the only displacement that changes sign as a function of $\chi_\alpha$, as illustrated in Fig. 3(b). The results for the low-temperature limit $\varepsilon \to 0$, and for finite electron temperatures $T_e = 5$ and 25 keV, are plotted as black, blue, and red lines, respectively. In the electron-only friction limit described by Eqs. (32) and (33), the perpendicular displacements, $\left|\Delta x_p\right|$ and $\Delta x_E$, and cross-gradient displacements, $-\Delta y_p$ and $-\Delta y_E$, depend monotonically on $\chi_\alpha$, as represented by the black solid and dashed lines, respectively. Correspondingly, the colored curves follow the same general trends but exhibit oscillations, which are especially pronounced for $\Delta x_p$, and reflect the loop structure of Fig. 3(b). The energy displacement exhibits small oscillations and is very well approximated by Eq. (33), even at high temperatures, except at low magnetization.

With the aid of Eq. (28), we also calculate the partition of the dissipated α-particle energy between the electrons and ions. Electron friction, Eq. (22), produces the differential work $dE_{\alpha e} = -m_\alpha \nu_{\alpha e} u^2 dt$, where $dt$ is expressed via $du$ from Eq. (29). Consequently, the energy partition fractions to plasma electrons and ions, $F_e$ and $F_i$, are

$$F_e = \frac{1}{E_{\alpha 0}} \int dE_{\alpha e} = \frac{2}{u_{\alpha 0}^2} \int_0^{u_{\alpha 0}} \frac{u^4 du}{u^3 + u_c^3} = 1 - \frac{\varepsilon^2}{3} L_- \equiv 1 - F_i , \tag{34}$$

where $L_-$ is defined by Eq. (31).

## D. Kinetic equation and balance equations for fast α-particles

To describe the equilibrium and transport of fast α-particles, we employ the kinetic equation written in the laboratory reference frame. The fast α-particle velocity is presented as

$$\mathbf{w} = \mathbf{v} + \mathbf{u}, \tag{35}$$

where, as before, $\mathbf{u}$ denotes the α-particle velocity with respect to the background plasma, and $\mathbf{v}$ is the plasma hydrodynamic velocity. The kinetic equation is

$$\frac{\partial f_\alpha}{\partial t} + \mathbf{w}\nabla_{\mathbf{r}} f_\alpha + \left[\frac{2e\mathbf{E}}{m_\alpha} + \omega_\alpha \mathbf{w}\times\mathbf{b}\right]\nabla_{\mathbf{w}} f_\alpha = \mathrm{St}\left(f_\alpha\right) + \Sigma_\alpha\left(\mathbf{u}\right). \tag{36}$$

Here, $f_\alpha\left(\mathbf{w},\mathbf{r},t\right)$ is the fast α-particle distribution function, $\mathbf{E}$ is the electric field in the laboratory reference frame, and $\Sigma_\alpha\left(\mathbf{u}\right)$ is the fusion source. If this source is isotropic, then it reduces to (11), but we do not require this at the moment. We assume merely that this source function depends only on the relative velocity $\mathbf{u}$ of the α-particle and the background plasma, and that it has central symmetry, which means that the fusion production of fast α-particles does not impart momentum in any direction.

We conventionally introduce the moments of the distribution function:

$$n_\alpha\left(\mathbf{r},t\right) = \int d^3w\, f_\alpha\left(\mathbf{w},\mathbf{r},t\right), \tag{37}$$

$$\mathcal{E}_\alpha\left(\mathbf{r},t\right) = \int d^3w\, \frac{1}{2} m_\alpha u^2 f_\alpha\left(\mathbf{w},\mathbf{r},t\right), \tag{38}$$

$$\mathbf{J}_\alpha\left(\mathbf{r},t\right) = \int d^3w\, \mathbf{u} f_\alpha\left(\mathbf{w},\mathbf{r},t\right), \tag{39}$$

$$\mathbf{j}_\alpha\left(\mathbf{r},t\right) = 2e\mathbf{J}_\alpha\left(\mathbf{r},t\right), \tag{40}$$

$$\mathbf{q}_\alpha\left(\mathbf{r},t\right) = \int d^3w\, \frac{1}{2} m_\alpha u^2 \mathbf{u} f_\alpha\left(\mathbf{w},\mathbf{r},t\right), \tag{41}$$

$$\mathsf{P}_\alpha(\mathbf{r},t)=\int d^3w\, m_\alpha \mathbf{uu} f_\alpha(\mathbf{w},\mathbf{r},t). \quad (42)$$

The scalar moments $n_\alpha$ and $\mathcal{E}_\alpha$ denote the number density and energy density of the fast α-particles, respectively. The vector moments $\mathbf{J}_\alpha$ and $\mathbf{q}_\alpha$ represent the particle and energy fluxes of fast α-particles, respectively, defined relative to the background plasma flow; $\mathbf{j}_\alpha$ is the current density associated with the α-particle flux. The tensor moment $\mathsf{P}_\alpha$ is the stress tensor associated with the fast α-particle motion relative to the background plasma; the dyad $\mathbf{uu}$ in the integrand denotes a tensor product.

Multiplying the kinetic equation (36) by 1, $m_\alpha \mathbf{w}$, and $m_\alpha w^2/2$, and integrating over α-particle velocities, we obtain the balance equations for the α-particle number, momentum, and energy density. These equations can be reduced to the following form:

$$\frac{dn_\alpha}{dt}+n_\alpha \nabla\cdot\mathbf{v}+\nabla\cdot\mathbf{J}_\alpha=-S_\alpha+\dot{n}_\alpha, \quad (43)$$

$$\begin{aligned} &m_\alpha n_\alpha \frac{d\mathbf{v}}{dt}+m_\alpha\left[\frac{d\mathbf{J}_\alpha}{dt}+(\nabla\cdot\mathbf{v})\mathbf{J}_\alpha+(\mathbf{J}_\alpha\nabla)\mathbf{v}\right] \\ &+\nabla\cdot\mathsf{P}_\alpha-\frac{1}{c}\mathbf{j}_\alpha\times\mathbf{B}-2en_\alpha\mathbf{E}^*=-\mathbf{R}_\alpha, \end{aligned} \quad (44)$$

$$\frac{d\mathcal{E}_\alpha}{dt}+\mathcal{E}_\alpha\nabla\cdot\mathbf{v}+(\mathsf{P}_\alpha\nabla)\mathbf{v}+\nabla\cdot\mathbf{q}_\alpha=-Q_\alpha+\dot{n}_\alpha\left(E_{\alpha0}+\Delta E_+\right)+\mathbf{j}_\alpha\mathbf{E}^*-m_\alpha\mathbf{J}_\alpha\frac{d\mathbf{v}}{dt}, \quad (45)$$

where $d/dt=\partial/\partial t+\mathbf{v}\nabla$ is the Lagrangian time derivative. Here and below, $\nabla$ without a subscript designates $\nabla_{\mathbf{r}}$, and we have denoted the electric field in the reference frame of the background plasma by

$$\mathbf{E}^*=\mathbf{E}+\frac{1}{c}\mathbf{v}\times\mathbf{B}. \quad (46)$$

In Eqs. (43)-(45), the terms describing the collisional coupling of the fast α-particle population with the background plasma are defined as

$$S_\alpha(\mathbf{r},t) = -\int d^3w \,\mathrm{St}(f_\alpha), \tag{47}$$

$$\mathbf{R}_\alpha(\mathbf{r},t) = -\int d^3w \, m_\alpha \mathbf{u} \mathrm{St}(f_\alpha), \tag{48}$$

$$Q_\alpha(\mathbf{r},t) = -\int d^3w \, \frac{1}{2} m_\alpha u^2 \mathrm{St}(f_\alpha). \tag{49}$$

The positive-definite scalar terms $S_\alpha$ and $Q_\alpha$ represent, respectively, the rate of helium ash production due to the friction that removes α-particles from the "fast" category, and the α-heating power generated by this deceleration. The vector $\mathbf{R}_\alpha$ denotes the force exerted upon the background plasma because of momentum imbalance associated with α-particle deceleration.

Equations (43)-(45) are exact, first-principles balance equations derived directly from the kinetic equation (36). They require no assumptions about the isotropy of the background plasma or the α-particle distribution, the form of the collision operator, or the properties of the fusion source beyond local central symmetry. Note that in a resting plasma ( $\mathbf{v} = 0$ ), if we omit the α-Ohmic heating term $\mathbf{j}_\alpha \cdot \mathbf{E}^*$ , which is insignificant for most conditions of interest, and neglect the small thermal addition $\Delta E_+$ (13) to the α-particle birth energy $E_{\alpha 0}$ , the exact energy balance equation (45) formally reduces to (2). However, the system of balance equations (43)-(45) is clearly not closed, as it involves more unknowns than equations. Usable expressions for the α-particle energy flux $\mathbf{q}_\alpha$ , the α-heating term $Q_\alpha$ , and other moments and coupling terms can only be obtained by solving the kinetic equation (36).

To construct our diffusion model, we use the Fokker-Planck collision operator, simplified by using the ordering (21):

$$\mathrm{St}\left(f_\alpha\right)=\nu_{\alpha e}\left\{\frac{1}{u^2}\frac{\partial}{\partial u}\left[\left(u^3+u_c^3\right)f_\alpha\right]+\frac{(1-\mu)u_c^3}{2u^3}\nabla_\Omega^2 f_\alpha+\frac{T_e}{m_\alpha u^2}\left[\frac{\partial}{\partial u}\left(u^2\frac{\partial f_\alpha}{\partial u}\right)+\nabla_\Omega^2 f_\alpha\right]\right\}, \quad (50)$$

where

$$\nabla_\Omega^2 f_\alpha=\frac{1}{\sin\theta}\frac{\partial}{\partial\theta}\left(\sin\theta\frac{\partial f_\alpha}{\partial\theta}\right)+\frac{1}{\sin^2\theta}\frac{\partial^2 f_\alpha}{\partial\varphi^2} \quad (51)$$

is the angular part of the Laplace operator (the Legendre operator, written here in spherical coordinates). As the fusion source term in Eq. (36), the operator (50) depends only on the relative velocity $\mathbf{u}$ of the α-particles and the background plasma.

The collisional operator (50) is linear in the distribution function $f_\alpha$, which means that it does not account for the interaction between fast α-particles – their transport is fully determined by the background plasma that produces and decelerates them. The first term in curly brackets in Eq. (50) describes slowing down of the fast α-particles determined by the first term in the right-hand side of the equation of motion (28). The second term accounts for the transverse spreading of the distribution function due to α-particles' pitch-angle scattering in collisions with plasma ions, which are presumed stationary in our approximation. The isotropic part of the third term represents the straggling, or longitudinal diffusive broadening of the distribution function caused by collisions with electrons. For example, if a monoenergetic isotropic beam of fast α-particles is injected into the plasma, as in the stochastic simulations of Ref. [54], which demonstrate the equivalence between the Fokker-Planck operator and the Langevin dynamics of individual test particles, the friction term slows down the beam while keeping it sharply peaked, whereas the diffusion term broadens its width.

## E. Equilibrium distribution of fast α-particles

### *1. Equilibrium distribution function*

Now let us assume that the background plasma can, to a first approximation, be considered uniform and at rest in its own reference frame: $\mathbf{v} = 0$. These are necessary conditions for using the Chapman-Enskog approximation. To employ this approximation, we begin with finding the quasi-equilibrium distribution function of fast α-particles. Of course, fast α-particles produced by the fusion source (1) are inherently non-equilibrium species relative to the background plasma. Their thermalization involves plasma heating and compositional changes that prevent the attainment of a true equilibrium.

However, the constant injection of α-particles at energies near $E_{\alpha 0}$ balanced by their energy loss establishes a quasi-steady distribution on the time scale of $\tau_{\mathrm{dec}}$, a behavior well-documented in simulations such as those in Ref. [55]. While fusion heating gradually changes the burn-up fraction and plasma parameters, this evolution proceeds on a timescale of $n / \dot{n}_\alpha$ that is large compared to the α-particle thermalization time given by Eq. (29). This ordering, expressed by the inequality $n \gg n_\alpha \sim \dot{n}_\alpha \tau_{\alpha e}$, was utilized in deriving Eq. (4) and is justified in this Section.

The equilibrium distribution function is conventionally derived from the Fokker-Planck equation. To elucidate its physical meaning, we obtain a very accurate approximation to it directly from the single-particle equation of motion (28) and the source spectrum (11). We assume the equilibrium distribution function to be isotropic in velocity space, although the background plasma is not necessarily isotropic; it can be strongly magnetized. The key conditions here are the isotropy of the fusion source (11), and the fact that the magnetic field affects only the shape of the α-particle path, not its deceleration along the path. Therefore, the equilibrium distribution function sought for depends on the absolute value $u$ of the α-particle velocity $\mathbf{u}$, not on its direction. The equation for $u$ stemming from Eq. (28) is

$$\frac{du}{dt} = -\nu_{\alpha e} u \left(1 + \frac{u_c^3}{u^3}\right). \tag{52}$$

Consider an infinitesimal spherical shell of radius $u$ in velocity space. Its surface area is $4\pi u^2$, thickness is $du$, and volume is $4\pi u^2 du$. The rate of fast α-particle production within this volume element is $4\pi u^2 \Sigma_\alpha(u) du$, where $\Sigma_\alpha(u)$ is given by Eq. (11). We calculate the inflow rate of the decelerating α-particles through the outer surface of the shell noting that the flux density in velocity space is $-f_{\alpha,\text{eq}}(du/dt)$. Hence, the inflow rate $-4\pi u^2 f_{\alpha,\text{eq}}(du/dt)$ with the aid of Eq. (52) is reduced to $4\pi\nu_{\alpha e}(u^3 + u_c^3) f_{\alpha,\text{eq}}$.

The differential of the latter quantity represents the difference between the inflow through the outer surface and the outflow through the inner surface of the infinitesimal shell, which in equilibrium is balanced by the fast α-particle production within its volume. The balance between the fusion source and the friction sink is, therefore, expressed by the equation

$$\frac{d}{du}\left[\left(u^3 + u_c^3\right) f_{\alpha,\text{eq}}\right] = \frac{\dot{n}_\alpha \tau_{\alpha e}}{(2\pi)^{3/2}} \frac{u}{u_{\alpha 0}} \left(\frac{M}{T_i}\right)^{1/2} \exp\left[-\frac{M\left(u^2 + u_{\alpha 0}^2\right)}{2T_i}\right] \sinh\left(\frac{M u u_{\alpha 0}}{T_i}\right). \tag{53}$$

Equation (53) is immediately integrated. For convenience, we omit the dimensional normalization factor $\dot{n}_\alpha \tau_{\alpha e} / u_{\alpha 0}^3$ of the distribution function $f_{\alpha,\text{eq}}$, denote the normalized ion temperature by

$$\theta_i = \frac{m_\alpha T_i}{\left(m_D + m_T\right) E_{\alpha 0}}, \tag{54}$$

and introduce the normalized α-particle velocity $z = u / u_{\alpha 0}$ (not to be confused with the Cartesian coordinate). Then the normalized distribution function satisfying Eq. (53) can with exponential accuracy be presented as

$$f_{\alpha,\mathrm{eq}}^{(0)}(z) = \frac{1}{8\pi\left(z^3+\varepsilon^3\right)}\left\{1-\mathrm{erf}\left(\frac{z-1}{\sqrt{\theta_i}}\right)+\sqrt{\frac{\theta_i}{\pi}}\exp\left[-\frac{(z-1)^2}{\theta_i}\right]\right\}. \tag{55}$$

The superscript (0) in Eq. (55) indicates that this solution represents the first approximation, not accounting for the electron scattering contribution, the third term in curly brackets in the collisional operator (50). The distribution function (55) is a slightly more accurate version of Eq. (39) in Ref. [39].

The ordering (21) implies that the parameter $\theta_i$ is small, varying between 0.0011 and 0.0057 as the ion temperature increases from 5 to 25 keV. At $\theta_i \ll 1$, the distribution (55) can be further simplified. Neglecting terms of $O(\sqrt{\theta_i})$, we reduce Eq. (55) to the approximate distribution, which we also write down in dimensional form:

$$f_{\alpha,\mathrm{eq}}^{(0)}(u) = \frac{\dot{n}_\alpha \tau_{\alpha e}}{u_{\alpha 0}^3} f_{\alpha,\mathrm{eq}}^{(0)}(z) = \frac{\dot{n}_\alpha \tau_{\alpha e}}{u_{\alpha 0}^3} \times \frac{\Theta(1-z)}{4\pi\left(z^3+\varepsilon^3\right)} = \frac{\dot{n}_\alpha \tau_{\alpha e}\Theta\left(u_{\alpha 0}-u\right)}{4\pi\left(u^3+u_c^3\right)}, \tag{56}$$

where $\Theta(z)$ is the Heaviside step function. Equation (56) represents the equilibrium distribution established with a monoenergetic source of fast α-particles, the well-known isotropic classical slowing-down distribution expressed by Eqs. (1.34), (12b), (51), and (28) in Refs. [29, 31, 42, 56], respectively.

To assess the accuracy of the approximation (55), we write down the kinetic equation for the normalized equilibrium distribution function using the full collisional operator (50):

$$\frac{d}{dz}\left[\left(z^3+\varepsilon^3\right)f_{\alpha,\mathrm{eq}}+\theta_e z^2\frac{df_{\alpha,\mathrm{eq}}}{dz}\right]=-\frac{z}{2\pi^{3/2}\sqrt{\theta_i}}\exp\left(-\frac{z^2+1}{\theta_i}\right)\sinh\left(\frac{2z}{\theta_i}\right), \tag{57}$$

where we have denoted

$$\theta_e=\frac{T_e}{2E_{\alpha 0}}, \tag{58}$$

which is also a small parameter. However, because this parameter appears as a coefficient of a higher-order derivative of $f_{\alpha,\mathrm{eq}}$, this term cannot be neglected solely due to the smallness of $\theta_e$.

The appropriate boundary condition for Eq. (57) is

$$f_{\alpha,\mathrm{eq}}\left(z\right)\to 0 \text{ at } z\to\infty \tag{59}$$

at least as fast as the fusion source function (11), $\sim\exp\left[-(z-1)^2/\theta_i\right]$. This boundary condition excludes the solutions of the homogeneous Eq. (57) making its left-hand side vanish. We use the smallness of $\theta_e$ to derive a perturbative analytic solution of Eq. (57), seeking it as a series in powers of this parameter:

$$f_{\alpha,\mathrm{eq}}\left(z\right)=f_{\alpha,eq}^{(0)}\left(z\right)+\theta_e f_{\alpha,\mathrm{eq}}^{(1)}\left(z\right)+\theta_e^2 f_{\alpha,\mathrm{eq}}^{(2)}\left(z\right)+\ldots \tag{60}$$

Here, the zero-order solution is given by Eq. (56). The higher-order terms in the expansion (60), which describe the effect of the diffusive broadening in velocity space, are determined from the requirement that the left-hand side of Eq. (57) remain unchanged in all orders:

$$f_{\alpha,\mathrm{eq}}^{(k)}\left(z\right)=-\frac{z^2}{z^3+\varepsilon^3}\frac{df_{\alpha,\mathrm{eq}}^{(k-1)}}{dz},\quad k=1,\ 2,\ \ldots\ . \tag{61}$$

The boundary condition (59) is automatically satisfied by (61) for all orders due to the fast exponential decay of the zero-order term (55). The series (60) rapidly converges. The results of these calculations are shown in Fig. 4.

Figure 4(a) plots the equilibrium distribution functions $f_{\alpha,\mathrm{eq}}(u)$ in physical units of cm$^{-6}$s$^{3}$ calculated from Eq. (57), in the log scale, vs. the normalized fast α-particle velocity, $z = u / u_{\alpha 0}$, for the number density $n = 5 \cdot 10^{23}$ cm$^{-3}$ and the temperatures $T_e = T_i = 10$, 15, 25, and 50 keV (symbols), and the respective approximate solutions (55) (lines). Figure 4(b) shows the same distribution functions normalized with respect to the fusion rate factor, $\dot{n}_\alpha \tau_{\alpha e} / u_{\alpha 0}^3$, in the linear scale. It illustrates how the shape of the distribution function changes with increased electron temperature, becoming flatter due to the increased relative role of the ion friction. Both figures demonstrate that Eq. (55) provides an excellent approximation to the exact solution of Eq. (57). Figure 4(c) evaluates the accuracy of this approximation, plotting the difference $\delta f_{\alpha,\mathrm{eq}} = f_{\alpha,\mathrm{eq}} - f_{\alpha,\mathrm{eq}}^{(0)}$ vs. the normalized α-particle velocity $z$. The difference is small for all temperatures, at all α-particle velocities, peaking where the variation of $f_{\alpha,\mathrm{eq}}$ is the fastest: near the birth velocity, $z \approx 1$, and at low α-particles velocities, the latter being prominent at low temperatures. Figure 4(d) illustrates the deviation of $f_{\alpha,\mathrm{eq}}$ (solid light blue line) and $f_{\alpha,\mathrm{eq}}^{(0)}$ (dotted black line) with the inverse-cubic approximation (56) to $f_{\alpha,\mathrm{eq}}$ (solid black line) for $T_e = T_i = 25$ keV. The deviation is only noticeable within the width of the fusion source (11), shown as a red line.

*2. Equilibrium quantities*

The condition $\mathbf{v} = 0$ implies that the integration over velocity space in Eqs. (37)-(39) reduces to integration over relative α-particle velocities, $d^3 u$. When the integrand is a function of its absolute value, $u$, the integration over solid angle denoted by $\langle ... \rangle_\Omega$ yields

$$\langle u_i \rangle_\Omega = 0, \quad \langle u_i u_j \rangle_\Omega = \frac{1}{3} u^2 \delta_{ij} , \tag{62}$$

where the indexes $i, j = 1,\ 2$, and 3 correspond to $x$, $y$, and $z$ coordinates, respectively.

We begin with calculating the scalar moments. Substituting the distribution function (55) into Eq. (37), we obtain the equilibrium number density of fast α-particles:

$$n_{\alpha,\mathrm{eq}} = \dot{n}_\alpha \tau_{\alpha e} \times 4\pi \int_0^\infty z^2 f_{\alpha,\mathrm{eq}}(z)\, dz = \dot{n}_\alpha \tau_{\alpha e} \left[ \frac{1}{3} \ln\left(1+\varepsilon^{-3}\right) + \frac{1+4\varepsilon^3}{4\left(1+\varepsilon^3\right)^2} \theta_i + O\left(\theta_i^2\right) \right]. \quad (63)$$

The leading term here reproduces Eq. (42) of Ref. [39]. It is consistent with the formula (29) for deceleration time: in equilibrium, $n_{\alpha,\mathrm{eq}} = \dot{n}_\alpha \tau_{\mathrm{dec}}$. Corrections due to the finite width of the fusion source (11) to the integral (63) are small, less than 1%, and will be neglected henceforth. With a finite $\varepsilon$, the value of $n_{\alpha,\mathrm{eq}}$ is also finite, even though we have not introduced a thermalization sink term into the Fokker-Planck collision operator (50), as done in Refs. [39, 42]. Such a sink is often used because the collisional operator (50) does not describe thermalization; the ordering (21) assumes background ions are much slower than α-particles, which fails near thermalization. Instead of a sink, we assume that the inverse cubic equilibrium distribution (56) remains valid down to $u = 0$. The error of this approximation is the difference between the additional thermal energy $\Delta E_+$ (13) that the α-particle receives at birth and the thermal energy $3T_i / 2$ that it returns to the background plasma at thermalization. Our model neglects this difference, which is less than 0.3% of the released fusion energy $E_{\alpha 0}$ (cf. Fig. 1), an approximation that is acceptable within the overall uncertainty of our diffusion model.

The ratio of the equilibrium fast α-particle number density (63) to the electron number density in the background plasma is found using Eqs. (1) and (4):

$$\frac{n_{\alpha,\mathrm{eq}}}{n} = \frac{(1-\phi)^2 \langle \sigma v \rangle_{DT} m_\alpha T_e^{3/2}}{64 (2\pi m_e)^{1/2} e^4 \ln \Lambda_{\alpha e}} \ln\left(1+\varepsilon^{-3}\right). \quad (64)$$

The ratio (64) is primarily a function of the plasma temperature and composition. It is plotted in Fig. 5(a) vs. $T_e$, assuming $\phi = 0$, $T_i = T_e$, $n = 5\times10^{23}$ cm$^{-3}$, and using Bosch and Hale's fit [43] for $\langle\sigma v\rangle_{DT}$. Consistently with (4), its value is small, $n_\alpha \ll n$, in the MagLIF-relevant temperature range.

Similarly, we calculate the equilibrium energy density (38) of fast α-particles:

$$\mathcal{E}_{\alpha,\mathrm{eq}} = \dot{n}_\alpha \tau_{\alpha e} \times 4\pi \int_0^\infty z^4 f_{\alpha,\mathrm{eq}}(z) \frac{m_\alpha u_{\alpha 0}^2}{2} dz = \frac{1}{2} \dot{n}_\alpha \tau_{\alpha e} E_{\alpha 0} \left[ F_e + \frac{3(1+2\varepsilon^3)}{2(1+\varepsilon^3)^2} \theta_i + O(\theta_i^2) \right], \quad (65)$$

where the energy partition fraction to plasma electrons $F_e$ is given by Eq. (34). As in (63), the correction due to the finite width of the fusion source in (55) is small and will be neglected below.

The anisotropic vector moments, the fluxes $\mathbf{J}_\alpha$ and $\mathbf{q}_\alpha$, and the current density $\mathbf{j}_\alpha$, vanish in the equilibrium, which follows from Eq. (62). As for the stress tensor (42), Eq. (62) demonstrates that it reduces to an isotropic α-pressure characterized by the scalar value $P_{\alpha,\mathrm{eq}} = 2\mathcal{E}_{\alpha,\mathrm{eq}}/3$:

$$\begin{aligned} \mathsf{P}_{\alpha,\mathrm{eq},ij} &= \int d^3u \times m_\alpha f_{\alpha,\mathrm{eq}}(u) u_i u_j = \int_0^\infty 4\pi u^2 du \times m_\alpha f_{\alpha,\mathrm{eq}}(u) \langle u_i u_j \rangle_\Omega \\ &= \frac{1}{3}\delta_{ij} \int_0^\infty 4\pi u^4 du \times m_\alpha f_{\alpha,\mathrm{eq}}(u) \equiv P_{\alpha,\mathrm{eq}}\delta_{ij} = \frac{2}{3}\mathcal{E}_{\alpha,\mathrm{eq}}\delta_{ij}, \end{aligned} \quad (66)$$

as derived from the Fokker-Planck equation in [57, 58] and used in the simulations of burning DT plasma in a z pinch [34]. The ratio of this pressure to the thermal pressure of the plasma, $P = nT_e + n_i T_i$, assuming $T_i = T_e$, is expressed as

$$\frac{P_{\alpha,\mathrm{eq}}}{P} = \frac{(1-\phi)^2 \langle Z_i \rangle \langle\sigma v\rangle_{DT} m_\alpha E_{\alpha 0} T_e^{1/2}}{64(1+\langle Z_i \rangle)(2\pi m_e)^{1/2} e^4 \ln\Lambda_{\alpha e}} F_e. \quad (67)$$

The pressure ratio (67) is much larger than the number density ratio (64) due to the large factor $E_{\alpha 0}/T_e \gg 1$. It is plotted vs. $T_e$ under the same assumptions as above and $\langle Z_i \rangle = 1$ in Fig. 5(a). We see that the equilibrium pressure of the fast α-particles is small compared to the thermal pressure of the fusion plasma near $T_e = 5$ keV. However, the ratio (67) is about 20%, and therefore cannot be neglected, even at the temperature of 13 keV achieved in the NIF shot N221204 [1]. It is even larger, approaching 50% at higher temperatures relevant for high-gain MagLIF. Its relevance has been confirmed in research related to laser fusion [40, 41].

Figure 5(b) presents the ion energy partition fraction $F_i$ (34) vs. the electron temperature for $\phi = 0$, $n = 5\times 10^{23}$ cm$^{-3}$, and for $\phi = 0.5$, $n = 10^{24}$ cm$^{-3}$. Also plotted are the fits from Refs. [36, 59, 60] and [37], labeled "Fraley" and "Atzeni", respectively: $F_i = T_e/(32+T_e)$ and $F_i = T_e/(25+T_e)$, where $T_e$ is expressed in keV. The agreement between our first-principles results and fits [36, 37] is reasonably good, closer to Fraley's fit. The fit of Refs. [36, 60] was obtained for solid DT ice density, whereas that of Ref. [37] targeted higher compressed densities of 10 to 100 g/cm$^3$. Another minor source of discrepancy with Fraley's fit is that they employed a quantum formula for the α-*i* Coulomb logarithm, whereas we stick to the classical one, as discussed in Section II.C. However, because our expression (34) depends on the background density only logarithmically through $\varepsilon$ [Eq. (24)], the close agreement is expected. Figure 5(c) plots the average equilibrium energy of a fast α-particle defined as $\mathcal{E}_{\alpha,\mathrm{eq}}/n_{\alpha,\mathrm{eq}}$ vs. $T_e$ for the same conditions. Although fast α-particles are born with the same center-of-mass energy $E_{\alpha 0}$, and the relative thermal addition (13) is small, $\Delta E_+/E_{\alpha 0} = 3\theta_i/2 \ll 1$, Fig. 5(c) illustrates that fast α-particles in a hotter background plasma have a noticeably higher average energy. This is because at higher electron temperatures the ions start participating in thermalization of α-particles while their energy is larger, making the equilibrium distribution function flatter in the low-energy range, as seen in Fig. 4(b).

### F. First-order non-equilibrium terms

Based on the findings in Section II.E, we henceforth neglect longitudinal velocity-space diffusion, the third term in curly brackets in the collisional term (50), and the finite width of the Maxwellian fusion source (11). The latter is replaced with a monoenergetic fusion source, $\Sigma_\alpha\left(u\right)\rightarrow\dot{n}_\alpha\delta\left(u-u_{\alpha 0}\right)/\left(4\pi u_0^2\right)$. Under these conditions, the equilibrium distribution (56) is an exact solution of the Fokker-Planck equation (36) for a uniform, steady background plasma at rest, in the absence of an electric field, as its left-hand side vanishes identically.

However, the Fokker-Planck equation for fast α-particles is no longer satisfied if the background plasma parameters entering the equilibrium distribution (56) – specifically $\dot{n}_\alpha\tau_{\alpha e}$ and $\varepsilon$ – exhibit spatial or temporal dependence, or if an electric field $\mathbf{E}$ is present. The diffusion approximation assumes that temporal variations are sufficiently slow, spatial length scales are sufficiently large, and the electric field is not excessively strong. This is the standard assumption in transport theory [61] upon which the Chapman-Enskog approximation is based [50, 62].

Specifically, we seek a solution for the distribution function $f_\alpha$ in the form:

$$f_\alpha\left(\mathbf{u};\ \mathbf{r},t\right)=f_{\alpha,\mathrm{eq}}\left(u\right)+f_{\alpha 0}\left(u\right)+\mathbf{f}_{\alpha 1}\left(u\right)\cdot\mathbf{u}\,. \tag{68}$$

Here, we have omitted the explicit spatial and temporal dependence of the terms in the right-hand side of Eq. (68). In this expression, $f_{\alpha,\mathrm{eq}}\left(u\right)$ is the local equilibrium distribution function (56). The indices 0 and 1 here denote the order of the expansion of the velocity angular dependence in Legendre polynomials, where the isotropic correction $f_{\alpha 0}\left(u\right)$ corresponds to the zero-order polynomial ($P_0$), and the anisotropic term $\mathbf{f}_{\alpha 1}(u)\cdot\mathbf{u}$ corresponds to the first-order polynomial ($P_1$). While representing different angular harmonics, both $f_{\alpha 0}$ and $\mathbf{f}_{\alpha 1}$ are of the same, first order of smallness relative to $f_{\alpha,\text{eq}}$. The isotropic term accounts for the time dependence of the α-particle production rate and collision parameters, whereas the anisotropic term represents the linear response to spatial gradients and the electric field $\mathbf{E}$. To solve the Fokker-Planck equation (36), we substitute the expansion (68) into it and retain the first-order non-vanishing terms.

The equation for the isotropic correction to the equilibrium distribution is

$$\begin{aligned}\frac{\partial f_{\alpha,\text{eq}}}{\partial t} &= \frac{\dot{n}_\alpha \tau_{\alpha e}}{4\pi}\times\left[\frac{1}{u^3+u_c^3}\frac{\partial \ln\left(\dot{n}_\alpha \tau_{\alpha e}\right)}{\partial t} - \frac{3u_c^3}{\left(u^3+u_c^3\right)^2}\frac{\partial \ln \varepsilon}{\partial t}\right] \\ &= \frac{\nu_{\alpha e}}{u^2}\frac{\partial}{\partial u}\left[\left(u^3+u_c^3\right) f_{\alpha 0}\right] .\end{aligned} \tag{69}$$

The boundary condition for the function $f_{\alpha 0}$ is derived from the requirements of continuity at $u = u_{\alpha 0}$ and sufficiently fast decay as $u \to \infty$. Because the solution of the homogeneous equation (57) decays with increasing $u$ as $1/\left(u^3+u_c^3\right)$ – which is much slower than the decay of the fusion source function, cf. the boundary condition (59) – these requirements imply the boundary condition $f_0\left(u_{\alpha 0}\right) = 0$. Integrating Eq. (69) with this boundary condition yields:

$$f_{\alpha 0} = \frac{\dot{n}_\alpha \tau_{\alpha e}^2}{4\pi\left(u^3+u_c^3\right)}\left[\frac{1}{3}\ln\left(\frac{u^3+u_c^3}{u_{\alpha 0}^3+u_c^3}\right)\frac{\partial \ln\left(\dot{n}_\alpha \tau_{\alpha e}\right)}{\partial t} + \frac{u_c^3\left(u_{\alpha 0}^3-u^3\right)}{\left(u_{\alpha 0}^3+u_c^3\right)\left(u^3+u_c^3\right)}\frac{\partial \ln \varepsilon}{\partial t}\right]. \tag{70}$$

The solution (70) is consistent with the diffusion-approximation ordering: corrections due to the unsteadiness of the background plasma are proportional to the ratio of the small α-*e* collision time, $\tau_{\alpha e}$, and the characteristic hydro times, which are assumed to be much larger.

The isotropic correction (70) modifies the equilibrium number and energy densities of the fast α-particles, as well as the rate of ash production (43) and the α-heating power (45); see details in Section IV.A.

For the anisotropic corrections, without loss of generality (cf. the beginning of Section III), we assume that the magnetic field is directed along the *z*-axis, while the gradients and the electric field are directed along the *x*-axis. The term responsible for the spatial gradients of the background plasma parameters is:

$$\mathbf{u}\nabla f_{\alpha,\text{eq}} = u_x \frac{\partial f_{\alpha,\text{eq}}}{\partial x} = u_x \times \frac{\dot{n}_\alpha \tau_{\alpha e}}{4\pi\left(u^3+u_c^3\right)}\left[\frac{\partial \ln\left(\dot{n}_\alpha \tau_{\alpha e}\right)}{\partial x} - \frac{u_c^3}{\left(u^3+u_c^3\right)}\frac{\partial \ln\left(\varepsilon^3\right)}{\partial x}\right], \tag{71}$$

The term describing the magnetization is:

$$\omega_\alpha \mathbf{u}\times\mathbf{b}\nabla_{\mathbf{u}}\left(\mathbf{f}_{\alpha 1}\cdot\mathbf{u}\right) = \omega_\alpha\left[f_{\alpha 1x}\left(u\right)u_y - f_{\alpha 1y}\left(u\right)u_x\right]. \tag{72}$$

The electric-field term is:

$$\begin{aligned}\frac{2e\mathbf{E}}{m_\alpha}\nabla_{\mathbf{u}} f_{\alpha,\text{eq}} &= \frac{2eE_x}{m_\alpha}\times\frac{\partial f_{\alpha,\text{eq}}}{\partial u}\times\frac{u_x}{u} = \frac{2eE_x u_x}{m_\alpha u}\times\frac{\partial}{\partial u}\left[\frac{\dot{n}_\alpha \tau_{\alpha e}}{4\pi\left(u^3+u_c^3\right)}\right] \\ &= -\frac{3\dot{n}_\alpha \tau_{\alpha e} eE_x}{2\pi m_\alpha}\times\frac{u u_x}{\left(u^3+u_c^3\right)^2}.\end{aligned} \tag{73}$$

In spherical coordinates $u_x = u\sin\theta\cos\varphi$, $u_y = u\sin\theta\sin\varphi$, hence $\nabla_\Omega^2 u_x = -2u_x$ and $\nabla_\Omega^2 u_y = -2u_y$; cf. Eq. (51). Then, introducing the complex dependent variable by

$$\hat{f}_{\alpha 1}\left(u\right) = f_{\alpha 1x}\left(u\right) + i f_{\alpha 1y}\left(u\right), \tag{74}$$

we arrive at the linearized kinetic equation for this variable. Here again, employing the normalized independent variable, $z = u / u_{\alpha 0}$, we present it as

$$\frac{z^3+\varepsilon^3}{z^2}\frac{\partial \hat{f}_{\alpha 1}}{\partial z}+\left(4+\frac{\mu\varepsilon^3}{z^3}-i\chi_\alpha\right)\hat{f}_{\alpha 1}$$
$$=\frac{\dot{n}_\alpha \tau_{\alpha e}^2}{4\pi u_0^3}\left[\frac{1}{z^3+\varepsilon^3}\frac{\partial \ln\left(\dot{n}_\alpha \tau_{\alpha e}\right)}{\partial x}-\frac{\varepsilon^3}{\left(z^3+\varepsilon^3\right)^2}\frac{\partial \ln\left(\varepsilon^3\right)}{\partial x}-\frac{z}{\left(z^3+\varepsilon^3\right)^2}\frac{6eE_x}{m_\alpha u_{\alpha 0}^2}\right]. \tag{75}$$

The nonhomogeneous differential equation (75) is equivalent to Eq. (19) of [31] and Eq. (52) of [39]. It is solved by the method of undetermined coefficients. Its solution with the boundary condition $\hat{f}_{\alpha 1}(z=1)=0$ is

$$\hat{f}_{\alpha 1}=\frac{\dot{n}_\alpha \tau_{\alpha e}^2}{4\pi u_{\alpha 0}^3}\left(z^3+\varepsilon^3\right)^{\frac{\mu-4+i\chi_\alpha}{3}} z^{-\mu}\left[A_p\varphi_{00}(z)+A_T\varphi_{10}(z)+A_E\varphi_{01}(z)\right]. \tag{76}$$

Here, we have labeled the driving-force terms determined by the gradients of $\dot{n}_\alpha \tau_{\alpha e}$ and $\varepsilon^3$, and the electric field by subscripts *p*, *T*, and *E*, respectively:

$$A_p=-\frac{\partial}{\partial x}\ln\left(\dot{n}_\alpha \tau_{\alpha e}\right),\quad A_T=\frac{\partial}{\partial x}\varepsilon^3,\quad A_E=\frac{3eE_x}{E_{\alpha 0}}. \tag{77}$$

All of the driving-force terms in Eq. (77) have the dimensionality of inverse length. For the first two terms, these are gradient lengths of the respective variables, and for the third – the length over which the work of the electric field $E_x$ equals 2/3 of the α-particle birth energy $E_{\alpha 0}$. The diffusion approximation is justified if these lengths are much longer than the fast α-particle displacement in the *x*-direction; see Sec. II.C. The signs in Eq. (77) are chosen considering that the fast α-particles flow in the direction opposite to $\nabla\left(\dot{n}_\alpha \tau_{\alpha e}\right)$, but in the same direction as $\nabla\varepsilon$ and $\mathbf{E}$; see Sec. III.

All corresponding functions $\varphi_{kl}(z)$ are cast into the universal form (78), parameterized by the indices $k$ and $l$, which specify the driving forces – the combinations 00, 10 and 01 refer to gradients of $\dot{n}_\alpha \tau_{\alpha e}$ and $\varepsilon$, and the electric field, respectively:

$$\varphi_{kl}(z) = \Phi_{kl}(z) - \Phi_{kl}(z=1),$$
$$\Phi_{kl}(z) = \frac{{}_2F_1\left(-\frac{\mu+l}{3},\ \frac{3k+2l-1+i\chi_\alpha}{3};\ \frac{3k+2l+2+i\chi_\alpha}{3};\ \frac{\varepsilon^3}{z^3+\varepsilon^3}\right)}{\left(3k+2l-1+i\chi_\alpha\right)\left(z^3+\varepsilon^3\right)^{\frac{3k+2l-1+i\chi_\alpha}{3}}}, \tag{78}$$

where ${}_2F_1(a,\ b;\ c;\ z)$ is the Gauss hypergeometric function.

Calculation of the transport coefficients involves integration of the corresponding terms constituting the function $\hat{f}_{\alpha 1}$ multiplied by $z^4$ or $z^6$, for particle and energy fluxes, respectively, over $z$ from 0 to 1. For each transport mechanism, this integration defines a complex function of the relevant dimensionless parameters: $\varepsilon$, primarily characteristic of the electron temperature but also dependent on the plasma composition; $\mu$, characteristic of the average ion mass; and the fast α-particle Hall parameter, $\chi_\alpha$. These functions are likewise described by the unified formula (79), differing only in their indices:

$$\hat{I}_{klm}(\varepsilon,\mu,\chi_\alpha) = \int_0^1 \left(z^3+\varepsilon^3\right)^{\frac{\mu-4+i\chi_\alpha}{3}} z^{4+2m-\mu}\varphi_{kl}(z)\,dz = \frac{1}{3k+2l-1+i\chi_\alpha}$$
$$\times\left[\int_0^1 \frac{z^{7+2m+l}}{\left(z^3+\varepsilon^3\right)^{2+k+l}}\ {}_2F_1\left(1,\ \frac{\mu+3k+3l+2+i\chi_\alpha}{3};\ \frac{3k+2l+2+i\chi_\alpha}{3};\ \frac{\varepsilon^3}{z^3+\varepsilon^3}\right)dz\right.$$
$$-\frac{{}_2F_1\left(1,\ \frac{\mu+3k+3l+2+i\chi_\alpha}{3};\ \frac{3k+2l+2+i\chi_\alpha}{3};\ \frac{\varepsilon^3}{1+\varepsilon^3}\right)}{\left(1+\varepsilon^3\right)^{2+k+l}\left(5+2m-\mu\right)} \tag{79}$$
$$\left.\times\, {}_2F_1\left(1,\ \frac{4-\mu-i\chi_\alpha}{3};\ \frac{8+2m-\mu}{3};\ \frac{1}{1+\varepsilon^3}\right)\right],$$

where the additional index $m$ equals 0 and 1 for particle and energy fluxes, respectively.

The anisotropic correction (76) determines the fluxes $\mathbf{J}_\alpha$ and $\mathbf{q}_\alpha$, and the current density $\mathbf{j}_\alpha$. To express these vector moments via the parameters of the background plasma, we need the transport coefficients calculated in the next Section.

## III. TRANSPORT COEFFICIENTS

In this section, we calculate the coefficients in the linear relations between the fast α-particle fluxes $\mathbf{J}_\alpha$ and $\mathbf{q}_\alpha$, the current density $\mathbf{j}_\alpha$, and the driving forces representing the corresponding gradients of background plasma parameters. In magnetized plasmas, these coefficients are tensor components. We will use vector notation, presenting the relation between a flux $\mathbf{J}$ and the driving force $\mathbf{A}$ in the form:

$$\mathbf{J} = D_\perp \mathbf{A} + \left(D_\parallel - D_\perp\right)\left(\mathbf{b}\cdot\mathbf{A}\right)\mathbf{b} + D_\wedge\left(\mathbf{b}\times\mathbf{A}\right), \tag{80}$$

where $D_\perp$, $D_\parallel$, and $D_\wedge$ are the perpendicular, parallel, and cross-gradient diffusion coefficients, respectively, $\mathbf{b} = \mathbf{B}/B$ is the unit vector in the direction of the magnetic field, and the driving forces listed in Eq. (77) are presented in vector form: for example, $\mathbf{A}_p = -\nabla\left(\dot{n}_\alpha \tau_{\alpha e}\right)$, etc. The three terms on the right-hand side of Eq. (80) represent the perpendicular, parallel, and cross-gradient fluxes, and are denoted by $\mathbf{J}_\perp$, $\mathbf{J}_\parallel$, and $\mathbf{J}_\wedge$, respectively. Below, we calculate the real-valued, positive-definite perpendicular and cross-gradient diffusion coefficients, which automatically give us the values of the corresponding parallel coefficients via $D_\parallel = D_\perp(B=0)$. In the absence of a magnetic field, $D_\perp = D_\parallel$ and $D_\wedge = 0$; thus, even though the unit vector $\mathbf{b}$ is undefined, the tensor relation (80) cleanly reduces to the simple proportionality $\mathbf{J} = D_\parallel \mathbf{A}$. The linearity of the Chapman-Enskog relations and the Fokker-Planck collisional operator (50) ensures that our results are applicable to arbitrary directions of the gradients, as well as the electric and magnetic fields, rather than being limited to those assumed in Eqs. (71)–(73).

### A. Diffusion and thermodiffusion

The flux of fast α-particles defined by Eq. (39) is expressed as

$$\mathbf{J}_\alpha = \int d^3u\ \mathbf{u} f_\alpha(\mathbf{u}). \tag{81}$$

Substituting the distribution function expansion (76) into (81), and integrating over the solid angle with the aid of Eq. (62), we calculate the fluxes $J_{\alpha x}$ and $J_{\alpha y}$:

$$\begin{aligned} J_{\alpha x} &= \int d^3u \times u_x \left[ f_{\alpha 1x}(u) u_x + f_{\alpha 1y}(u) u_y \right] = \frac{4\pi u_{\alpha 0}^5}{3} \int_0^1 z^4 f_{\alpha 1x}(z) dz, \\ J_{\alpha y} &= \frac{4\pi u_{\alpha 0}^5}{3} \int_0^1 z^4 f_{\alpha 1y}(z) dz\ . \end{aligned} \tag{82}$$

The results of integration over $z$ are given by Eq. (79). It is convenient to introduce the complex flux by $\hat{J}_\alpha = J_{\alpha x} + iJ_{\alpha y}$. Quote Ref. [62]: "The coefficients are represented in a complex fashion, the real part giving a flow in the direction and the imaginary part giving a flow at right angles to the driving force":

$$\begin{aligned} \hat{J}_\alpha &= \frac{1}{3} \dot{n}_\alpha \tau_{\alpha e}^2 u_{\alpha 0}^2 \int_0^1 \left( z^3 + \varepsilon^3 \right)^{\frac{\mu - 4 + i\chi_\alpha}{3}} z^{4-\mu} \left[ A_p \varphi_{00}(z) + A_T \varphi_{10}(z) \right] dz \\ &= \frac{1}{3} \dot{n}_\alpha \tau_{\alpha e}^2 u_{\alpha 0}^2 \left[ -\hat{I}_{000} \frac{\partial}{\partial x} \ln\left( \dot{n}_\alpha \tau_{\alpha e} \right) + 3\varepsilon^2 \hat{I}_{100} \frac{\partial \varepsilon}{\partial x} \right]. \end{aligned} \tag{83}$$

Here, we have retained only two terms from the solution (76) because the fast α-particle flux generated by the electric field (the current density) is determined by their conductivity, which is calculated separately in Section III.C. Let us clarify the physical meaning of these fluxes and the effect of magnetization.

Clearly, the equilibrium number density of fast α-particles (63) is higher where the product $\dot{n}_\alpha \tau_{\alpha e}$ is larger at constant $\varepsilon$ (for instance, if the ion temperature is higher at constant electron temperature and density); cf. Fig. 4(a). Therefore, there should be a diffusive flux $\mathbf{J}_{\alpha p}$ of α-particles proportional to $-\nabla\left(\dot{n}_\alpha \tau_{\alpha e}\right)$, determined by a positive particle diffusion coefficient. Conversely, if the value of $\varepsilon$ is increased while keeping $\dot{n}_\alpha \tau_{\alpha e}$ constant, then the number density $n_\alpha$ of fast α-particles decreases, as demonstrated by Eq. (63). Hence, the non-uniformity of $\varepsilon$ – that is, of the electron temperature – creates an α-particles flux proportional to $+\nabla\varepsilon$. This is identified as thermodiffusion, or the Soret effect [63]. The direction of this flux indicates "thermophilicity" (particles moving toward a hotter region), which corresponds to a negative value of the Soret coefficient. For this model, we instead define a positive thermodiffusion coefficient, $D_T$, and denote its corresponding flux by $\mathbf{J}_{\alpha T}$.

The effect of magnetization on the fast α-particle fluxes is qualitatively illustrated in Fig. 6, following the analysis in §3 of Ref. [50]. Assuming the α-particles are magnetized in a field $\mathbf{B} = B\hat{\mathbf{e}}_z$, the positively charged particles gyrate clockwise (in the negative direction). The α-particles arriving at a surface element perpendicular to the *y* axis, shown as a thick black segment in Fig. 6, from the right and left move in the positive and negative *y*-directions, respectively. With a positive gradient of $\dot{n}_\alpha \tau_\alpha$, more α-particles arrive from the right, making the net particle flux $J_{\alpha py}$ positive. Conversely, a positive gradient of the electron temperature (and thus, of $\varepsilon$, not shown in Fig. 6) produces a negative cross-gradient flux, $J_{\alpha Ty} < 0$.

The diffusion α-particle fluxes are related to the gradient of $\dot{n}_\alpha \tau_{\alpha e}$ by

$$J_{\alpha p\perp} = -D_{p\perp} \frac{\partial}{\partial x}\left(\dot{n}_\alpha \tau_{\alpha e}\right), \quad J_{\alpha p\wedge} = D_{p\wedge} \frac{\partial}{\partial x}\left(\dot{n}_\alpha \tau_{\alpha e}\right), \tag{84}$$

where

$$D_{p\perp} = \frac{1}{3} u_{\alpha 0}^2 \tau_{\alpha e} \operatorname{Re}\left(\hat{I}_{000}\right), \quad D_{p\wedge} = -\frac{1}{3} u_{\alpha 0}^2 \tau_{\alpha e} \operatorname{Im}\left(\hat{I}_{000}\right). \tag{85}$$

The signs in Eqs. (84) and (85) are chosen consistently with the above discussion, to make the diffusion coefficients positive. The small-$\varepsilon$ expansion of the complex integral (79) determining the α-particle diffusion coefficients is given by

$$\hat{I}_{000} = \frac{1}{1+\chi_\alpha^2}\left[\frac{1-i\chi_\alpha}{2} - \frac{\Gamma\left(\frac{5-\mu}{3}\right)\Gamma\left(\frac{2-i\chi_\alpha}{3}\right)}{\Gamma\left(\frac{4-\mu-i\chi_\alpha}{3}\right)} \varepsilon^{1+i\chi_\alpha}\right] + O\left(\varepsilon^2\right), \tag{86}$$

where $\Gamma(z)$ is the Euler gamma function. Here, the first term in the small-$\varepsilon$ asymptotic expansion (86) corresponds to the low-temperature approximation [31] used to calculate the diffusion coefficients in Ref. [35]:

$$D_{p\perp} \cong \frac{u_{\alpha 0}^2 \tau_{\alpha e}}{6\left(1+\chi_\alpha^2\right)}, \quad D_{p\wedge} \cong \frac{u_{\alpha 0}^2 \tau_{\alpha e} \chi_\alpha}{6\left(1+\chi_\alpha^2\right)}. \tag{87}$$

Suppression of the perpendicular diffusion coefficient by a factor of $1+\chi_\alpha^2$, as in (87) was assumed in the pioneering paper on magnetically assisted laser fusion, Ref. [64]. When only the electron friction is taken into account, Eq. (87) is consistent with the standard gas-kinetic formula for the diffusion coefficient: $D = \bar{v}\lambda/3$, where $\bar{v}$ and $\lambda$ are the mean value of the particle velocity and free path; see Eq. (3.17) of Ref. [65]. The velocity of an α-particle decelerated by the electron friction force (22) decreases linearly with the distance traveled;

hence, its mean value is $\bar{v} = u_{\alpha 0} / 2$. Substituting it into this formula with $\lambda = \Delta x_p$ and $\lambda = |\Delta y_p|$ given by Eq. (32), we reproduce the expressions of Eq. (87) for the perpendicular and cross-gradient diffusion coefficients, $D_{p\perp}$ and $D_{p\wedge}$, respectively, as should occur.

The first-order asymptotic expressions for these diffusion coefficients in the limit of strong magnetization, $\chi_\alpha \to \infty$ (see Appendix) are

$$D_{p\perp} \cong \frac{u_{\alpha 0}^2 \tau_{\alpha e}}{6\chi_\alpha^2}\left[1 + \frac{1}{3}(1-\mu)\varepsilon^2 L_- + 2C_{\text{osc}}\right], \quad D_{p\wedge} \cong \frac{u_{\alpha 0}^2 \tau_{\alpha e}}{6\chi_\alpha}\left(1 - \frac{1}{3}\varepsilon^2 L_-\right), \tag{88}$$

where the function $L_-(\varepsilon)$ is defined by Eq. (31), and the oscillatory correction $C_{\text{osc}}$ is given by Eq. (A13) in the Appendix. As shown in the Appendix, the period of these oscillations in the Hall parameter is $\Delta\chi_\alpha = 2\pi\tau_{\alpha e} / \tau_{\text{dec}}$ [cf. Eq. (29)]. For given local plasma conditions, this corresponds to a gyrofrequency period $\Delta\omega_\alpha = 2\pi / \tau_{\text{dec}}$, which has a transparent physical meaning: consecutive oscillation peaks occur when the total α-particle gyration angle accumulated during deceleration increments by 2π, so that stopping terminates at the identical gyrophase after each additional full gyration. The relative amplitude of the oscillations decays with increasing magnetization as $\chi_\alpha^{-(2-\mu)/3} \approx \chi_\alpha^{-0.543}$ (for an equimolar DT plasma with $\mu = 0.372$ ). On the background of the perpendicular diffusion coefficient decaying as $\chi_\alpha^{-2}$, the absolute oscillation amplitude scales as $\chi_\alpha^{-(8-\mu)/3} \approx \chi_\alpha^{-2.54}$ , which represents a relatively slow decay that remains discernible up to large values of $\chi_\alpha$. Averaging over these oscillations makes $C_{\text{osc}} \to 0$, recovering the smooth monotonic asymptotic behavior. In the low-temperature limit $\varepsilon \to 0$, the formulas (88) reduce to the strong-magnetization asymptotics of the baseline coefficients given by Eq. (87).

Figure 7 displays the diffusion coefficients, $D_{p\perp}$ (a) and $D_{p\wedge}$ (b), normalized with respect to the factor $u_{\alpha 0}^2 \tau_{\alpha e} / 3$, for an equimolar DT plasma vs. the α-particle Hall parameter, $\chi_\alpha$. The oscillatory dependence observed in Fig. 7, which is not captured by the first approximation (87) but appears in the next term of the small-$\varepsilon$ expansion (86), is a direct consequence of the "hard stop" near thermalization caused by ion dynamic friction, as illustrated in Figs. 3(b) and 3(c). Unlike the statistical oscillations of transport coefficients predicted for collisionless 2D plasmas [66], which were later shown to be non-physical [67], the oscillations described here are a purely single-particle ballistic effect arising from the gyrophase sensitivity of the stopping point in a collisional plasma. The oscillations of $D_{p\perp}(\chi_\alpha)$ directly mirror loops in the stopping trajectory shown in Fig. 3(b) and the periodic structure of the displacement $\Delta x_p$ in Fig. 3(c). In particular, as $\chi_\alpha$ ranges from 0 to 40 in Fig. 7(a), the oscillation period $\Delta\chi_\alpha = 2\pi\tau_{\alpha e} / \tau_{\text{dec}}$ yields nine completed oscillation cycles for $T_e = 5\,\text{keV}$ and four cycles for $T_e = 25\,\text{keV}$. This corresponds precisely to the nine and four trajectory loops traced out in the $(x, y)$ plane before stopping, matching the local minima in Fig. 7(a) as well as the displacement extrema in Fig. 3(c). Averaging over the isotropic α-particle velocity distribution makes the oscillations in Fig. 7 less pronounced than those of the single-particle displacement in Fig. 3(c). Furthermore, the perpendicular diffusion coefficient $D_{p\perp}$ exhibits noticeably stronger oscillations than the cross-gradient coefficient $D_{p\wedge}$, reflecting the fact that $\Delta x_p$ oscillates around zero while $-\Delta y_p$ remains one-sided, a property explained analytically in the Appendix.

At low magnetization, the normalized perpendicular diffusion coefficient $D_{p\perp}$ decreases significantly with increasing electron temperature. Conversely, at high magnetization, the smooth trend of perpendicular diffusion suppression – scaling as $\chi_\alpha^{-2}$ when averaged over the oscillations – is universal across all temperatures. This behavior is confirmed by the first relation in Eq. (88), where the leading monotonic term in brackets increases by only 17% as $T_e$ increases from 5 to 25 keV. Physically, the perpendicular diffusion coefficient scales with the mean-square transverse displacement. In the unmagnetized limit, this step length is the full deceleration path $l_\alpha$, which decreases relative to the electron-only path $l_{\alpha e}$ as ion drag grows with temperature [cf. Eq. (30) and Fig. 3(a)]. Under strong magnetization, however, the transverse step length is governed by the α-particle gyroradius $r_L$, which is independent of the electron temperature.

Figure 8 compares the particle diffusion coefficients evaluated using the exact formula (85) with approximations from the first terms in the small-$\varepsilon$ expansion (86), as well as with the strong-magnetization asymptotics (88), shown vs. $\chi_\alpha$ for the same conditions as in Fig. 7, for $T_e = 5$ and 20 keV. At 5 keV, even the lowest order of the small-$\varepsilon$ expansion (87) provides decent approximations for $D_{p\perp}$ and $D_{p\wedge}$, and the second order of (86) is highly accurate, capturing the amplitude, frequency, and phase of the oscillations. At 20 keV, the lowest-order approximation (87) for $D_{p\perp}$ is still quite accurate, except at low magnetization, and the second order of (86) captures the amplitude and the frequency, but not the phase of the oscillations. However, the same approximation is no longer reasonable for $D_{p\wedge}$. The strong-magnetization formula (88), with the oscillatory term $C_{\text{osc}}$ included, correctly captures the asymptotic behavior, amplitude, frequency, and phase of the oscillations at large $\chi_\alpha$ for all temperatures.

Similarly, the thermodiffusion fluxes are related to the gradient of $\varepsilon \sim T_e^{1/2}$ by

$$J_{\alpha T\perp} = \dot{n}_\alpha \tau_{\alpha e} D_{T\perp} \frac{\partial \varepsilon}{\partial x}, \quad J_{\alpha T\wedge} = -\dot{n}_\alpha \tau_{\alpha e} D_{T\wedge} \frac{\partial \varepsilon}{\partial x}, \tag{89}$$

where

$$D_{T\perp} = \frac{1}{3} u_{\alpha 0}^2 \tau_{\alpha e} \times 3\varepsilon^2 \operatorname{Re}\left(\hat{I}_{100}\right), \quad D_{T\wedge} = -\frac{1}{3} u_{\alpha 0}^2 \tau_{\alpha e} \times 3\varepsilon^2 \operatorname{Im}\left(\hat{I}_{100}\right). \tag{90}$$

The small-$\varepsilon$ expansion of this result is

$$3\varepsilon^2 \hat{I}_{100} = \frac{10\pi\sqrt{3}}{27\left(2+i\chi_\alpha\right)} {}_3F_2\left(\frac{1}{3},\ 1,\ \frac{5+\mu+i\chi_\alpha}{3};\ 3,\ \frac{5+i\chi_\alpha}{3};\ 1\right)\varepsilon - \frac{3}{1+i\chi_\alpha}\varepsilon^2 + O\left(\varepsilon^3\right), \tag{91}$$

and ${}_3F_2\left(a, b, c; p, q; 1\right)$ is a generalized hypergeometric function of argument unity.

Figure 9 displays the fast α-particle thermodiffusion coefficients, $D_{T\perp}$ (a) and $D_{T\wedge}$ (b), normalized as in Fig. 7, vs. $\chi_\alpha$, showing the same suppression factors of $\chi_\alpha^2$ and $\chi_\alpha$ for perpendicular and cross-gradient thermodiffusion, respectively. Similarly to Fig. 7, these coefficients are well approximated by the first two terms of the small-$\varepsilon$ expansion (91) at $T_e = 5$ keV, but the approximation becomes inaccurate at 20 keV. Comparing Figs. 9 and 7, we observe that the diffusion coefficients $D_p$ and $D_T$ are of the same order of magnitude. The gradient $\left|\partial \ln \varepsilon / \partial x\right|$ is typically smaller than $\left|\partial \ln\left(\dot{n}_\alpha \tau_{\alpha e}\right) / \partial x\right|$ – see the detailed comparison in Section IV.B – so thermodiffusion can be neglected at $T_e = 5$ keV. However, this term is no longer negligible at 20 keV, indicating that thermodiffusion should be accounted for in high-yield MagLIF parameter range.

Without magnetization, all cross-gradient fluxes and their corresponding coefficients vanish. Figure 10 shows the only diffusion coefficients that remain finite: parallel diffusion (a) and thermodiffusion (b) coefficients vs. $T_e$. A comparison of the exact results shown as lines with symbols with the small-$\varepsilon$ expansion demonstrates that this approximation does not work effectively in the relevant temperature range due to the peculiar analytical structure of our expansion as $\chi_\alpha$ approaches zero; see Sec. III.B. The normalized parallel diffusion and thermodiffusion coefficients rapidly decrease with increasing electron temperature because the α-particle's mean free path (30) drops below the normalization value, $l_{\alpha e}$, as the influence of ion friction grows.

## B. Diffusive energy flux

Similarly to Eq. (81), we express the energy flux (41) of fast α-particles as

$$\mathbf{q}_\alpha = \int d^3u \times \frac{1}{2} m_\alpha u^2 \ \mathbf{u} f_\alpha(\mathbf{u}). \tag{92}$$

As in Sec. III.A, we introduce the complex energy flux by $\hat{q}_\alpha = q_{\alpha x} + i q_{\alpha y}$. In the same manner as above, we obtain:

$$\begin{aligned} \hat{q}_\alpha &= \frac{1}{6} \dot{n}_\alpha \tau_{\alpha e}^2 m_\alpha u_{\alpha 0}^4 \int_0^1 \left(z^3 + \varepsilon^3\right)^{\frac{\mu - 4 + i\chi_\alpha}{3}} z^{6-\mu} \left[ A_p \varphi_{00}(z) + A_T \varphi_{10}(z) \right] dz \\ &= \frac{1}{3} \dot{n}_\alpha \tau_{\alpha e}^2 u_{\alpha 0}^2 E_{\alpha 0} \left[ -\hat{I}_{001} \frac{\partial}{\partial x} \ln\left(\dot{n}_\alpha \tau_{\alpha e}\right) + 3\varepsilon^2 \hat{I}_{101} \frac{\partial \varepsilon}{\partial x} \right]. \end{aligned} \tag{93}$$

Here again, we have retained only two terms from the solution (76) because the fast α-particle energy flux generated by the electric field is determined by the Peltier coefficient, which is calculated separately in Sec. III.C. We retain the energy fluxes proportional to the gradients of $\dot{n}_\alpha \tau_{\alpha e}$ and $\varepsilon$. The former is similar to the Dufour effect [63], the energy flux driven by the concentration gradient of fast α-particles. The latter is the energy flux driven by the temperature gradient, which resembles thermal conduction, but with a twist. The gradient of $\varepsilon$ drives the α-particle energy flux to a hotter region, where the average α-particle energy is higher, cf. Fig. 5(c). This is a direct consequence of the "thermophilicity" of the corresponding Soret effect, translating into a negative effective thermal conduction coefficient. For the present model, we introduce positive coefficients of energy diffusivity corresponding to these energy fluxes, driven by the α-particle number density and electron temperature gradients, labeled $\kappa_p$ and $\kappa_T$, respectively. The perpendicular and cross-gradient energy fluxes are expressed as

$$\begin{aligned} q_{\alpha p\perp} &= -\kappa_{p\perp}\frac{\partial}{\partial x}\left(\dot{n}_\alpha \tau_{\alpha e} E_{\alpha 0}\right), \quad q_{\alpha p\wedge} = \kappa_{p\wedge}\frac{\partial}{\partial x}\left(\dot{n}_\alpha \tau_{\alpha e} E_{\alpha 0}\right), \\ q_{\alpha T\perp} &= \dot{n}_\alpha \tau_{\alpha e} E_{\alpha 0}\kappa_{T\perp}\frac{\partial \varepsilon}{\partial x}, \quad q_{\alpha T\wedge} = -\dot{n}_\alpha \tau_{\alpha e} E_{\alpha 0}\kappa_{T\wedge}\frac{\partial \varepsilon}{\partial x}, \end{aligned} \tag{94}$$

where the corresponding positive diffusion coefficients are

$$\begin{aligned} \kappa_{p\perp} &= \frac{1}{3}u_{\alpha 0}^2 \tau_{\alpha e}\,\mathrm{Re}\left(\hat{I}_{001}\right), \quad \kappa_{p\wedge} = -\frac{1}{3}u_{\alpha 0}^2 \tau_{\alpha e}\,\mathrm{Im}\left(\hat{I}_{001}\right), \\ \kappa_{T\perp} &= \frac{1}{3}u_{\alpha 0}^2 \tau_{\alpha e}\times 3\varepsilon^2\,\mathrm{Re}\left(\hat{I}_{101}\right), \quad \kappa_{T\wedge} = -\frac{1}{3}u_{\alpha 0}^2 \tau_{\alpha e}\times 3\varepsilon^2\,\mathrm{Im}\left(\hat{I}_{101}\right). \end{aligned} \tag{95}$$

The small-$\varepsilon$ expansions of these terms are, respectively,

$$\begin{aligned} \hat{I}_{001} &= \frac{1}{4\left(3+i\chi_\alpha\right)} + i\frac{4-\mu+i\chi_\alpha}{\chi_\alpha\left(3+i\chi_\alpha\right)}\varepsilon^3 \\ &-i\frac{\left(4-\mu-i\chi_\alpha\right)\Gamma\left(\frac{7-\mu}{3}\right)\Gamma\left(1-\frac{i\chi_\alpha}{3}\right)}{\chi_\alpha\left(3+i\chi_\alpha\right)\left(1-i\chi_\alpha\right)\Gamma\left(\frac{7-\mu-i\chi_\alpha}{3}\right)}\varepsilon^{3+i\chi_\alpha} + O\left(\varepsilon^4\right), \end{aligned} \tag{96}$$

$$\begin{aligned} 3\varepsilon^2\hat{I}_{101} &= \frac{3}{3+i\chi_\alpha}\varepsilon^2 - \frac{28\pi\sqrt{3}}{27\left(2+i\chi_\alpha\right)}\,{}_3F_2\left(-\frac{1}{3},\ 1,\ \frac{5+\mu+i\chi_\alpha}{3};\ 3,\ \frac{5+i\chi_\alpha}{3};\ 1\right)\varepsilon^3 \\ &+O\left(\varepsilon^5\right). \end{aligned} \tag{97}$$

The first term in the small-$\varepsilon$ asymptotic expansion of (96) corresponds to the low-temperature approximation of Ref. [31], which predicted suppression of the perpendicular diffusion coefficient by a factor of $1+\left(\chi_\alpha/3\right)^2$:

$$\kappa_{p\perp} \cong \frac{u_{\alpha 0}^2 \tau_{e\alpha}}{4\left(9+\chi_\alpha^2\right)}, \quad \kappa_{p\wedge} \cong \frac{u_{\alpha 0}^2 \tau_{e\alpha} \chi_\alpha}{12\left(9+\chi_\alpha^2\right)}. \tag{98}$$

Like the particle diffusion coefficients (87), the gas-kinetic estimate based on the single-particle displacement is consistent with the baseline diffusion model. Substituting the mean velocity $\bar{v} = u_{\alpha 0}/2$ and the displacement $\lambda = \Delta x_E$ or $\lambda = |\Delta y_E|$ from Eq. (33) into the standard formula $D = \bar{v}\lambda/3$ yields the baseline diffusion coefficients with denominators differing from those in Eq. (98) by a factor of 2. This difference arises from the definition of the driving thermodynamic force. In the baseline model (2), the energy flux is driven by the gradient of the α-particle energy density $\mathcal{E}_\alpha$ and the corresponding diffusion coefficient $\kappa_\perp$ given by Eq. (3) reproduces the gas-kinetic estimate. In Eq. (94), the energy flux is driven by the gradient of $\dot{n}_\alpha \tau_{\alpha e} E_{\alpha 0}$. Because $\dot{n}_\alpha \tau_{\alpha e} E_{\alpha 0} = 2\mathcal{E}_{\alpha,\mathrm{eq}}$ in the low-temperature limit [Eq. (65)], the coefficient $\kappa_{p\perp}$ in Eq. (98) is smaller by a factor of 2, satisfying $\kappa_{p\perp} = \kappa_\perp / 2$. The difference between the magnetic suppression factors in Eqs. (87) and (98) is readily explained by inspecting Eqs. (32) and (33).

Note that both higher-order terms in Eq. (96) – the monotonic $O\left(\varepsilon^3\right)$ and the oscillatory $O\left(\varepsilon^{3+i\chi_\alpha}\right)$ – diverge in the unmagnetized limit $\chi_\alpha \to 0$, but their sum remains finite, reducing to the asymptotic expansion at $\chi_\alpha = 0$:

$$\hat{I}_{001} = \frac{1}{12} + \frac{4-\mu}{3}\varepsilon^3 \ln\varepsilon + O\left(\varepsilon^3\right). \tag{99}$$

The third derivative of this expression monotonically diverges at $\varepsilon \to 0$, which explains the failure of the small-$\varepsilon$ asymptotic expansion in this limit; see Figs. 10 and 13.

The first-order asymptotic expressions for these diffusion coefficients in the limit of strong magnetization, $\chi_\alpha \to \infty$ (see Appendix) are

$$\kappa_{p\perp} \cong \frac{u_{\alpha0}^2 \tau_{\alpha e}}{4\chi_\alpha^2}\left[1+\frac{4}{3}\varepsilon^3(1-\mu)\left(1-\frac{\varepsilon}{6}L_+\right)\right], \quad \kappa_{p\wedge} \cong \frac{u_{\alpha0}^2 \tau_{\alpha e}}{12\chi_\alpha}\left[1-4\varepsilon^3\left(1-\frac{\varepsilon}{6}L_+\right)\right]. \qquad (100)$$

The asymptotic formulas (100) represent the energy-density diffusion coefficients averaged over the small oscillations. In the low-temperature limit, the formulas (100) reduce to strong-magnetization asymptotics of the coefficients given by Eq. (98).

Figures 11(a) and (b) display the energy-density diffusion coefficients, $\kappa_{p\perp}$ and $\kappa_{p\wedge}$, respectively, normalized with respect to the factor $u_{\alpha0}^2 \tau_{\alpha e}/3$, vs. $\chi_\alpha$. Unlike the particle diffusion coefficients in Fig. 7, and consistently with the single-particle energy displacement shown in Fig. 3(d), these coefficients exhibit virtually no discernible oscillations. The physical reason for this suppression becomes clear when considering the relative weighting along the deceleration trajectory: the energy flux moment weights the velocity distribution by $u^3$, concentrating energy transport in the early, high-velocity portion of the deceleration path where α-particles slow down almost entirely on electrons. By the time the α-particles reach the low-velocity "hard stop" governed by ion friction, most of their kinetic energy has already been deposited, greatly attenuating the gyrophase sensitivity of the remaining energy displacement. Mathematically, as established in the Appendix, the oscillatory correction in the strong-magnetization expansion for energy transport ($m=1$) decays rapidly as $\chi_\alpha^{-(4-\mu)/3} \approx \chi_\alpha^{-1.21}$ (compared to the much slower $\sim \chi_\alpha^{-0.54}$ decay for particle transport, $m=0$, rendering the oscillatory term smaller relative to the smooth $O\left(\chi_\alpha^{-1}\right)$ cross-gradient and $O\left(\chi_\alpha^{-2}\right)$ perpendicular monotonic trends. In the small-$\varepsilon$ expansion, this suppression is reflected by the fact that the oscillatory term in Eq. (96) is of high order, $O\left(\varepsilon^{3+i\chi_\alpha}\right)$, compared to the dominant $O\left(\varepsilon^{1+i\chi_\alpha}\right)$ oscillatory term in Eq. (86). Consequently, all curves in Fig. 11(a) follow the monotonic lowest-order analytic approximation, Eq. (98), for $\kappa_{p\perp}$, except at low magnetization. As in Fig. 7(a), and for the same reasons, the normalized perpendicular diffusion coefficient $\kappa_{p\perp}$ decreases with electron temperature in the unmagnetized limit, whereas across the magnetized range all curves collapse onto essentially the same universal line. This behavior is confirmed by the first relation in Eq. (100), where the bracketed expression increases by only 5% as $T_e$ is increased from 5 to 25 keV. As illustrated in Fig. 11(b), the monotonic suppression of cross-gradient energy transport by a factor of $\chi_\alpha^{-1}$ likewise persists across all temperatures, in complete agreement with the second relation in Eq. (100).

Figure 12 compares these coefficients evaluated using the exact formula, Eq. (95), with approximations from the first terms in the small-$\varepsilon$ expansion (96), as well as with the strong-magnetization asymptotics (100), for $T_e = 5$ and 20 keV. The lowest order of the small-$\varepsilon$ expansion, Eq. (98), provides a good approximation for $\kappa_{p\perp}$ up to 20 keV, except at low magnetization. As for $\kappa_{p\wedge}$, it is well approximated by Eq. (98) at 5 keV, but even the next-order term in Eq. (96) does not improve accuracy for $\kappa_{p\wedge}$ at 20 keV. The strong-magnetization formula (100) correctly captures the asymptotic behavior of both diffusion coefficients at large $\chi_\alpha$ for all temperatures.

The energy-density thermodiffusion coefficients $\kappa_{T\perp}$ and $\kappa_{T\wedge}$ are evaluated in the same manner using Eq. (95) and (97). The contribution of those terms must be accounted for at higher temperatures, similar to the thermodiffusion contribution to the α-particle flux. At finite $\varepsilon$, the relative magnitudes of these terms can be compared in Fig. 13, which shows the parallel energy-density diffusion coefficients, $\kappa_{p\|}$ and $\kappa_{T\|}$, vs. $T_e$. As with Fig. 10, the small-$\varepsilon$ expansion does not work effectively in the relevant temperature range for the reasons explained following Eq. (99).

### C. α-particle conductivity and the Peltier effect

The flux $\mathbf{J}_\alpha$ of fast α-particles (39) driven by any force generates a current density $\mathbf{j}_\alpha$ (40). Here, we focus on the contribution to the current density produced by the electric field $\mathbf{E}$, which is not accounted for in Eq. (83). While the background electron conductivity dominates the plasma electrical response owing to the much higher electron density – as demonstrated below – we calculate the fast α-particle conductivity to complete the transport matrix. Introducing the complex current density, $\hat{j}_\alpha = j_{\alpha x} + i j_{\alpha y}$, the corresponding complex conductivity $\hat{\sigma}_\alpha$, and utilizing Eqs. (76) and (79), we arrive at the Ohm's law for fast α-particles:

$$\begin{aligned}\hat{j}_\alpha &= 2e \times \frac{4\pi}{3} u_{\alpha 0}^5 \times \frac{\dot{n}_\alpha \tau_{\alpha e}^2}{4\pi u_{\alpha 0}^3} \times \frac{6eE_x}{m_\alpha u_{\alpha 0}^2} \times \int_0^1 \left(z^3+\varepsilon^3\right)^{\frac{\mu-4+i\chi_\alpha}{3}} z^{4-\mu} \varphi_{01}(z)\,dz \\ &= \hat{\sigma}_\alpha E_x,\end{aligned} \tag{101}$$

where the complex conductivity is expressed as

$$\hat{\sigma}_\alpha = \frac{4e^2 \dot{n}_\alpha \tau_{\alpha e}^2}{m_\alpha} \times \hat{I}_{010} = \frac{4e^2 n_{\alpha,\mathrm{eq}} \tau_{\alpha e}}{m_\alpha} \times \hat{g}, \tag{102}$$

Here, the small-$\varepsilon$ expansion of the corresponding integral (79) is

$$\begin{aligned}\hat{I}_{010} &= \frac{1}{1+i\chi_\alpha} \ln\left(\frac{1}{\varepsilon}\right) - \frac{3\mu + 7 - \chi_\alpha^2 + 5i\chi_\alpha}{3\left(\mu+2+i\chi_\alpha\right)\left(1+i\chi_\alpha\right)^2} \\ &+ \frac{(2-\mu)(1+\mu)\left[\psi\left(\frac{5-\mu}{3}\right) - \psi\left(\frac{4+i\chi_\alpha}{3}\right)\right]}{3\left(\mu+2+i\chi_\alpha\right)\left(\mu-1+i\chi_\alpha\right)\left(1+i\chi_\alpha\right)} + O(\varepsilon),\end{aligned} \tag{103}$$

where $\psi(z)=\Gamma'(z)/\Gamma(z)$ is the digamma function. We have defined

$$\hat{g}=\hat{I}_{010}\times\left[\frac{1}{3}\ln\left(1+\varepsilon^{-3}\right)\right]^{-1}=\frac{1}{1+i\chi_{\alpha}}\left\{1+O\left[\frac{1}{\ln(1/\varepsilon)}\right]\right\}. \tag{104}$$

The logarithmic factor (63) has been introduced into the definition (102) to demonstrate explicitly that the contribution of the fast α-particles to conductivity is proportional to their equilibrium number density $n_{\alpha,\mathrm{eq}}$. This term logarithmically diverges in the limit $\varepsilon\to 0$, but so does the integral (103), while the factor $\hat{g}$ tends to the finite value of $\left(1+i\chi_{\alpha}\right)^{-1}$.

For a DT plasma, $\ln\Lambda_{\alpha e}=\ln\Lambda_{ei}$, and, therefore, $\tau_{\alpha e}=m_{\alpha}\tau_{ei}/\left(4m_{e}\right)$; cf. our Eq. (19) and Eq. (2.8e) of Ref. [50]. Substituting this into Eq. (102), we find:

$$\hat{\sigma}_{\alpha}=\sigma_{e\perp}\frac{n_{\alpha,\mathrm{eq}}}{n}\hat{g}\,, \tag{105}$$

where the perpendicular electron conductivity, $\sigma_{e\perp}=e^{2}n\tau_{ei}/m_{e}$, is given by Eq. (2.7) of Ref. [50]. This translates into

$$\sigma_{\alpha\perp}=\sigma_{e\perp}\frac{n_{\alpha,\mathrm{eq}}}{n}\mathrm{Re}\left(\hat{g}\right),\quad \sigma_{\alpha\wedge}=-\sigma_{e\perp}\frac{n_{\alpha,\mathrm{eq}}}{n}\mathrm{Im}\left(\hat{g}\right), \tag{106}$$

where the perpendicular and cross-gradient conductivities, $\sigma_{\alpha\perp}$ and $\sigma_{\alpha\wedge}$, are both positive, in agreement with Fig. 6: the positive α-particle flux in the *x* direction generates a negative cross-gradient flux in the *y* direction.

Observe that the fast α-particle conductivity is much smaller than the electron conductivity because there are fewer α-particles, see Eq. (63) and Fig. 5(a), whereas their mobility is about the same as that of the electrons.

The effect of magnetization on the conductivity $\hat{\sigma}_\alpha$ determined by the function $\hat{g}$ is illustrated in Fig. 14. Oscillations are clearly visible in the perpendicular conductivity $\sigma_{\alpha\perp}$ in Fig. 14(a). Because electrical conduction represents particle transport ( $m = 0$ ), the oscillatory correction $C_{\rm osc}$ in Eq. (A20) retains the exact same oscillation period and power-law decay as in perpendicular particle diffusion, Eq. (A18). However, the temperature dependence of the relative oscillation amplitude differs noticeably between the two coefficients. In $D_{p\perp}$, the prefactor multiplying $C_{\rm osc}$ in Eq. (A18) decreases by 15% as $T_e$ increases from 5 to 25 keV. In contrast, the corresponding prefactor in the asymptotic expression (A20) for $\sigma_{\alpha\perp}$ increases by 71% over the same temperature range. Consequently, the oscillations in Fig. 14(a) become more pronounced at higher temperatures, in contrast to Fig. 7(a). The overall trend of perpendicular conductivity suppression by a factor of $\chi_\alpha^{-2}$, averaged over the oscillations, remains universal across all temperatures, with significant deviations confined to low magnetization. While the lowest-order approximation from Eq. (103) provides only logarithmic accuracy and is insufficient even at low temperatures, the second-order expansion performs well for both $\sigma_{\alpha\perp}$ and $\sigma_{\alpha\wedge}$.

The flux of fast α-particles driven by the electric field carries energy in the same direction, which is similar to the Peltier effect. The corresponding energy flux, labeled by the subscript *E*, is given by

$$\begin{aligned} \hat{q}_{\alpha E} &= \frac{4\pi}{3} u_{\alpha 0}^5 E_{\alpha 0} \times \frac{\dot{n}_\alpha \tau_{\alpha e}^2}{4\pi u_{\alpha 0}^3} \times \frac{6eE_x}{m_\alpha u_{\alpha 0}^2} \times \int_0^1 \left(z^3 + \varepsilon^3\right)^{\frac{\mu - 4 + i\chi_\alpha}{3}} z^{6-\mu} \varphi_{01}(z)\, dz \\ &= \hat{\Pi}_\alpha E_x, \end{aligned} \qquad (107)$$

Here, the perpendicular and cross-gradient Peltier coefficients are found from

$$\Pi_{\alpha\perp} = \frac{2e\dot{n}_\alpha \tau_{\alpha e}^2 E_{\alpha 0}}{m_\alpha} \mathrm{Re}\left(\hat{I}_{011}\right), \quad \Pi_{\alpha\wedge} = -\frac{2e\dot{n}_\alpha \tau_{\alpha e}^2 E_{\alpha 0}}{m_\alpha} \mathrm{Im}\left(\hat{I}_{011}\right), \qquad (108)$$

with the sign choice ensuring that both coefficients are positive. The small-$\varepsilon$ expansion of the corresponding integral (79) is

$$\hat{I}_{011} = \frac{1}{2(3+i\chi_\alpha)} - \frac{40\pi\sqrt{3}}{81(3+i\chi_\alpha)} {}_3F_2\left(-\frac{2}{3},\ 1,\ \frac{\mu+5+i\chi_\alpha}{3};\ 3,\ \frac{4+i\chi_\alpha}{3};\ 1\right)\varepsilon^2 + O\left(\varepsilon^3\right) . \tag{109}$$

The effect of magnetization on the α-particles Peltier coefficients is illustrated in Fig. 15 and is similar to its effect on α-particle energy-density diffusion coefficients, Fig. 11.

## IV. FAST α-PARTICLE CONTRIBUTIONS TO MHD EQUATIONS

### A. Conservation equations

The effect of fusion on the background DT plasma manifests itself in two ways. First, there is fuel depletion, accompanied by a net loss of mass due to neutron escape. Second, there is feedback from the fast α-particles. Thermalization returns some mass to the plasma in the form of helium ash – α-particles thermalized to the ion temperature $T_i$ and slowed down to the velocity $\mathbf{v}$ of the background plasma. Each of these particles carries a corresponding average momentum $m_\alpha \mathbf{v}$, kinetic energy $m_\alpha v^2/2$, and thermal energy $3T_i/2$.

The fusion reaction removes pairs of deuterium and tritium ions from the background plasma at the rate $\dot{n}_\alpha$ cm$^{-3}$s$^{-1}$. The mass of each pair is $m_D + m_T$. Its average momentum, lost from the background plasma, is $\left(m_D + m_T\right)\mathbf{v}$. The average loss of kinetic energy per reaction is $\left(m_D + m_T\right)v^2/2$, whereas the average loss of thermal energy is $\Delta E_-$, which exceeds the thermal energy of the two ions, $3T_i$, due to the Gamow shift, given by Eq. (14).

Since in our model the positively charged fast α-particles are a separate species, their presence leaves the background plasma with a net negative charge, carrying a charge density $-2en_\alpha$ to maintain overall quasineutrality. Consequently, the total current density of the background plasma is

$$\mathbf{j}_{\mathrm{pl}} = \mathbf{j}_i + \mathbf{j}_e = e(n - 2n_\alpha)\mathbf{v} - en(\mathbf{v} + \mathbf{u}_e) = -2en_\alpha \mathbf{v} - en\mathbf{u}_e, \tag{110}$$

where the first term is the convective current associated with the transport of the net negative charge density of the background plasma moving at hydrodynamic velocity $\mathbf{v}$, and the second term,

$$\mathbf{j}_e = -en\mathbf{u}_e, \tag{111}$$

is the regular current density associated with the electron drift velocity $\mathbf{u}_e$ relative to the heavy ion species.

The fast α-particle population supplies the ash at the rate $S_\alpha$ cm$^{-3}$s$^{-1}$, adding mass at the rate $m_\alpha S_\alpha$. The rates of momentum, kinetic, and thermal energy addition are $m_\alpha S_\alpha \mathbf{v}$, $m_\alpha v^2 S_\alpha / 2$, and $(3T_i/2)S_\alpha$, respectively. The bulk force exerted upon the background plasma from the fast α-particle population is $\mathbf{R}_\alpha$, which is equal and opposite to the drag force (48) exerted by the background plasma upon the fast α-particle population. The power corresponding to the work done by this force on the background plasma moving at velocity $\mathbf{v}$ is $\mathbf{v} \cdot \mathbf{R}_\alpha$. The inertial work performed by the background plasma on the relative α-particle flux is represented by the last term on the right-hand side of Eq. (45). The α-heating power, that is, the power dissipated by the friction force acting upon the α-particles per unit volume, is $Q_\alpha$.

Collecting these contributions, we obtain the feedback represented by the terms on the right-hand sides of the balance equations written for the background plasma in the conservative form:

$$\frac{\partial \rho}{\partial t} + \nabla \cdot (\rho \mathbf{v}) = -(m_D + m_T)\dot{n}_\alpha + m_\alpha S_\alpha , \tag{112}$$

$$\frac{\partial}{\partial t}(\rho \mathbf{v}) + \nabla \cdot \bar{\mathsf{P}}_{\text{pl}} = \left[-(m_D + m_T)\dot{n}_\alpha + m_\alpha S_\alpha\right]\mathbf{v} + \mathbf{R}_\alpha - 2en_\alpha \mathbf{E}^* + \frac{1}{c}\mathbf{j}_e \times \mathbf{B} , \tag{113}$$

$$\begin{aligned} &\frac{\partial \bar{\mathcal{E}}_{\text{pl}}}{\partial t} + \nabla \cdot \bar{\mathsf{q}}_{\text{pl}} = -\left(\frac{m_D + m_T}{2} v^2 + \Delta E_-\right)\dot{n}_\alpha \\ &+\left(\frac{1}{2} m_\alpha v^2 + \frac{3}{2} T_i\right) S_\alpha + m_\alpha \mathbf{J}_\alpha \frac{d\mathbf{v}}{dt} + \mathbf{v} \cdot \mathbf{R}_\alpha + Q_\alpha + \mathbf{j}_e \cdot \mathbf{E} - 2en_\alpha \mathbf{E} \cdot \mathbf{v}. \end{aligned} \tag{114}$$

In Eqs. (113), (114), $\rho$ , $\bar{\mathsf{P}}_{\text{pl}}$ , $\bar{\mathcal{E}}_{\text{pl}}$ , and $\bar{\mathsf{q}}_{\text{pl}}$ denote, respectively, the mass density, stress tensor (including inertial terms $\rho \mathbf{v}\mathbf{v}$ and viscous stresses $\hat{\pi}$ ), energy density (including internal and kinetic energy), and energy flux density of the background plasma in the laboratory reference frame. On the right-hand side of Eq. (113), the electromagnetic (Lorentz) force density has been grouped using our self-consistent relations:

$$\begin{aligned} \mathbf{f}_{L,\text{pl}} &= -2en_\alpha \mathbf{E}^* + \frac{1}{c}\mathbf{j}_e \times \mathbf{B} = -2en_\alpha \mathbf{E} + \frac{1}{c}(-2en_\alpha \mathbf{v} + \mathbf{j}_e) \times \mathbf{B} \\ &= -2en_\alpha \mathbf{E} + \frac{1}{c}\mathbf{j}_{\text{pl}} \times \mathbf{B}, \end{aligned} \tag{115}$$

where $\mathbf{j}_{\text{pl}}$ is given by Eq. (110).

Correspondingly, in the separate energy balance equations for the electron and ion plasma species, we partition the electromagnetic work $\mathbf{j}_{\text{pl}} \cdot \mathbf{E}$ and the mechanical feedback terms in Eq. (114) according to which species physically carries the associated energy flow. Because the background ions carry virtually all of the background plasma mass, they absorb the convective electrostatic work $-2en_\alpha \mathbf{v} \cdot \mathbf{E}$ associated with the charge imbalance. Thus, the first five terms on the right-hand side of Eq. (114) – representing fuel depletion, ash addition, the inertial work performed on the alpha relative flux, and the mechanical work $\mathbf{v} \cdot \mathbf{R}_\alpha$ belong to the right-hand side of the ion energy balance equation, along with the term $-2en_\alpha \mathbf{v}\cdot\mathbf{E}$. Conversely, the highly mobile electrons carry the conduction current. The work done on this relative drift produces the power density $\mathbf{j}_e \cdot \mathbf{E} = -en\mathbf{u}_e \cdot \mathbf{E}$, which belongs to the right-hand side of the electron energy equation, governing Ohmic heating, thermal transport, and electron pressure work. The collisional α-heating power $Q_\alpha$ is split between the electron and ion species, which receive $F_e Q_\alpha$ and $F_i Q_\alpha$, respectively, where the partition fractions $F_e$ and $F_i$, are given by Eq. (34).

While the overall energy balance equations themselves remain exact consequences of the conservation laws, this specific partition represents the closure within our Chapman-Enskog diffusion framework, distinct from alternative closures such as multi-group diffusion or Monte Carlo simulations.

In our diffusion approximation, we also do not take into account the thermal energy $3T_i/2$ per ash ion returned to the background plasma at the thermalization rate $S_\alpha$ because the integration determining both the equilibrium quantities and transport coefficients extends down to $u=0$ (no sink, no cutoff). This omission is partly compensated by also neglecting the thermal energy $\Delta E_+$ added to the fusion energy $E_{\alpha 0}$ at birth, which stems from replacing the finite-width fusion source $\Sigma_\alpha(u)$ with the monoenergetic source $\dot{n}_\alpha \delta(u-u_{\alpha 0})/(4\pi u_{\alpha 0}^2)$, as explained in Section II.

The results of Sections II.F and III are used to evaluate the fast α-particle coupling terms in the Chapman-Enskog approximation. In the approximation used to derive our transport coefficients, the equilibrium values are:

$$S_{\alpha,\mathrm{eq}} = \dot{n}_\alpha, \quad Q_{\alpha,\mathrm{eq}} = \dot{n}_\alpha E_{\alpha 0}, \tag{116}$$

and there is no equilibrium contribution to the bulk force $\mathbf{R}_\alpha$. The contributions of the isotropic correction (70) account for the non-steadiness of the fast α-particle population:

$$S_{\alpha 0} = -\int d^3u \, \mathrm{St}\left(f_{\alpha 0}\right) = -\frac{\partial n_{\alpha,\mathrm{eq}}}{\partial t}, \tag{117}$$

$$Q_{\alpha 0} = -\int d^3u \frac{m_\alpha u^2}{2} \mathrm{St}\left(f_{\alpha 0}\right) = -\frac{\partial \mathcal{E}_{\alpha,\mathrm{eq}}}{\partial t}, \tag{118}$$

where $n_{\alpha,\mathrm{eq}}$ and $\mathcal{E}_{\alpha,\mathrm{eq}}$ are the equilibrium densities given by the main terms in Eqs. (63) and (65), respectively, and the subscript 0 refers to the isotropic contributions from the zero-order Legendre polynomial in Eq. (68). These results have a clear physical interpretation: a positive rate of change $S_{\alpha 0}$ indicates that the equilibrium number density of fast α-particles is decreasing over time. Consequently, some of the α-particles thermalizing at the present moment were born earlier when the fusion source was stronger. This implies that the number density of thermalized α-particles increases slightly faster than it would if present conditions were maintained indefinitely. Similar logic applies to the time variation of the α-heating power, $Q_{\alpha 0}$.

The anisotropic correction determines – to the first order captured by the Chapman-Enskog method – the particle flux $\mathbf{J}_\alpha$, the energy flux $\mathbf{q}_\alpha$, and the current density $\mathbf{j}_\alpha$. These approximate the corresponding terms in the exact balance equations. Substituting these expressions allows the exact balance equations to be used for the approximate evaluation of the coupling terms $S_\alpha$, $Q_\alpha$, and $\mathbf{R}_\alpha$ in the fluid equations (112)-(114).

Specifically, the exact particle balance equation (43) is transformed into:

$$\frac{dn_{\alpha,\mathrm{eq}}}{dt} + n_{\alpha,\mathrm{eq}} \nabla \cdot \mathbf{v} + \nabla \cdot \mathbf{J}_\alpha = -S_\alpha + \dot{n}_\alpha, \tag{119}$$

where the equilibrium number density $n_{\alpha,\mathrm{eq}}$ is given by Eq. (63), and the flux $\mathbf{J}_\alpha$ is expressed via the gradients of the background plasma parameters by the formulas of Section III.A. Equation (119) is now interpreted as an algebraic equation for $S_\alpha$. The first term on the left-hand side accounts for the non-steadiness of the fast α-particle population; the second represents the reaction of this population to the compression or expansion of the background plasma; and the third term captures the non-locality of α-particle thermalization.

The magnitudes of these terms scale as the hydrodynamic velocity $|\mathbf{v}|$ and the α-diffusion velocity $|\mathbf{J}_\alpha|/n_{\alpha,\mathrm{eq}} \sim u_{\alpha 0}(l_\alpha / L_{\mathrm{hydro}})$, respectively, where $L_{\mathrm{hydro}}$ is the hydrodynamic gradient length. At stagnation, $|\mathbf{v}| < v_{Ti} \ll u_{\alpha 0}$, but the diffusion approximation requires that $l_\alpha / L_{\mathrm{hydro}} \ll 1$, meaning the ordering of these contributions is not immediately obvious. However, if these terms are of the same order, it implies that the fast α-particles move together with the background plasma and are thereby well-confined.

The same argument applies to the fast α-particle energy balance equation, which, with the aid of Eq. (66), is expressed as:

$$\frac{d\mathcal{E}_{\alpha,\mathrm{eq}}}{dt} + \frac{5}{3}\mathcal{E}_{\alpha,\mathrm{eq}}\nabla\cdot\mathbf{v} + \nabla\cdot\mathbf{q}_\alpha = -Q_\alpha + \dot{n}_\alpha E_{\alpha 0}, \tag{120}$$

where the equilibrium energy density $\mathcal{E}_{\alpha,\mathrm{eq}}$ is given by Eq. (65), and the energy flux $\mathbf{q}_\alpha$ is expressed via the gradients of the background plasma parameters by the formulas of Section III.B. Equation (120) is interpreted as an algebraic equation for $Q_\alpha$. The coefficient 5/3 accounts for the work done by the fast α-particle population as the background plasma expands or compresses. Here, the Ohmic heating term $\mathbf{j}_\alpha \cdot \mathbf{E}^*$ has been omitted because, as demonstrated in Section III.C, the fast α-particle current density generated by the local field $\mathbf{E}^*$ is much smaller than the electron current density, and the classic electron Ohmic heating is negligible under fusion conditions. For the same reason, the corresponding flux contribution can be neglected when evaluating $\mathbf{J}_\alpha$ in Eq. (119), as can the Peltier contribution to the energy flux $\mathbf{q}_\alpha$. Also omitted is the term representing the work performed by the accelerated background plasma on the relative α-particle mass flux.

In the low-temperature limit $\varepsilon \to 0$, Eq. (65) reduces to $\mathcal{E}_{\alpha,\text{eq}} = \dot{n}_\alpha \tau_{\alpha e} E_{\alpha 0} / 2$. The equilibrium value of the α-heating term is then given by $Q_{\alpha,\text{eq}} = 2\nu_{\alpha e}\mathcal{E}_{\alpha,\text{eq}}$, and the perpendicular flux of the fast α-particle energy reduces to: $\mathbf{q}_{\alpha\perp} = -2\kappa_{p\perp}\nabla\mathcal{E}_{\alpha,\text{eq}}$. The diffusion approximation requires that the terms on the right-hand side of Eq. (120), $\dot{n}_\alpha E_{\alpha 0}$ and $2\nu_{\alpha e}\mathcal{E}_{\alpha,\text{eq}}$, remain close to each other. Treating the α-energy density $\mathcal{E}_\alpha$ as an unknown quantity to be determined by the diffusion equation itself (thereby removing the "eq" subscript) and assuming a plasma at rest ( $\mathbf{v} = 0$ ), Eq. (120) transforms into the heuristic Eq. (2). Without the $\mathbf{v} = 0$ assumption, Eq. (120) transforms into a generalized version of Eq. (2) containing proper convection and $PdV$ work terms.

A key distinction exists between the balance equation (120) derived strictly from the diffusion approximation and its heuristic counterpart, Eq. (2). In the former case, the fast α-particle energy density $\mathcal{E}_\alpha = \mathcal{E}_{\alpha,\text{eq}}$ is a known function of the background plasma parameters. However, the heuristic Eq. (2) offers distinct advantages when the diffusion approximation formally breaks down. For instance, if the source term $\dot{n}_\alpha E_{\alpha 0}$ features a sharp boundary – such as in a uniform cylinder of magnetized DT plasma – the diffusion equation (120) fails due to diverging gradients at the interface. Equation (2) remains non-singular and provides a good approximation to the Monte Carlo simulations in Ref. [33].

Because a sharp boundary invalidates the strict diffusion approximation, formulations like Eq. (120) require flux limiters incorporated into the diffusion coefficients to prevent the predicted drift velocity of the fast α-particles from artificially approaching $u_{\alpha 0}$. While this heuristic approach successfully introduces an effective limiter for the α-particle energy flux, it is unsuitable for cases beyond the low-temperature limit $\varepsilon \to 0$, where ion friction must be explicitly taken into account.

Finally, neglecting the Ohmic term, the momentum balance equation reads:

$$m_\alpha n_{\alpha,\mathrm{eq}} \frac{d\mathbf{v}}{dt} + m_\alpha \left[ \frac{d\mathbf{J}_\alpha}{dt} + (\nabla \cdot \mathbf{v})\mathbf{J}_\alpha + (\mathbf{J}_\alpha \cdot \nabla)\mathbf{v} \right] + \nabla P_{\alpha,\mathrm{eq}} - \frac{1}{c} \mathbf{j}_\alpha \times \mathbf{B} = -\mathbf{R}_\alpha . \tag{121}$$

The first two terms on the left-hand side of Eq. (121) represent the inertia of the fast α-particles, which are advected with the background plasma at velocity $\mathbf{v}$ and move relative to it at the diffusive velocity $\mathbf{J}_\alpha / n_{\alpha,\mathrm{eq}}$. The relative magnitudes of these terms scale identically to those in Eqs. (119) and (120). These diffusive-velocity terms are of the order of $\tau_{\alpha e} / \tau_{\mathrm{hydro}} \ll 1$ compared to the $\nabla P_{\alpha,\mathrm{eq}}$ term, where $\tau_{\mathrm{hydro}}$ is the characteristic hydrodynamic timescale. Hence, the first two terms are much smaller than the pressure-gradient force. However, the left-hand side of Eq. (121) also contains the Ampère force, $\mathbf{j}_\alpha \times \mathbf{B} / c$, which can be comparable to the pressure-gradient force, balancing it, as shown in the following subsection.

### B. Pressure gradient and Ampère forces

The pressure gradient force term in the left-hand side of Eq. (121) is significant in the high-temperature, high-yield laser-fusion and MagLIF parameter range, as demonstrated in Refs. [40, 41] and illustrated in Fig. 5(a). The next term represents the Ampère force associated with the α-particle current, which is specific to MIF. If the fast α-particles are strongly magnetized, $\chi_\alpha \gg 1$, then it is commensurate with the pressure gradient term, as shown below. These terms in the left-hand side of Eq. (121) can be comparable to the pressure gradient of the background plasma, and, hence, affect its force balance.

While MHD pressure gradient and Ampère forces are generally independent, the forces $-\nabla P_{\alpha,\mathrm{eq}}$ and $\mathbf{j}_\alpha \times \mathbf{B}/c$ in Eq. (121) are driven by the same background plasma gradients. Figure 6 illustrates that the directions of these forces are opposite. Let us compare their magnitudes. Using Eqs. (34), (65), (66), (84), (85), and (89), we evaluate both forces in the direction of the gradient:

$$\frac{\partial P_{\alpha,\mathrm{eq}}}{\partial x} = \frac{1}{3}\frac{\partial}{\partial x}\left(\dot{n}_\alpha \tau_{\alpha e} E_{\alpha 0}\right)\left[F_e - 2H\left(\frac{1}{1+\varepsilon^3} - F_e\right)\right], \tag{122}$$

$$\frac{1}{c} j_{\alpha\wedge} B = \frac{1}{3}\frac{\partial}{\partial x}\left(\dot{n}_\alpha \tau_{\alpha e} E_{\alpha 0}\right)\left[-2\chi_\alpha \operatorname{Im}\left(\hat{I}_{000}\right) + 6\varepsilon^3 \chi_\alpha \operatorname{Im}\left(\hat{I}_{100}\right) H\right]. \tag{123}$$

Here, we have introduced a dimensionless parameter

$$H = \frac{\partial \ln \varepsilon / \partial x}{\partial \ln\left(\dot{n}_\alpha \tau_{\alpha e}\right)/\partial x}. \tag{124}$$

To estimate the relative magnitudes of gradients in Eq. (124), let us assume that the background plasma at stagnation is both isothermal and isobaric: $T_i = T_e$, $nT_e = const$. Then both $\dot{n}_\alpha \tau_{\alpha e}$ and $\varepsilon$ are functions of temperature $T_e$, with the latter given by Eq. (27), $\varepsilon \propto T_e^{1/2}$, and, therefore,

$$H = \frac{\partial \ln \varepsilon / \partial \ln T_e}{\partial \ln\left(\dot{n}_\alpha \tau_{\alpha e}\right)/\partial \ln T_e} = \left[1 + 2\frac{d \ln\langle \sigma v\rangle_{DT}}{d \ln T_i} - \frac{3}{\ln \Lambda_{\alpha e}}\right]^{-1} \approx \left[2.9 \ln\left(\frac{66\ \mathrm{keV}}{T_e}\right)\right]^{-1}, \tag{125}$$

where the Bosch and Hale fit for fusion reactivity [43] was utilized.

Using Eqs. (122), (123), and (125), we define the force balance factor $\Psi_\alpha$ by

$$-\frac{\partial P_{\alpha,\text{eq}}}{\partial x}+\frac{1}{c}j_{\alpha\wedge}B=-\frac{1}{3}\frac{\partial}{\partial x}\left(\dot{n}_{\alpha}\tau_{\alpha e}E_{\alpha 0}\right)\times\Psi_{\alpha}, \tag{126}$$

where

$$\begin{aligned}\Psi_{\alpha}&=F_e+2\chi_{\alpha}\,\text{Im}\left(\hat{I}_{000}\right)-2H\left[\frac{1}{1+\varepsilon^3}-F_e+3\varepsilon^3\chi_{\alpha}\,\text{Im}\left(\hat{I}_{100}\right)\right]\\&=\frac{1}{\chi_{\alpha}^2+1}+O(\varepsilon).\end{aligned} \tag{127}$$

In the low-temperature limit $\varepsilon\to 0$, the oppositely directed pressure gradient and Ampère forces almost balance each other at strong magnetization, $\chi_{\alpha}\gg 1$, with the resulting force in the same direction as $-\nabla P_{\alpha,\text{eq}}$ but reduced by a factor of $\chi_{\alpha}^2+1$. In Figure 16, the force balance factor $\Psi_{\alpha}$ is plotted vs. $\chi_{\alpha}$ for finite plasma temperatures $T_e=5$, …, 25 keV, with the first-order approximation (127) shown as a dashed black line. At large magnetization, $\Psi_{\alpha}$ is finite and negative, demonstrating that the Ampère force is larger than the α-particle pressure gradient and dominates the resulting force (126). The latter does not vanish in the limit $\chi_{\alpha}\to\infty$, but it is still much smaller than the normalization value, $(1/3)\partial\left(\dot{n}_{\alpha}\tau_{\alpha e}E_{\alpha 0}\right)/\partial x$ – by a factor of ~6 at $T_e=25$ keV and $\chi_{\alpha}\gg 1$ – so, in the first approximation, the resulting radial force can be neglected in MagLIF. On the other hand, magnetization of the fast α-particles does not affect the axial pressure-gradient force, $-(\partial/\partial z)P_{\alpha,\text{eq}}$, which still needs to be accounted for in 2D simulations.

## V. EVALUATION OF MODEL TERMS AND PARAMETER RANGES IN HIGH-YIELD MAGLIF SIMULATION

In this section, we explore the relevant parameter regimes of the transport model and evaluate the relative magnitudes of its constituent terms by postprocessing simulation data from a facility-class, high-yield MagLIF target. The underlying background hydrodynamic profiles are generated using Pacific Fusion's internal branch of the FLASH code [17], driven by a circuit model representative of the Demonstration System (DS) [68]. For this analysis, we consider the one-dimensional dynamics of an ice-layered cryogenic target assembly, sketched in Fig. 17 (a). The initial assembly comprises a central DT vapor region with a radius of 3.669 mm surrounded by a 304 μm-thick DT ice layer and contained within an aluminum liner extending to an outer radius of 4.734 mm (initial aspect ratio AR = 6.22). Rather than assessing whether a specific diffusion model fully captures the target performance, our objective here is to evaluate the spatial and temporal variations of the characteristic transport lengths, nonlocality indicators, deposited $\alpha$-energy partitions, and force balance terms across the distinct burning plasma regimes realized in such high-gain implosions.

The simulation is carried out using Pacific Fusion's internal branch of the FLASH code [17]. An initial, homogeneous axial magnetic field of 7 T is imposed throughout the computational domain. Coupling the target dynamics to the DS circuit model yields the load current waveform shown in Fig. 17 (b). It reaches above 60 MA prior to the implosion stagnation event, which occurs at 164.8 ns, shortly before the dip in the current trace. Laser preheat is represented by an energy deposition source applied to the plasma electrons as a proxy for inverse bremsstrahlung absorption, depositing 30 kJ/cm over an 8 ns pulse beginning at 130 ns. At the onset of preheat, the target has already undergone substantial compression, corresponding to a convergence ratio of 1.35. This late preheat timing minimizes conduction losses prior to stagnation (see Ref. [69] for an analytical description of preheat propagation), though at the cost of reduced $PdV$ heating of the fuel.

The baseline α-transport model implemented in Pacific Fusion's branch of FLASH adopts the heuristic formulation of Ref. [31], in which the α-particle energy density field $\mathcal{E}_\alpha$ obeys the convective-diffusive equation (2). This model neglects collisions between α-particles and background ions, using the diffusion coefficient given by Eq. (3). While the implementation of the upgraded transport closures developed here is currently underway in FLASH (to be reported in a future publication), the present simulation provides a realistic background plasma state to evaluate the relevant transport parameters and assess the relative magnitudes of the newly derived terms.

The burn history for the target considered in this section is shown in Fig. 18 (a). The predicted yield exceeds 2 GJ/cm. Previous 1D implosion simulations driven by the DS used a target height of 6 mm to accommodate circuit-modeling considerations [68]. In light of this, to infer an equivalent 2D yield from the present 1D results, an effective hot spot length of 4 mm (with 2 mm of dud fuel at the ends) is assumed. The extrapolated yield would exceed the DS's 80 MJ of stored electrical energy, leaving a margin to accommodate yield penalties due to end losses and hydrodynamic instabilities. The evolution of the mass in the hot spot (defined as the central region where the electron temperature exceeds 1.2 times the radiation temperature) is depicted in Fig. 18 (b). We identify a regime of robust burn propagation beginning at $t = 165$ ns ($t_{\text{prop}}$), when the rate of ice-mass engulfment, $d \ln m_{\text{hs}} / dt$, doubles, and ending at the fuel exhaustion time $t = 165.8$ ns ($t_{\text{exh}}$), when the burn wave has propagated through and exhausted the entire DT ice mass. During this period, the burn-averaged ion temperature remains relatively constant before increasing steeply for $t > t_{\text{exh}}$, once the entire fuel mass begins burning volumetrically. The subsequent decrease in DT mass within the hot spot for $t > t_{\text{exh}}$, is due to fuel burnup depletion. We focus on the time at which burn has propagated through half of DT ice mass, $t = 165.4$ ns ($t_{\text{half}}$), as representative of this propagation regime to examine the parameter ranges and evaluate the magnitudes of the relevant transport terms.

Radial profiles of temperature, axial magnetic field, and plasma density at $t = t_{\text{half}}$ are shown in Fig. 19 (a). The hot spot encompasses a central region of low, relatively uniform density and an outer region containing fuel mass ablated from the ice. It is surrounded by the denser DT ice layer (now in a plasma state) that extends to the fuel–liner interface. The magnetic field profile exhibits cavitation in the central region due to Nernst convection [69, 70], which accumulates the magnetic flux and causes the field to peak in the outer region of the hot spot. Figure 19 (b) shows the α-energy production and deposition rates, along with the α-particle characteristic number density. The peak of the energy deposition profile is shifted toward denser fuel layers relative to the peak of the energy production profile, reflecting the nonlocal nature of α-particle transport [71]. In Eq. (120), this outward shift arises from the divergence term on the left-hand side, which enhances deposition where the gradient of the α-production rate is negative. Because α-particles take longer to deposit their energy in hotter, lower-density regions, their number density peaks at the central region of the hot spot.

To assess the degree of nonlocality in α-particle transport, radial profiles of the deceleration path length $l_\alpha$ and alpha gyroradius $r_L$ are shown in Fig. 20 (a). Both characteristic lengths are large in the central low-density core, where α-particle motion is essentially ballistic and the diffusion approximation is not valid. However, this is not a major concern because macroscopic gradients are small across this plateau. In the outer hot spot region containing the ablated fuel mass, gradients steepen and transport becomes more relevant. Here, while the deceleration length remains long, the gyroradius decreases and remains below the hot spot radius, providing effective magnetic confinement of the α-particles. Near the hot spot boundary, $l_\alpha$ significantly decreases to become the dominant confining length scale.

Although both characteristic lengths exceed the α-energy gradient scale length $L_{\mathcal{E}}$, their ratios to $L_{\mathcal{E}}$ remain of order unity. While the formal asymptotic validity of the diffusion approximation requires these ratios to be much less than unity, values of order unity indicate that the transport is not far from the diffusive limit, well short of collisionless free streaming. Nevertheless, because the ablation front is driven primarily by α-heating, its characteristic width $L_\rho = \rho / |\nabla \rho|$ naturally adjusts to the scale of the alpha deceleration path, so that $l_\alpha / L_\rho \sim 1$ is an inherent feature of such burn wave. Under these conditions, the diffusion approximation – which models net transport as a small difference between nearly isotropic counter-streaming fluxes – cannot be fully accurate for a propagating burn front dominated by strongly directed, outward-streaming alphas. Incorporating a flux limiter (beyond the scope of the present work) would be a necessary regularizer to cap these directional fluxes. Figure 20 (b) shows the temporal evolution of the ratios $r_L / L_\rho$ and $l_\alpha / L_\rho$ evaluated at the ablation front, which lies adjacent to the hot spot boundary and is located where the density-gradient scale length $L_\rho$ reaches its minimum. During the steady propagation phase ($t \approx 165.0$ to 165.6 ns, represented in our examples by $t_{\text{half}}$), the ratio of the deceleration length to $L_\rho$ remains between 1.5 and 2, before rising sharply as the burn wave approaches complete fuel exhaustion.

In Sec. IV, the differences between the heuristic formulation of Eq. (2) and the balance equation (120) were discussed. In the baseline heuristic model of Ref. [31], the α-energy density $\mathcal{E}_\alpha$ is evolved dynamically as an independent field governed by the convective-diffusive equation (2). In contrast, the updated model evaluates the local equilibrium α-energy density $\mathcal{E}_{\alpha,\mathrm{eq}}$ calculates the resulting α-particle fluxes directly from the background plasma parameters and their gradients.

Figure 21 (a) compares both approaches by displaying the dynamically evolved $\mathcal{E}_\alpha$ alongside $\mathcal{E}_{\alpha,\mathrm{eq}}$ at mid-burn propagation ($t = t_{\mathrm{half}}$). Although both profiles follow similar spatial trends, $\mathcal{E}_{\alpha,\mathrm{eq}}$ is systematically higher across most of the hot spot, differing by approximately 20% in the core. In the heuristic model [Eq. (2)], the gradient of the evolved field $\mathcal{E}_\alpha$ drives the flux divergence that shifts the heating peak in Fig. 19 (b). By contrast, in the updated theory, $\mathcal{E}_{\alpha,\mathrm{eq}}$ is determined entirely by the local background plasma parameters, independent of any energy flux down the gradient of the characteristic energy density $\dot{n}_\alpha \tau_{\alpha e} E_{\alpha 0}$.

Figure 21 (b) displays the ratios of the equilibrium alpha pressure and number density to the background plasma thermal pressure and electron density, respectively. While the density ratio remains small ($n_{\alpha,\mathrm{eq}} / n_e \lesssim$ 3%), in accordance with the asymptotic ordering of this model, the pressure exerted by the fast α-particles exceeds 50% of the thermal pressure in the core – a significant mechanical effect accounted for in the present model; cf. Fig. 5(a).

The model derived in this article also improves upon the previous heuristic model by incorporating Coulomb friction with the background ions. This extends its range of applicability to temperatures well above the ignition threshold characteristic of high-gain MagLIF [15]. Ion friction is quantified by the dimensionless parameter $\varepsilon$ [Eq. (27)], which directly determines the α-energy partition fraction to electrons ($F_e$), shown in Fig. 22 (a). The electron partition fraction falls significantly below unity in the core (dropping to $\approx 0.4 - 0.5$) and increases steadily toward the ablation front. This behavior contrasts sharply with the fixed 90% electron–10% ion energy partition assumed in the simulation. Ion friction should therefore be accounted for not only in the local energy deposition but also in the transport coefficients to model burn propagation accurately. Figure 22 (b) shows the alpha Hall parameter $\chi_\alpha$ during burn propagation. It peaks at large values ( $\chi_\alpha \gg 1$ ) in the outer region of the hot spot, where α-particles are strongly magnetized, and decreases steeply toward the ablation front as the slowing-down path shrinks and the magnetic field drops beyond the magnetic flux-pileup layer shown in Fig. 19(a).

Using the plasma background frozen at $t = t_{\mathrm{half}}$ , we post-process and evaluate the alpha-particle flux and momentum balance terms derived in Eqs. (43) and (44). Figure 23 (a) displays the resulting alpha-current streamline in the $\left(r,\theta\right)$ plane, obtained by integrating $dr/d\theta = rJ_{\alpha r}\left(r\right)/J_{\alpha\theta}\left(r\right)$ with initial condition $r_{\theta=0}$ = 65 μm, over a composite background that combines the azimuthally symmetric maps of equilibrium α-particle number density $n_{\alpha,\mathrm{eq}}$ in the central burning core (labeled "Hot spot") and background plasma density $\rho$ in the dense peripheral shell (labeled "Ice"). Here, considering both α-particle diffusion and thermodiffusion, the radial and azimuthal flux components correspond to $J_{\alpha\perp}$ and $J_{\alpha\wedge}$, respectively. Because radial transport is inhibited by the axial magnetic field while the azimuthal cross-field flux remains uninhibited, the collective drift of the fast α-particles traces an outward spiral. Viewed from the positive $z$ direction with $\mathbf{B}$ pointing out of the page, the diamagnetic α-current density $j_{\alpha\wedge} = 2eJ_{\alpha\wedge}$ flows clockwise.

Finally, Figure 23 (b) shows the radial profiles of the α-pressure gradient force, $-\nabla P_{\alpha,\mathrm{eq}}$ and the Ampère force $\mathbf{j}_\alpha \times \mathbf{B}/c$ . Although governed by independent kinetic integrals, both forces are driven by the same background plasma gradients and act in opposite directions: the pressure gradient accelerates outward fuel expansion, reducing the confinement time, whereas the inward Ampère force aids confinement. In accordance with the asymptotic closure in Eq. (127), the two contributions almost exactly cancel across the strongly magnetized region.

## VI. DISCUSSION AND CONCLUSIONS

In this paper, we have revisited and extended the diffusion theory of fast α-particle transport in dense magnetized DT fusion plasmas, establishing a self-consistent kinetic framework tailored to Magneto-Inertial Fusion (MIF) conditions. Starting from the Fokker–Planck kinetic equation and utilizing the Chapman–Enskog method, we have generalized earlier low-temperature diffusion models to incorporate the full dynamic Coulomb friction from both electrons and ions of the background plasma. The primary results and physical insights obtained in this work are summarized below.

1) We have derived exact closed-form analytic formulas – expressed via Gauss hypergeometric integrals [Eq. (79)] – for the entire set of twelve independent transport coefficients: the six perpendicular and six cross-gradient coefficients describing particle diffusion and thermodiffusion, ($D_{p\perp}$, $D_{p\wedge}$, $D_{T\perp}$, $D_{T\wedge}$), energy-density diffusion and thermodiffusion ($\kappa_{p\perp}$, $\kappa_{p\wedge}$, $\kappa_{T\perp}$, $\kappa_{T\wedge}$), α-particle electrical conductivities ($\sigma_{\alpha\perp}$, $\sigma_{\alpha\wedge}$), and Peltier coefficients ($\Pi_{\alpha\perp}$, $\Pi_{\alpha\wedge}$) . The corresponding parallel coefficients are automatically obtained from the unmagnetized limit, $D_{\parallel} = D_{\perp}\left(B = 0\right)$ [Eq. (80)].

2) In the low-temperature limit ($\varepsilon = u_c / u_{\alpha 0} \to 0$), our formulas recover the baseline heuristic model of Ref. [31] as the zeroth-order approximation [Eqs. (87), (98)]. At temperatures near ignition (~5 keV), the small-$\varepsilon$ expansion accurately reproduces the transport coefficients. However, at higher temperatures characteristic of high-gain MagLIF ($T_e \sim 15$ to 25 keV), ion drag significantly reduces the deceleration length $l_\alpha$ below the electron-only deceleration path $l_{\alpha e}$ and flattens the quasi-equilibrium distribution function $f_{\alpha,eq}$, rendering the baseline electrons-only model inadequate.

3) We have established asymptotic formulas for all transport coefficients in the strong-magnetization limit ($\chi_\alpha \gg 1$) providing simple closed-form expressions that bypass the need to evaluate hypergeometric functions and integrals. Unlike the monotonic suppression $\sim\left(1+\chi_\alpha^2\right)^{-1}$ predicted by electron-only friction, including velocity-dependent ion friction introduces a "hard stop" at the end of the α-particle deceleration trajectory. This phase-sensitive ballistic stopping produces damped oscillations in the transport coefficients as functions of the Hall parameter $\chi_\alpha$. Our asymptotic expansion in the Appendix captures the exact oscillation frequency $\ln\left(1+\varepsilon^{-3}\right)/3$ and demonstrates that for particle transport ($m=0$), the relative oscillation amplitude decays slowly as $\chi_\alpha^{-(2-\mu)/3} \approx \chi_\alpha^{-0.54}$ for a DT plasma, keeping the oscillatory structure clearly discernible up to large $\chi_\alpha$. For energy-density diffusion ($m=1$), the oscillations decay much faster as $\chi_\alpha^{-(4-\mu)/3} \approx \chi_\alpha^{-1.21}$, explaining why energy transport is dominated by its smooth monotonic trend.

4) We evaluated the momentum exchange between the fast α-particles and the background plasma. In high-yield magnetized DT targets, the α-particle pressure $P_{\alpha,eq}$ can reach a substantial fraction of the thermal plasma pressure ($P_{\alpha,eq}/P \sim 0.2$ to 0.5). Although MIF plasmas in general, and high-yield MagLIF targets in particular, operate at high plasma $\beta$ where the fluid Ampère force $\mathbf{j}\times\mathbf{B}/c$ is negligible compared to the bulk thermal pressure gradient, the diffusive fast α-particle currents produce remarkably large Ampère forces. These α-driven

Ampère forces are commensurate with the α-particle pressure gradient itself. However, because the diamagnetic cross-gradient α-flux driven mostly by $-\nabla\left(\dot{n}_\alpha \tau_{\alpha e}\right)$ generates an Ampère force $\mathbf{j}_\alpha \times \mathbf{B} / c$ directed opposite to $-\nabla P_{\alpha,eq}$, the two terms largely cancel under strong magnetization ( $\chi_\alpha \gg 1$), reducing the net transverse α-force density by a factor of ~ $\chi_\alpha^{-2}$ [Eq. (127)]. Consequently, net radial momentum feedback from fast α-particles is strongly suppressed in the magnetized core of MagLIF targets, whereas axial α-pressure gradient force $-\partial P_{\alpha,eq} / \partial z$ remains unaffected by magnetic confinement and must be retained in 2D/3D simulations.

As an illustrative example, the post-processed diagnostics of a representative 1D high-yield MagLIF implosion simulated with FLASH (Section V) demonstrate the direct operational relevance of these theoretical results. Because the electron temperature in the burning core reaches up to 30 keV [Fig. 19(a)], the fast α-pressure is not negligible compared to the thermal pressure of the background plasma, exceeding 50% on axis [Fig. 21(b)]. Furthermore, background ions actively participate in α-particle stopping at these multi-keV temperatures, causing a sizable fraction of the α-energy to be deposited directly to the ions [Fig. 22(a)]. While the radial α-pressure gradient tends to accelerate outward fuel expansion and degrade inertial confinement, it is almost exactly counterbalanced by the inward α Ampère force, meaning that the mechanical deconfinement effect of the α-pressure gradient can be significant only in the axial direction. It is a rare treat in fusion physics when a more complete and realistic theoretical model predicts improved rather than degraded performance: here, incorporating ion friction enhances fuel reactivity by heating ions directly, while the self-consistent α-driven Ampère force largely neutralizes the destructive radial expansion driven by the suprathermal pressure gradient.

This example specifically addresses the propagation of a thermonuclear burn wave into dense DT ice. Magnetic α-confinement remains effective near the burn front. However, because propagating burn waves are dominated by strongly directed outward fluxes rather than small differences between counter-streaming populations, the diffusion approximation applies here only qualitatively. In practical radiation-MHD code implementations, regions where background plasma gradients become too steep to satisfy the Chapman–Enskog ordering $l_\alpha / L_{\text{hydro}} \ll 1$ – such as sharp fuel-liner interfaces or steep burn fronts, as in Fig. 20(a) – will require an α flux limiter to cap unphysical drifts below the free-streaming ballistic limit, following the approach validated for inertial fusion plasmas in Ref. [72].

Beyond primary DT α-heating, the present kinetic theory and transport matrix describe the diffusion and magnetic confinement of any fast MeV-energy charged fusion products – such as tritons, protons, and $^3$He ions from DD reactions, or α-particles born at other energies via secondary and tertiary branches – in either pure deuterium or DT fuel mixtures. Modeling the spatial transport and magnetic confinement of these minority products is essential for interpreting nuclear diagnostics [73, 74]. Operationally, this requires only substituting the fast ion's mass $m_f$, charge $Z_f e$, and birth energy $E_{f0}$, while evaluating $\mu = 1 - m_i / m_f$ with the appropriate background ion mass (e.g., $m_i = m_D$ for DD experiments). However, because their production rates are orders of magnitude lower than primary DT $\alpha$-production, their macroscopic feedback on the background plasma – from self-heating to pressure gradients and Ampère forces – remains negligible, allowing them to be treated strictly as non-perturbing test particles.

Earlier studies that calculated electron currents induced by fast α-particles under stagnated MagLIF conditions, Refs. [75, 35], relied upon simplified α-transport models, including the baseline electrons-only model of Ref. [31]. This description must now be generalized using the complete, high-temperature $\alpha$-transport matrix developed here. Fully self-consistent modeling requires formulating a generalized, Braginskii-style Ohm's law that incorporates the Ohkawa electron current induced by collisional friction with fast $\alpha$-particle fluxes [75–77] alongside the electron pressure gradient and thermal forces. Under the constraint of quasineutrality, which permits cross-field electron currents via the Hall electric field [cf. Braginskii, Eq. (6.22)], this generalized Ohm’s law will enable rigorous evaluations of: (a) induced-current and $\alpha$-Nernst modifications to magnetic flux advection [35], which could enhance axial cavitation and peripheral flux accumulation to potentially affect burn propagation; (b) corrections to the net $\alpha$-Ampère force and total plasma force balance; and (c) fast-$\alpha$-driven electrostatic plasma polarization and macroscopic rotational shear, concepts originally explored in early magnetized target studies by Linhart [78].

Addressing these coupled electrodynamic feedback mechanisms will complete the self-consistent alpha-diffusive-MHD description of magnetized burning DT plasmas across the full MIF parameter space.

## AUTHOR DECLARATIONS

### Conflict of Interest

There is no conflict of interest to disclose.

## DATA AVAILABILITY

The data that support the findings of this study are available from the corresponding author upon reasonable request.

## APPENDIX. ASYMPTOTIC FORMULAS FOR HIGH MAGNETIZATION LIMIT

We are seeking the asymptotics of the expression (79) at large $\chi_\alpha \gg 1$. The hypergeometric function in the integrand in the first term in square brackets can be presented as ${}_2F_1\left(1,\ \zeta+\lambda;\ \lambda;\ \xi\right)$, with both second and the third indexes, $\zeta+\lambda$ and $\lambda = \left(3k+2l+2+i\chi_\alpha\right)/3$ considered large parameters, whereas the constant $\zeta = \left(\mu+l\right)/3$ and the argument $\xi = \varepsilon^3/\left(z^3+\varepsilon^3\right)$ are finite. Note that to derive the strong-magnetization asymptotic formulas, we need to assume $\chi_\alpha \gg 3$, which explains the form of denominator in Eqs. (3) and (98). We utilize the transformation formula, Eq. (15.3.4) of Ref. [79], to reduce this hypergeometric function to one where only the third index is large, so that the asymptotic formula (15.7.1) applies:

$$\begin{aligned}
&{}_2F_1\left(1,\ \zeta+\lambda;\ \lambda;\ \xi\right) = \frac{1}{1-\xi}\,{}_2F_1\left(1,\ -\zeta;\ \lambda;\ -\frac{\xi}{1-\xi}\right) \\
&= \frac{1}{1-\xi} + \frac{\zeta}{\lambda}\frac{\xi}{\left(1-\xi\right)^2} + O\left(\lambda^{-2}\right)
\end{aligned} \tag{A1}$$

Therefore, the lowest-order terms of the asymptotic expression of the hypergeometric function in the integrand are

$$\begin{aligned}
&{}_2F_1\left(1,\ \frac{\mu+3k+3l+2+i\chi_\alpha}{3};\ \frac{3k+2l+2+i\chi_\alpha}{3};\ \frac{\varepsilon^3}{z^3+\varepsilon^3}\right) \\
&= \frac{z^3+\varepsilon^3}{z^3} + \left(\frac{\mu+l}{i\chi_\alpha}\right)\frac{\varepsilon^3(z^3+\varepsilon^3)}{z^6} + O\left(\chi_\alpha^{-2}\right).
\end{aligned} \tag{A2}$$

Integration of the first term in the expansion (A2) results in a real-valued, positive-definite function of $\varepsilon$:

$$G_{klm}\left(\varepsilon\right) = \int_0^1 \frac{z^{4+2m+l}}{\left(z^3+\varepsilon^3\right)^{1+k+l}}\,dz\,. \tag{A3}$$

Integration of the second term yields

$$\varepsilon^3\left(\frac{\mu+l}{i\chi_\alpha}\right)\int_0^1 \frac{z^{1+2m+l}}{\left(z^3+\varepsilon^3\right)^{1+k+l}}\,dz = -i\frac{\mu+l}{\chi_\alpha}K_{klm}\left(\varepsilon\right), \tag{A4}$$

where we have defined another real-valued, positive-definite function of $\varepsilon$:

$$K_{klm}\left(\varepsilon\right) = \varepsilon^3\int_0^1 \frac{z^{1+2m+l}}{\left(z^3+\varepsilon^3\right)^{1+k+l}}\,dz. \tag{A5}$$

The results of integration in (A3) and (A5) are expressed via elementary functions of $\varepsilon$, including the auxiliary functions $L_\pm\left(\varepsilon\right)$ defined by Eq. (31):

$$G_{000} = \frac{1}{2} - \frac{1}{6}\varepsilon^2 L_- = \frac{1}{2}F_e, \qquad K_{000} = \frac{1}{6}\varepsilon^2 L_- = \frac{1}{2}F_i \tag{A6}$$

where $F_e$ and $F_e$ are the energy partitions to plasma electrons and ions given by Eq. (34);

$$G_{100} = -\frac{1}{3\left(1+\varepsilon^3\right)} + \frac{1}{9\varepsilon}L_-, \quad K_{100} = \frac{1}{3\left(1+\varepsilon^3\right)} + \frac{1}{18\varepsilon}L_-; \tag{A7}$$

$$G_{010} = \frac{1}{3}\left[\ln\left(1+\varepsilon^{-3}\right) - \frac{1}{1+\varepsilon^3}\right], \quad K_{010} = \frac{1}{3\left(1+\varepsilon^3\right)}; \tag{A8}$$

$$G_{001} = \frac{1}{4} - \varepsilon^3 + \frac{1}{6}\varepsilon^4 L_+, \quad K_{001} = \varepsilon^3 - \frac{1}{6}\varepsilon^4 L_+ \;; \tag{A9}$$

$$G_{101} = \frac{3+4\varepsilon^3}{3\left(1+\varepsilon^3\right)} - \frac{2}{9}\varepsilon L_+, \quad K_{101} = -\frac{\varepsilon^3}{3\left(1+\varepsilon^3\right)} + \frac{1}{18}\varepsilon L_+; \tag{A10}$$

$$G_{011} = \frac{3+5\varepsilon^3}{6\left(1+\varepsilon^3\right)} - \frac{5}{18}\varepsilon^2 L_-, \qquad K_{011} = -\frac{\varepsilon^3}{3\left(1+\varepsilon^3\right)} + \frac{1}{9}\varepsilon^2 L_-. \tag{A11}$$

Note that the integrals $G_{100}$ and $K_{100}$ in Eq. (A7) diverge as $1/\varepsilon$ in the low-temperature limit $\varepsilon \to 0$. However, the calculation of the corresponding thermodiffusion coefficients includes multiplying the function (79) by $3\varepsilon^2$, as in Eq. (90), so these coefficients not only remain finite but vanish in this limit, as in Eq. (91).

Similarly, we calculate the asymptotics of the second term in square brackets on the right-hand side of Eq. (79). The leading term of one hypergeometric function is given by Eq. (A1) with $\xi = \varepsilon^3 / \left(1+\varepsilon^3\right)$ and equals to $1+\varepsilon^3$. The other hypergeometric function has a large second index: ${}_2F_1\left(1,\ \lambda;\ c;\ \xi\right)$, where the large parameter is $\lambda = \left(4-\mu-i\chi_\alpha\right)/3$, whereas the third index $c=\left(8+2m-\mu\right)/3$ and the argument $\xi = 1/\left(1+\varepsilon^3\right)$ are finite. The corresponding asymptotic expression (see Section 2.3.2 of Ref. [80]) is

$$\begin{aligned} {}_2F_1\left(1,\ \lambda;\ c;\ \xi\right) &= \frac{\Gamma(c)}{\Gamma(c-1)\left(-\lambda\xi\right)} + O\left(\chi_\alpha^{1-c}\right) \\ &= -\frac{i}{\chi_\alpha}\left(5+2m-\mu\right)\left(\varepsilon^3+1\right) + O\left(\chi_\alpha^{-\frac{5+2m-\mu}{3}}\right), \end{aligned} \tag{A12}$$

The decaying amplitude, frequency, and phase of the oscillations at large $\chi_\alpha$ are captured by retaining the second term in the asymptotic formula for ${}_2F_1\left(1,\ \lambda;\ c;\ \xi\right)$. Multiplying the leading monotonic term of Eq. (A12) by the full complex correction factor produces an oscillatory addition proportional to $1+\hat{C}$, where $\hat{C}$ is nominally complex. However, the strong ballistic stopping oscillations are exhibited primarily by the perpendicular (diagonal) transport coefficients, which are determined by the real parts $\mathrm{Re}\left(\hat{I}_{klm}\right)$, rather than the cross-gradient coefficients governed by the leading imaginary parts $\mathrm{Im}\left(\hat{I}_{klm}\right)$; see below. Furthermore, this oscillatory correction is of order $O\left(\chi_\alpha^{-(2+2m-\mu)/3}\right)$ relative to the leading monotonic term. For particle transport and electrical conductivity ($m=0$), this relative correction scales as $\chi_\alpha^{-(2-\mu)/3} \approx \chi_\alpha^{-0.543}$ (for the equimolar DT value of $\mu = 0.372$), which decays slower than $\chi_\alpha^{-1}$ and thus dominates over the $O\left(\chi_\alpha^{-2}\right)$ monotonic corrections neglected in Eq. (A1). Conversely, for the energy-diffusion coefficients ($m = 1$), the oscillatory correction decays much faster, as $\chi_\alpha^{-(4-\mu)/3} \approx \chi_\alpha^{-1.209}$, explaining why the energy transport coefficients exhibit negligible oscillations in Figs. 11 and 12.

Restricting our attention to the particle and charge transport components ($m = 0$), we account for these oscillations by multiplying the leading term of Eq. (A12) by the real factor $1 + C_{\text{osc}}$, where

$$C_{\text{osc}} = -\varepsilon\left(1+\varepsilon^3\right)^{\frac{1-\mu}{3}} \Gamma\left(\frac{5-\mu}{3}\right)\left(\frac{\chi_\alpha}{3}\right)^{-\frac{2-\mu}{3}} \cos\left[\frac{\chi_\alpha}{3}\ln\left(1+\epsilon^{-3}\right)-\frac{\pi(2-\mu)}{6}\right]\delta_{m0}, \qquad \text{(A13)}$$

and the Kronecker delta $\delta_{m0}$ ensures that the correction (A13) is only accounted for at $m = 0$. Equation (A13) explicitly captures both the oscillation period of the diffusion coefficients as functions of the Hall parameter, $\Delta\chi_\alpha = 2\pi\tau_{\alpha e} / \tau_{\text{dec}} = 6\pi / \ln\left(1+\varepsilon^{-3}\right)$ [cf. Eq. (29)], and the algebraic power-law decay $\sim \chi_\alpha^{-(2-\mu)/3}$ of those oscillations. At fixed density and temperature, this corresponds to the gyrofrequency period $\Delta\omega_\alpha = \Delta\chi_\alpha / \tau_{\alpha e} = 2\pi / \tau_{\text{dec}}$, reflecting the fact that stopping terminates at the identical gyrophase after each additional full gyration. In the low-temperature limit $\varepsilon \to 0$, the oscillation frequency asymptotically tends to $\ln\left(1/\varepsilon\right)$. Applying Stirling's formula to the ratio of Gamma functions multiplying the oscillatory term $\varepsilon^{1+i\chi_\alpha}$ in square brackets on the right-hand side of Eq. (86) directly reproduces Eq. (A13) in this limit. These oscillation characteristics – both the frequency and the power-law decay – are identical for the α-particle diffusion and thermodiffusion coefficients, as well as for the fast α-particle electrical conductivity, as functions of $\chi_\alpha$.

Combining these results, we obtain the leading-order term for the second expression in Eq. (79):

$$-\frac{{}_2F_1\left(1,\ \dfrac{\mu+3k+3l+2+i\chi_\alpha}{3};\ \dfrac{3k+2l+2+i\chi_\alpha}{3};\ \dfrac{\varepsilon^3}{1+\varepsilon^3}\right)}{\left(1+\varepsilon^3\right)^{2+k+l}\left(5+2m-\mu\right)} \times {}_2F_1\left(1,\ \frac{4-\mu-i\chi_\alpha}{3};\ \frac{8+2m-\mu}{3};\ \frac{1}{1+\varepsilon^3}\right)=\frac{i}{\chi_\alpha}\left(1+\varepsilon^3\right)^{-k-l}\left(1+C_{\text{osc}}\right)+O\left(\chi_\alpha^{-2}\right). \tag{A14}$$

Therefore, the main terms in square brackets in Eq. (79) are

$$[...]\cong G_{klm}+\frac{i}{\chi_\alpha}\left[\left(1+\varepsilon^3\right)^{-k-l}\left(1+C_{\text{osc}}\right)-\left(\mu+l\right)K_{klm}\right]. \tag{A15}$$

The constant prefactor in the expression for $\hat{I}_{klm}$ can be presented as

$$\frac{1}{3k+2l-1+i\chi_\alpha}=-\frac{i}{\chi_\alpha}+\frac{3k+2l-1}{\chi_\alpha^2}+O\left(\chi_\alpha^{-3}\right). \tag{A16}$$

Multiplying (A15) by (A16), separating the real and imaginary parts, and retaining only the leading terms, we arrive at the asymptotic expressions sought for:

$$\begin{aligned}\operatorname{Re}\left(\hat{I}_{klm}\right)&\cong\frac{1}{\chi_\alpha^2}\left[\left(1+\varepsilon^3\right)^{-k-l}\left(1+C_{\text{osc}}\right)+\left(3k+2l-1\right)G_{klm}-\left(\mu+l\right)K_{klm}\right];\\ \operatorname{Im}\left(\hat{I}_{klm}\right)&\cong-\frac{1}{\chi_\alpha}G_{klm}.\end{aligned} \tag{A17}$$

Equations (A17) clearly elucidate why the cross-gradient transport coefficients (such as $D_{p\wedge}$ and $\sigma_{\alpha\wedge}$) exhibit far weaker oscillations than the corresponding perpendicular coefficients $D_{p\perp}$ and $\sigma_{\alpha\perp}$ in Figs. 7, 8, and 14. The leading imaginary part $\operatorname{Im}\left(\hat{I}_{klm}\right)=-G_{klm}/\chi_\alpha$, which determines the cross-gradient fluxes, arises entirely from the monotonic integral $G_{klm}$ and is completely unaffected by the second hypergeometric branch in Eq. (79) that generates $C_{osc}$. Any oscillatory contribution to the imaginary part appears only at the higher order $O\left(\chi_\alpha^{-(8+2m-\mu)/3}\right)$ – suppressed by an additional factor of $\chi_\alpha^{-1}$ relative to the leading drift $\sim \chi_\alpha^{-1}$ In contrast, for the perpendicular transport coefficients determined by $\mathrm{Re}\left(\hat{I}_{klm}\right)$, the relative contributions of the oscillatory terms decay as $O\left(\chi_\alpha^{-(2+2m-\mu)/3}\right)$, making the oscillations discernible.

For example, the high-magnetization asymptotic expression for the perpendicular α-particle diffusion (85) is

$$\begin{aligned} D_{p\perp} &= \frac{1}{3} u_{\alpha 0}^2 \tau_{\alpha e} \,\mathrm{Re}\left(\hat{I}_{000}\right) \cong \frac{1}{3\chi_\alpha^2} u_{\alpha 0}^2 \tau_{\alpha e} \left(1 + C_{\mathrm{osc}} - G_{000} - \mu K_{000}\right) \\ &= \frac{1}{6\chi_\alpha^2} u_{\alpha 0}^2 \tau_{\alpha e} \left[1 + \frac{1}{3}(1-\mu)\varepsilon^2 L_- + 2C_{\mathrm{osc}}\right]. \end{aligned} \tag{A18}$$

Similarly, the high-magnetization asymptotics of the perpendicular energy-density diffusion coefficient (95) is

$$\begin{aligned} \kappa_{p\perp} &= \frac{1}{3} u_{\alpha 0}^2 \tau_{\alpha e} \,\mathrm{Re}\left(\hat{I}_{001}\right) \cong \frac{1}{3\chi_\alpha^2} u_{\alpha 0}^2 \tau_{\alpha e} \left(1 - G_{001} - \mu K_{001}\right) \\ &= \frac{1}{4\chi_\alpha^2} u_{\alpha 0}^2 \tau_{\alpha e} \left[1 + \frac{4}{3}\varepsilon^3 (1-\mu)\left(1 - \frac{1}{6}\varepsilon L_+\right)\right]. \end{aligned} \tag{A19}$$

The asymptotic formula for the perpendicular α-particle conductivity (102) is

$$\sigma_{\alpha\perp} = \frac{4e^2 \dot{n}_\alpha \tau_{\alpha e}^2}{m_\alpha} \mathrm{Re}\left(\hat{I}_{010}\right) \cong \frac{4e^2 n_\alpha \tau_{\alpha e}}{m_\alpha \chi_\alpha^2}\left[1 + \frac{1-\mu+3C_{\mathrm{osc}}}{\left(1+\varepsilon^3\right)\ln\left(1+\varepsilon^{-3}\right)}\right]. \tag{A20}$$

In the low-temperature range $\varepsilon \to 0$, the asymptotic expressions (A18) and (A19) match the high-magnetization limits of the explicit formulas (87) and (98), respectively. For arbitrary values of $\varepsilon$, Eqs. (A18)-(A20) provide the leading terms of the asymptotic expansions at large $\chi_\alpha$. We see that for any plasma temperature, the real part of the integral (79) responsible for the perpendicular diffusion scales as $\chi_\alpha^{-2}$, whereas the imaginary part describing the gyro-drift scales as $\chi_\alpha^{-1}$, consistent with the expected scaling for magnetized transport.

FIGURES

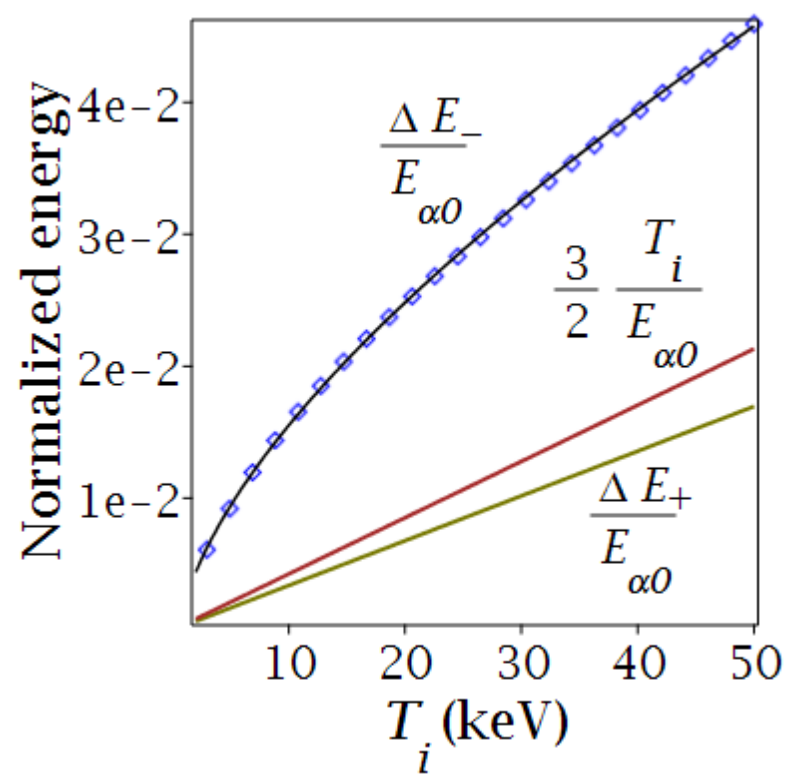

Figure 1. Average thermal addition to the energy of the born fast α-particle due to the thermal motion of the D and T ions, $\Delta E_+$, given by Eq. (13), the loss of the thermal energy from the fused ions affected by the Gamow shift, $\Delta E_-$, given by Eq. (14) (symbols), and the ion thermal energy $3T_i/2$ normalized to $E_{\alpha 0}$ vs. the ion temperature. The black line represents the numerical fit provided by Eq. (15).

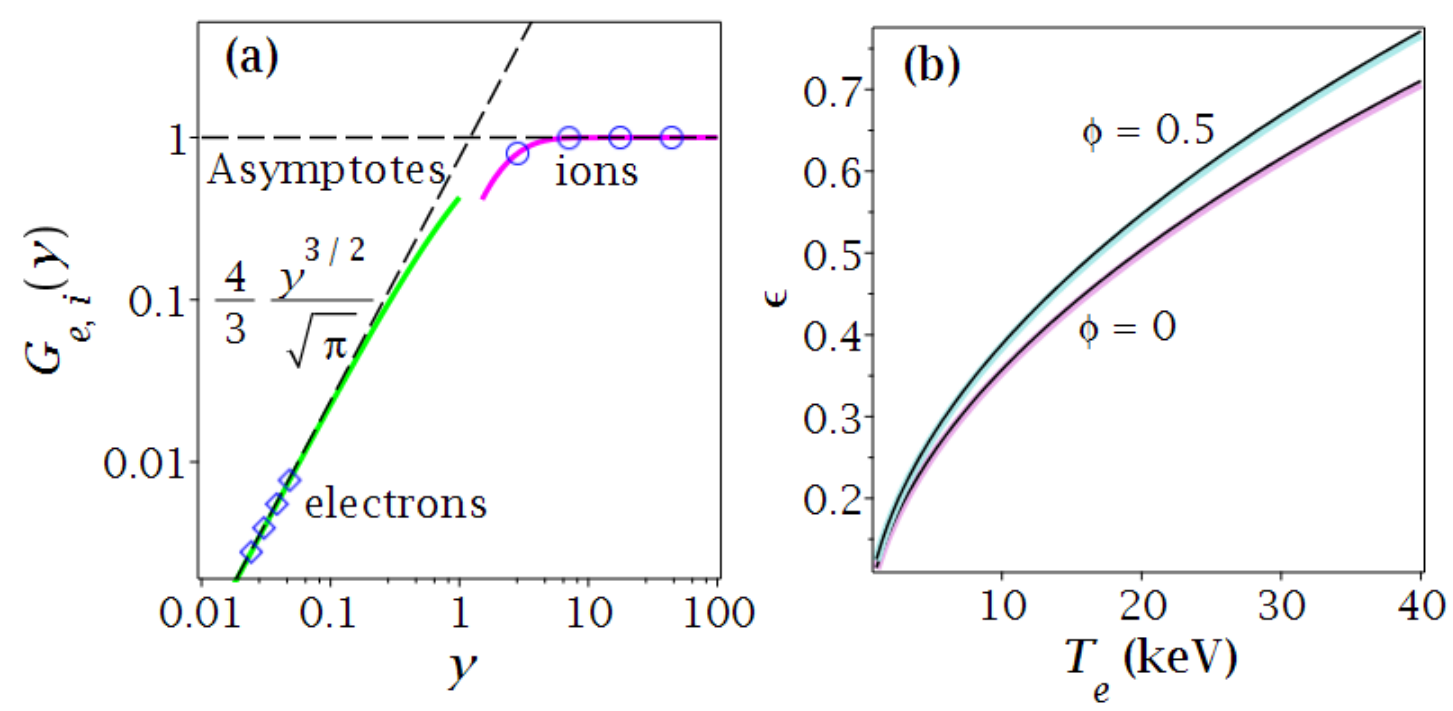

Figure 2. (a) Maxwell integrals $G_e(y)$ and $G_i(y)$ Eq. (18), vs. $y$ (solid green and magenta lines, respectively), the low-$y_e$ and high-$y$ asymptotes (dashed lines), and symbols representing MagLIF plasma conditions. (b) Temperature dependence of the parameter $\varepsilon$, Eq. (24), for $\phi = 0$ and 0.5 and the corresponding values of $\mu$, Eq. (6), $\langle Z_i \rangle$, and $\langle Z_i^2 \rangle$, Eq. (7). Colored and black lines correspond to electron densities $n = 5\times10^{23}$ cm$^{-3}$ and $10^{24}$ cm$^{-3}$, respectively.

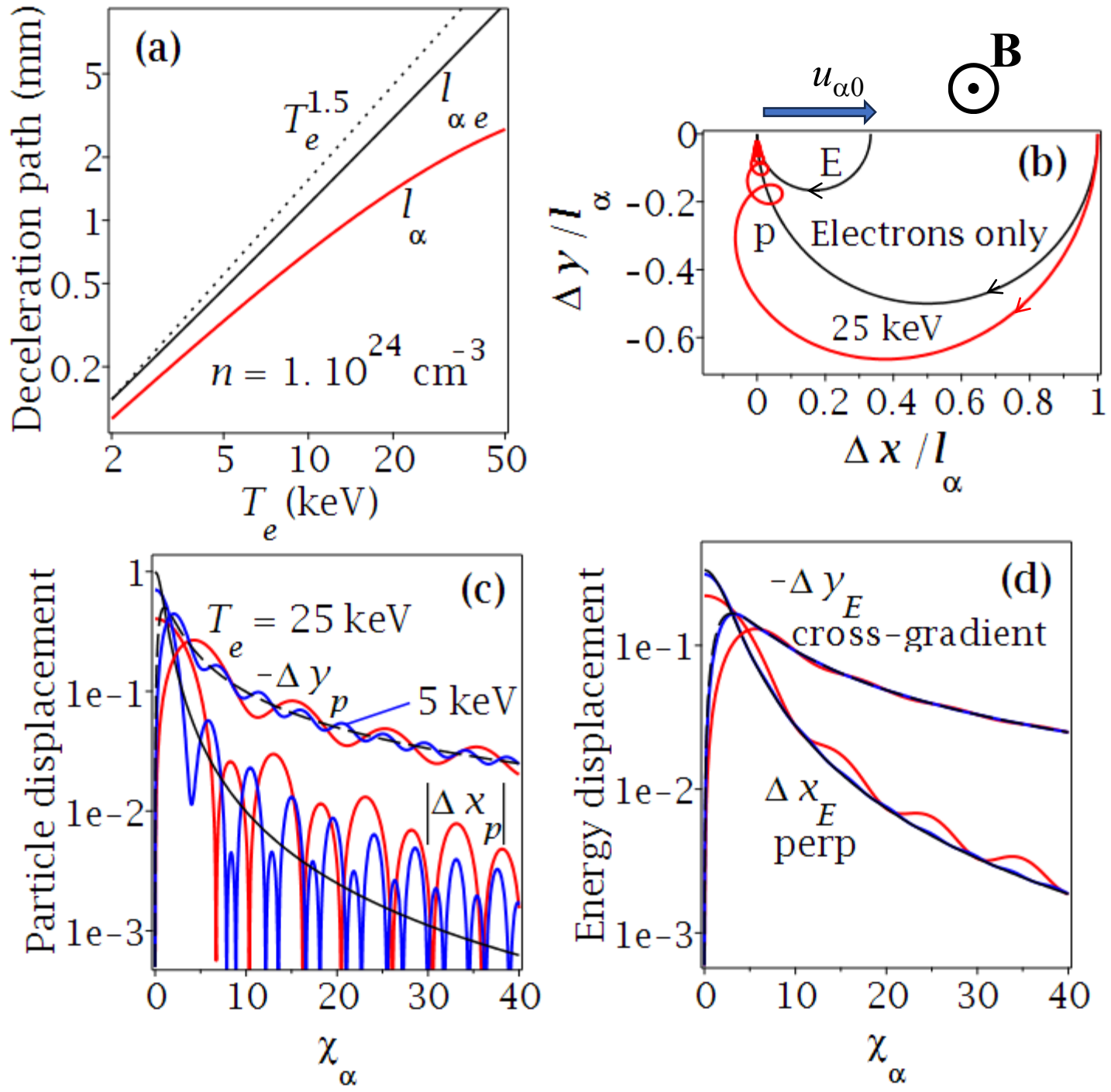

Figure 3. (a) Deceleration path $l_\alpha$ of an α-particle versus electron temperature $T_e$ for a density $n = 10^{24}$ cm$^{-3}$. The black line represents the path with ion friction neglected, while the red line accounts for ion friction effects. (b) Loci of the particle and energy displacement positions, $(\Delta x_p, \Delta y_p)$ and $(\Delta x_E, \Delta y_E)$, indexed by symbols *p* and *E*, respectively, for an α-particle born at the origin with initial velocity $u_{\alpha 0}$ in the *x*-direction. Black semicircles represent analytical solutions (32) and (33) for electron-only friction. The red line shows the locus at $T_e = 25$ keV, where ion friction causes the particle to approach the origin in an oscillatory manner as the Hall parameter $\chi_\alpha$ increases in the direction indicated by arrows. (c) Particle and (d) energy displacement positions, $\left(|\Delta x_p|, -\Delta y_p\right)$ and $\left(\Delta x_E, -\Delta y_E\right)$, respectively, in units of $l_{\alpha e}$, vs. $\chi_\alpha$. Analytical solutions (32) and (33) are shown as solid and dashed black lines, respectively. Numerical solutions for $T_e = 5$ and 25 keV are shown as blue and red lines, respectively.

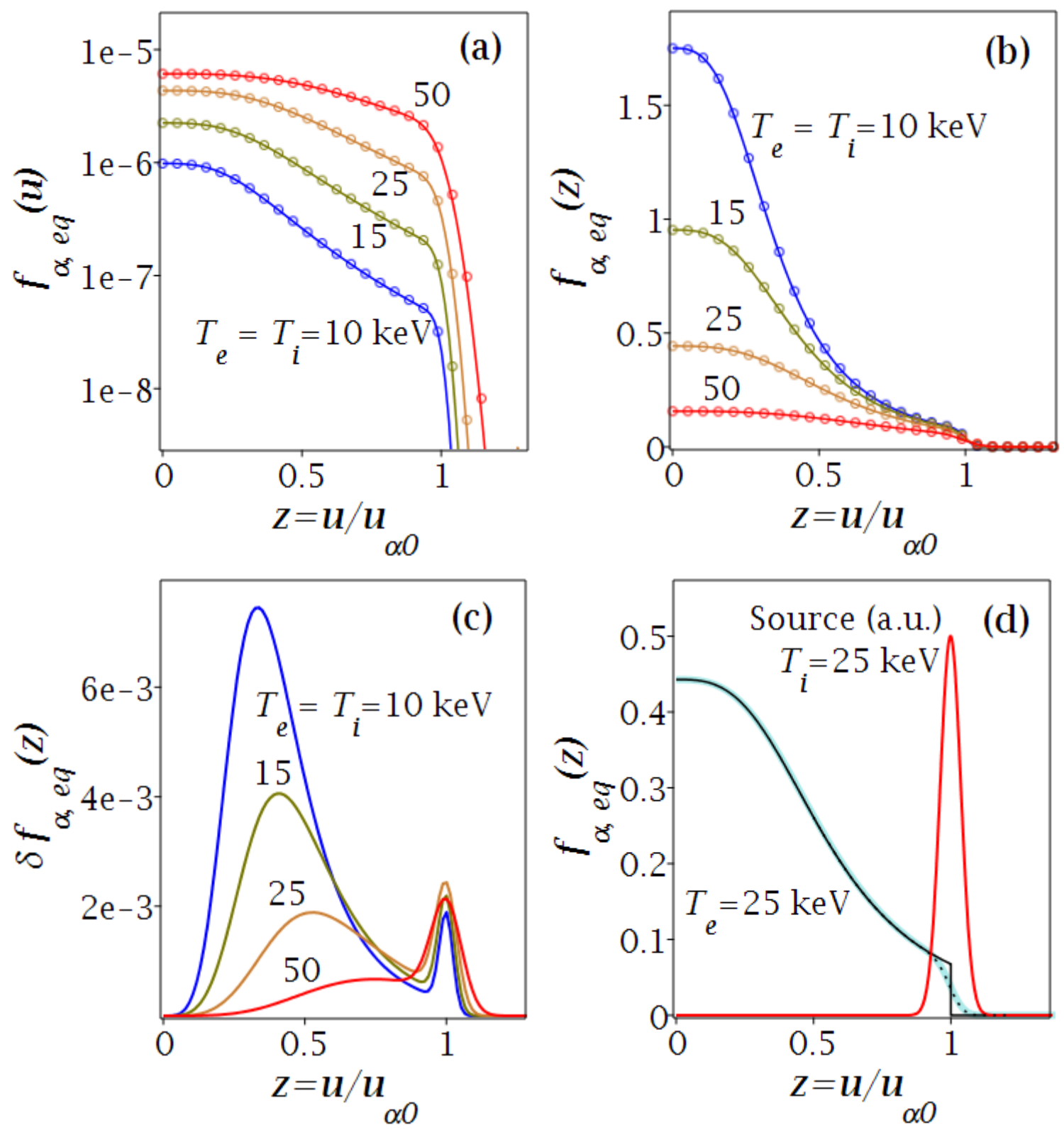

Figure 4. (a) The dimensional equilibrium distribution functions, in units of cm$^{-6}$s$^3$, calculated for $n = 5\cdot10^{23}\,\text{cm}^{-3}$, $T_e = T_i = 10$, 15, 25, and 50 keV. Symbols correspond to the solutions of Eq. (57); lines display the respective zero-order solutions (55). (b) The same distribution functions normalized with respect to the fusion rate factor, $\dot{n}_\alpha \tau_{\alpha e} / u_{\alpha 0}^3$, in the linear scale. (c) The difference $\delta f_{\alpha,\text{eq}} = f_{\alpha,\text{eq}} - f_{\alpha,\text{eq}}^{(0)}$ vs. the normalized α-particle velocity $z$. (d) The exact, normalized equilibrium distribution function, shown in the linear scale, for $T_e = T_i = 25$ keV (light blue line) and the monoenergetic-source approximate solution (56) (black line), deviating from it in the vicinity of $u = u_{\alpha 0}$. Dotted line represents the approximation (55). Red line shows the source spectrum (11) (not to scale) for $T_i = 25$ keV.

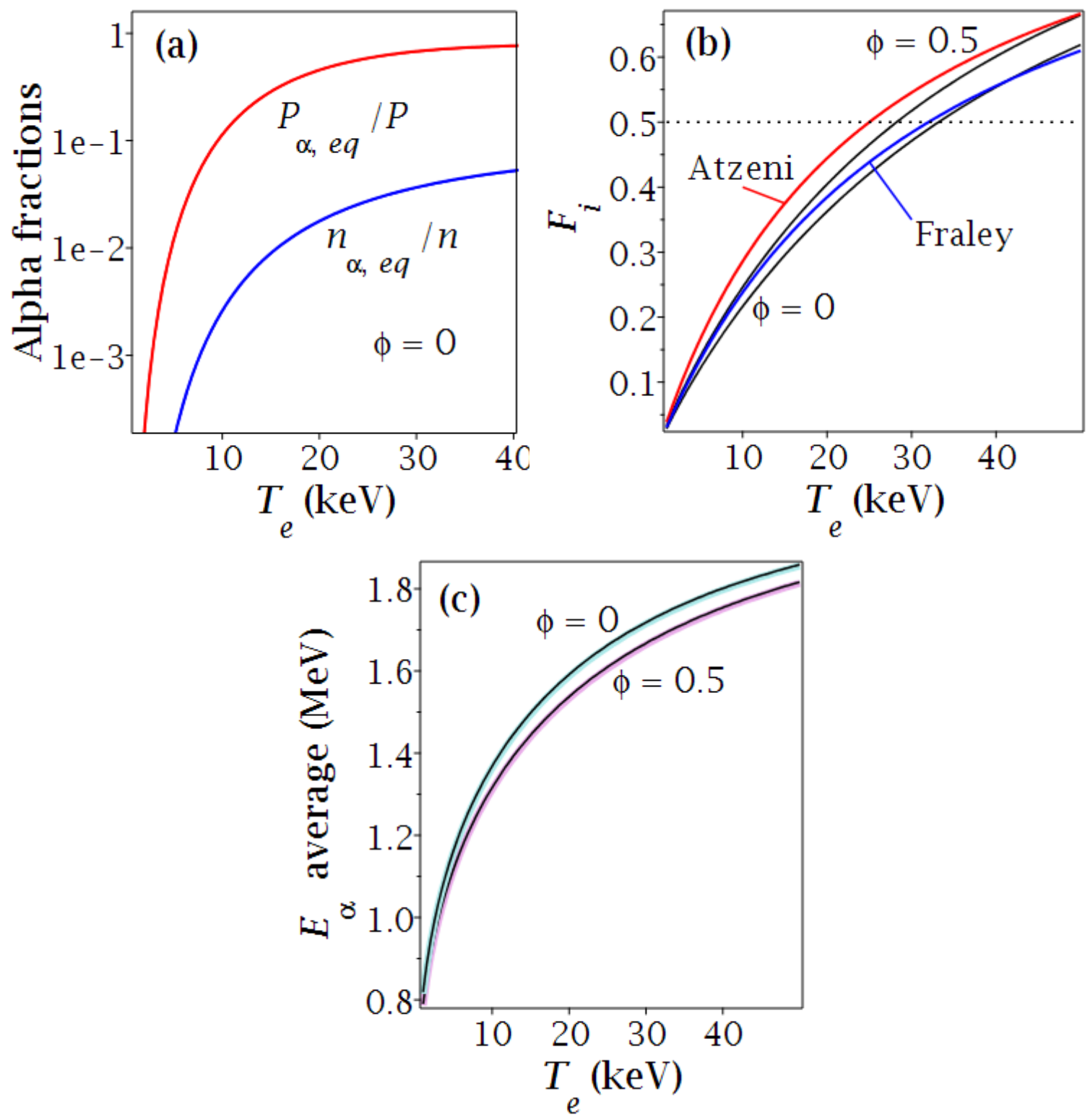

Figure 5. (a) The density and pressure ratios $n_\alpha / n$ and $P_\alpha / P$ vs. $T_e$ for an isothermal DT plasma at burn-up fraction $\phi = 0$ and $n = 5\times10^{23}$ cm$^{-3}$. (b) The ion energy partition fraction $F_i$ vs. $T_e$ for $\phi = 0$, $n = 5\times10^{23}$ cm$^{-3}$, and $\phi = 0.5$, $n = 10^{24}$ cm$^{-3}$ (black lines). Colored lines represent fits for $F_i$ from Refs. [36] and [37], labeled "Fraley" and "Atzeni", respectively. (c) Average energy of a fast α-particle vs. $T_e$ for the same conditions as in (b). Here, colored and black lines correspond to electron densities of $n = 5\times10^{23}$ cm$^{-3}$ and $10^{24}$ cm$^{-3}$, respectively, as in Fig. 2(b).

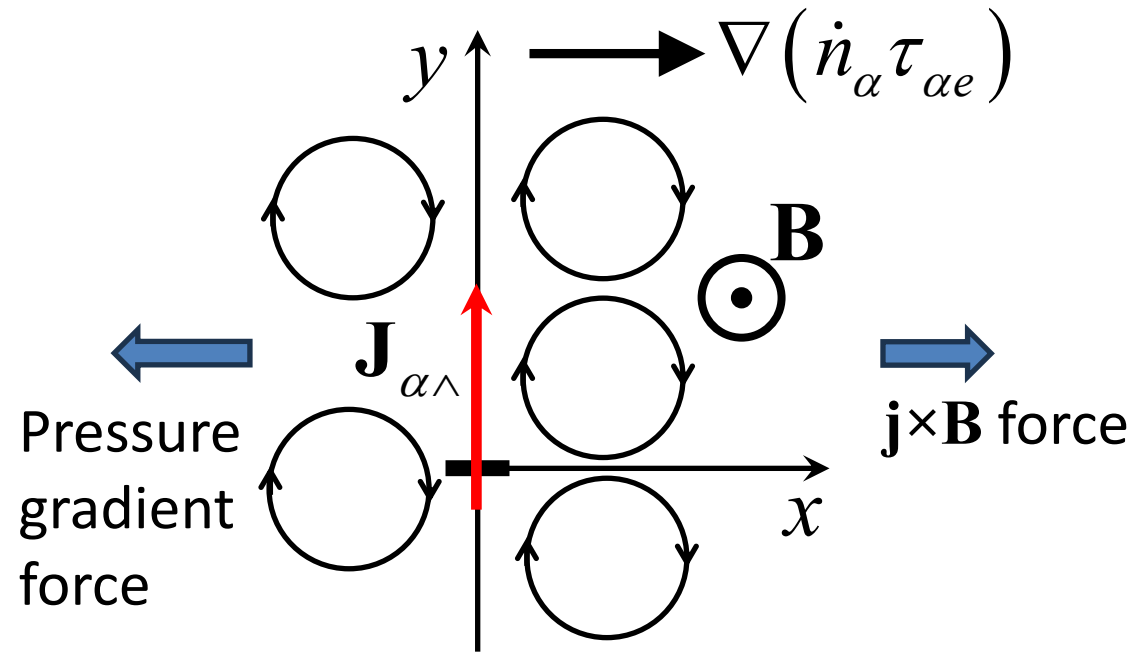

Figure 6. Schematic of the fast α-particle cross-gradient flux and force formation due to magnetization.

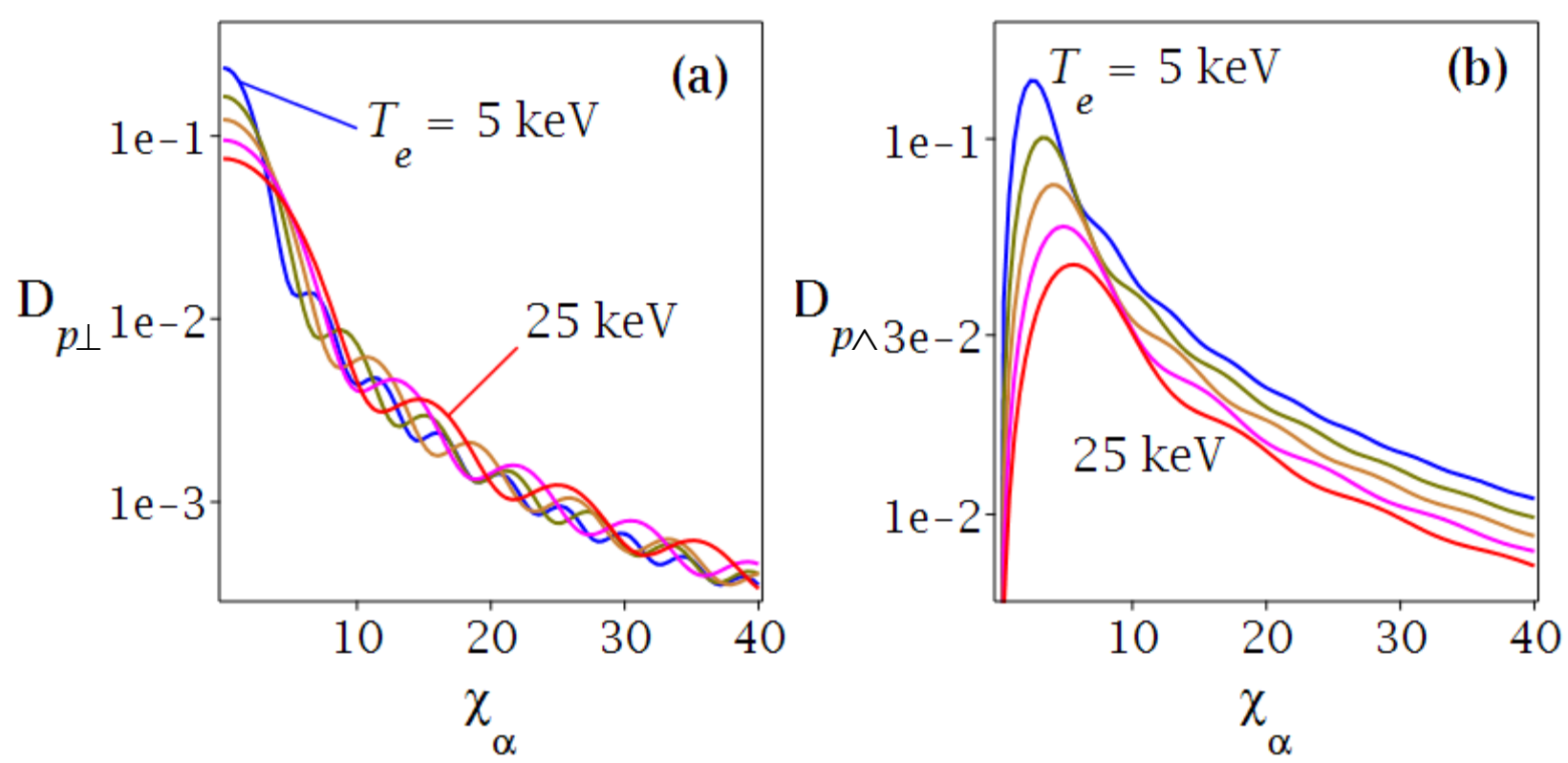

Figure 7. Perpendicular (a) and cross-gradient (b) diffusion coefficients, $D_{p\perp}$ and $D_{p\wedge}$, normalized with respect to the factor $u_{\alpha 0}^2 \tau_{\alpha e} / 3$, for a DT plasma with $\phi = 0$, $n = 5\times 10^{23}$ cm$^{-3}$, and electron temperature $T_e = 5$, 10, 15, 20, and 25 keV, plotted vs. the α-particle Hall parameter, $\chi_\alpha = \omega_\alpha \tau_{\alpha e}$.

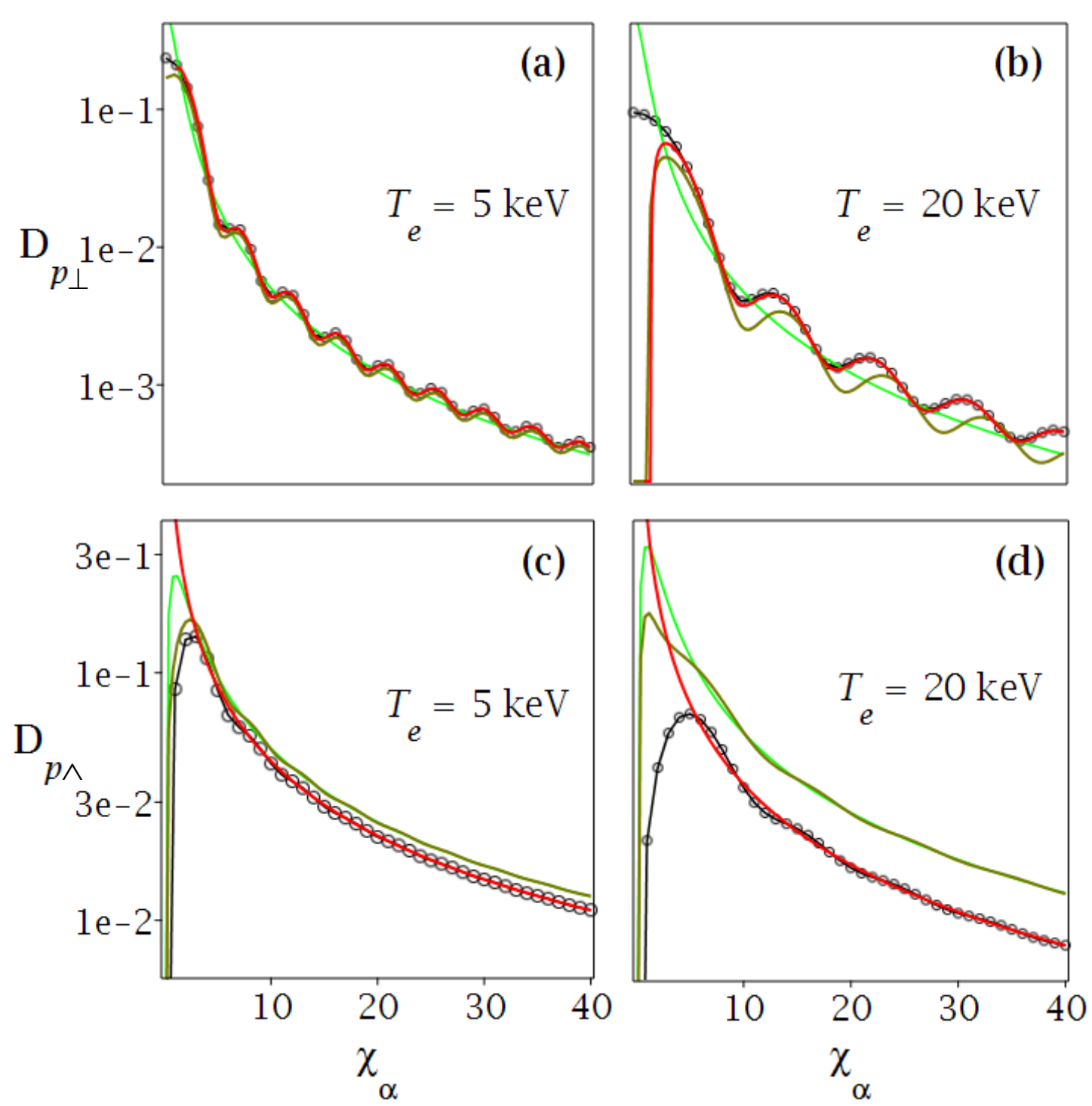

Figure 8. Perpendicular (a), (b) and cross-gradient (c), (d) diffusion coefficients, $D_{p\perp}$ and $D_{p\wedge}$, plotted vs. $\chi_\alpha$ at $T_e = 5$ keV (a), (c) and 20 keV (b), (d); normalization of these coefficients and all other conditions are the same as in Fig. 7. Black lines with symbols represent exact values, light green and olive lines correspond to the first and second approximations, respectively, in Eq. (86), and the red line represents the strong-magnetization asymptotics (88).

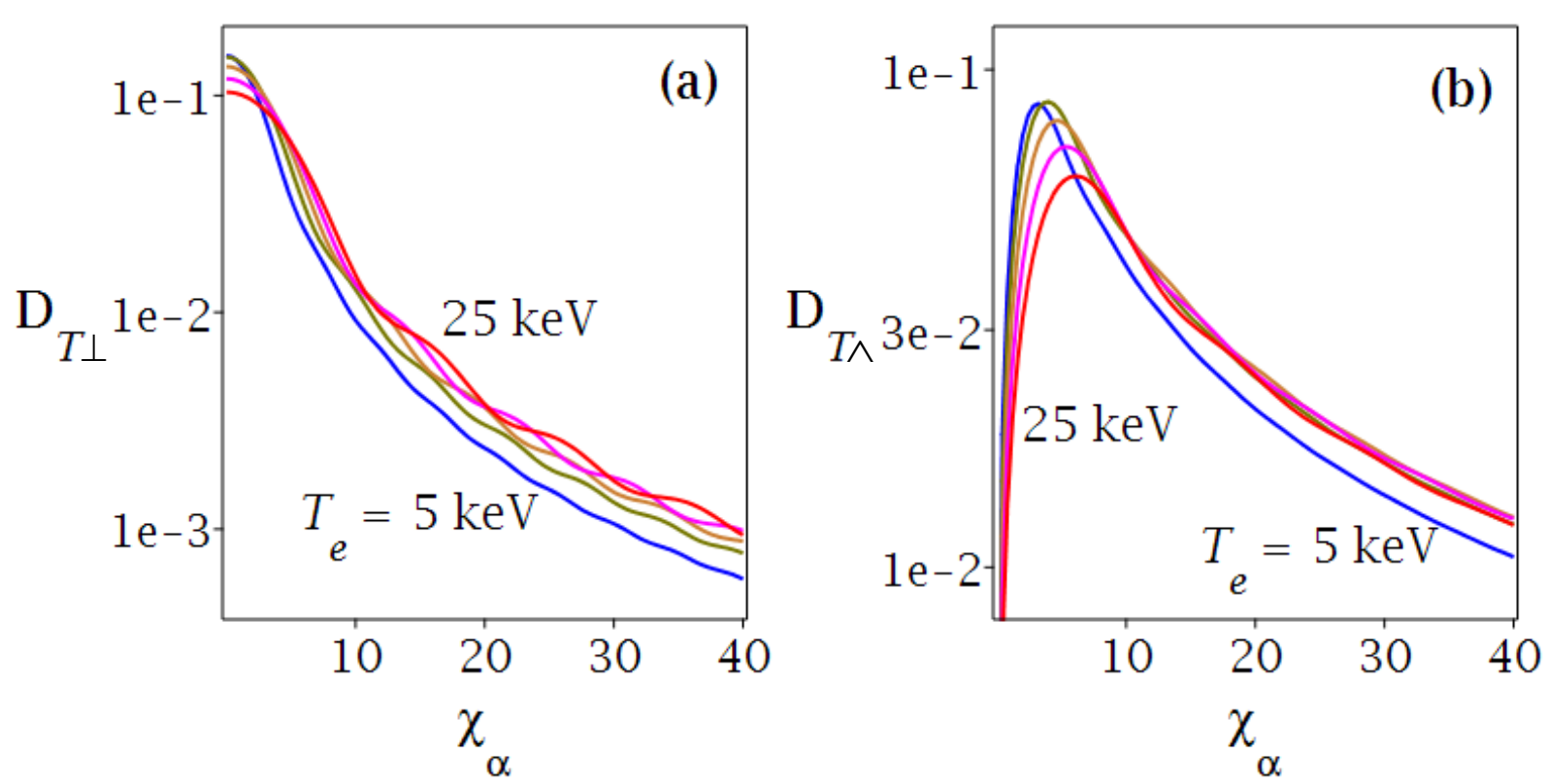

Figure 9. Perpendicular, $D_{T\perp}$ (a), and cross-gradient, $D_{T\wedge}$ (b), coefficients of fast α-particle thermodiffusion plotted vs. $\chi_\alpha$ for the same conditions as in Fig. 7.

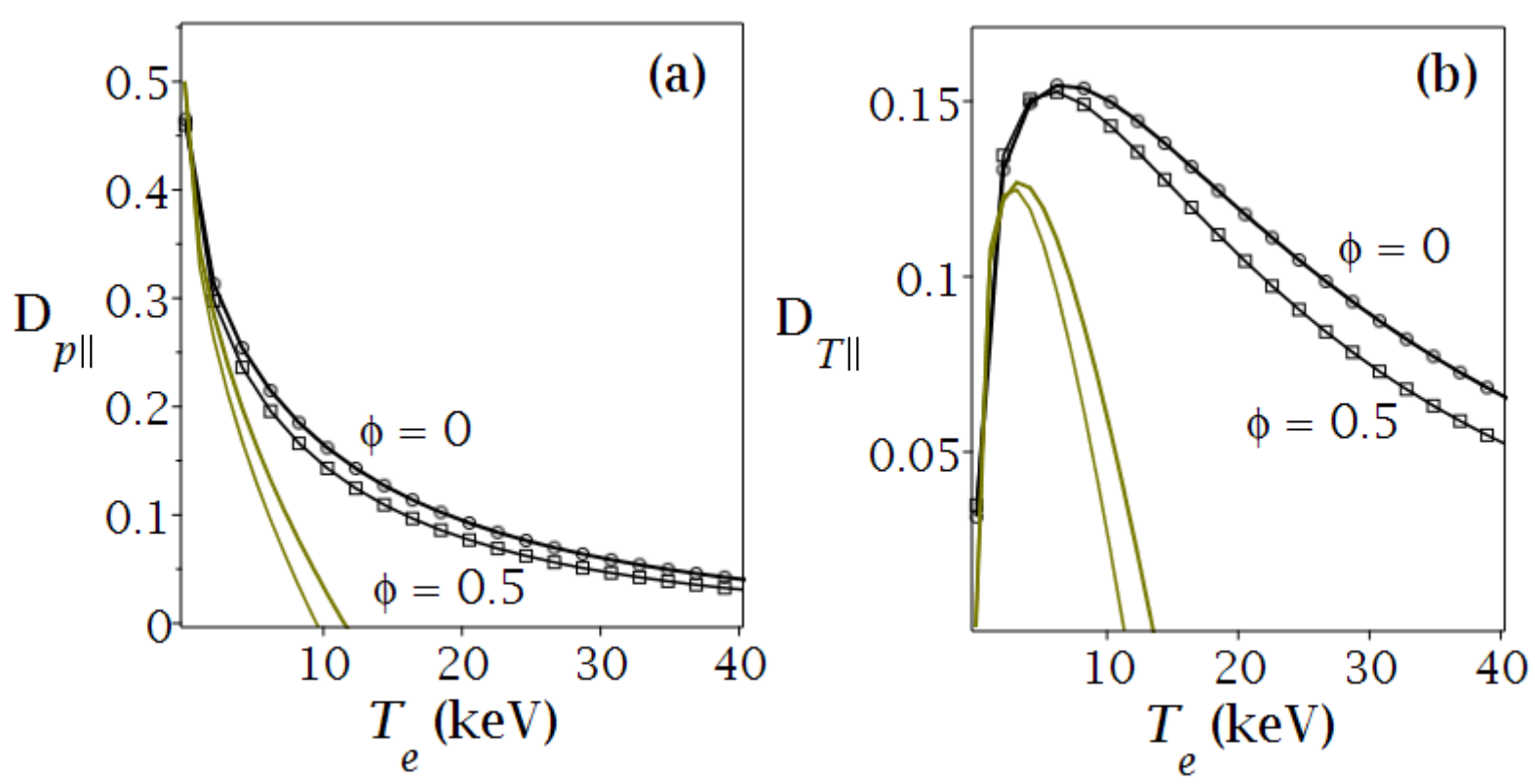

Figure 10. Parallel diffusion, $D_{p\|}$ (a), and thermodiffusion, $D_{T\|}$ (b), coefficients, normalized as in Fig. 7, plotted vs. $T_e$ for $\phi = 0$ and 0.5 and the corresponding values of $\mu$, $\langle Z_i \rangle$, and $\langle Z_i^2 \rangle$, at $n = 5\times10^{23}$ cm$^{-3}$ and $10^{24}$ cm$^{-3}$, respectively. Lines with symbols represent exact values; thin and thick lines correspond to the second-order approximation in the small-$\varepsilon$ expansions for the two cases.

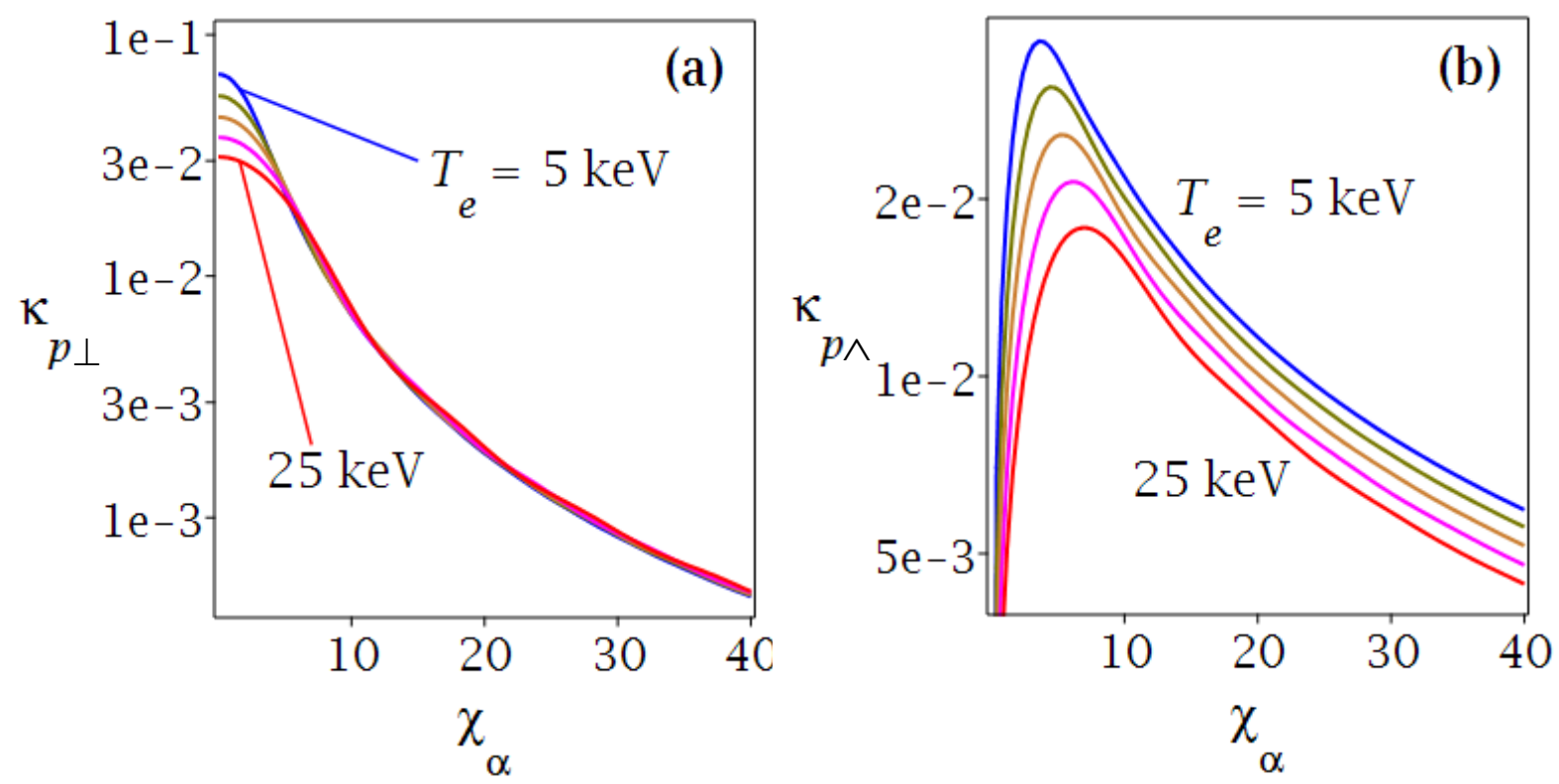

Figure 11. Perpendicular (a) and cross-gradient (b) energy-density diffusion coefficients, $\kappa_{p\perp}$ and $\kappa_{p\wedge}$, normalized as in Fig. 7, plotted vs. $\chi_\alpha$ for a DT plasma with $\phi = 0$, $n = 5\times10^{23}$ cm$^{-3}$, and electron temperature $T_e = 5$, 10, 15, 20, and 25 keV.

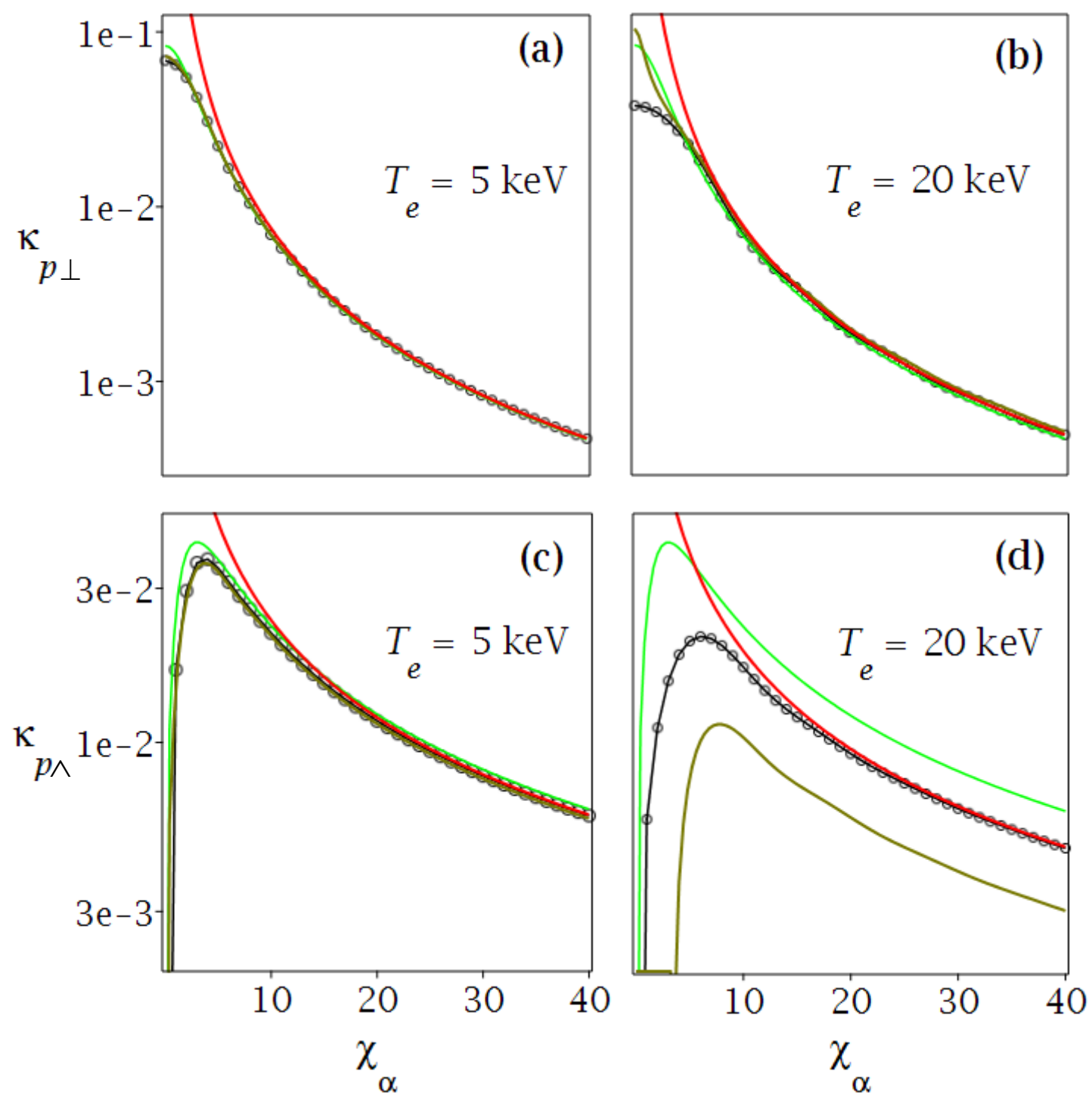

<u>Figure 12.</u> Perpendicular (a, b) and cross-gradient (c, d) energy-density diffusion coefficients, normalized as in Fig. 7, $\kappa_{p\perp}$ and $\kappa_{p\wedge}$, plotted vs. $\chi_\alpha$ for $T_e = 5$ keV (a, c) and 20 keV (b, d). All other conditions are the same as in Figs. 7 and 11. Black lines with symbols represent exact values; light green and olive lines correspond to the first and second approximations, respectively, from Eq. (96), and the red line shows the strong-magnetization asymptotics (100).

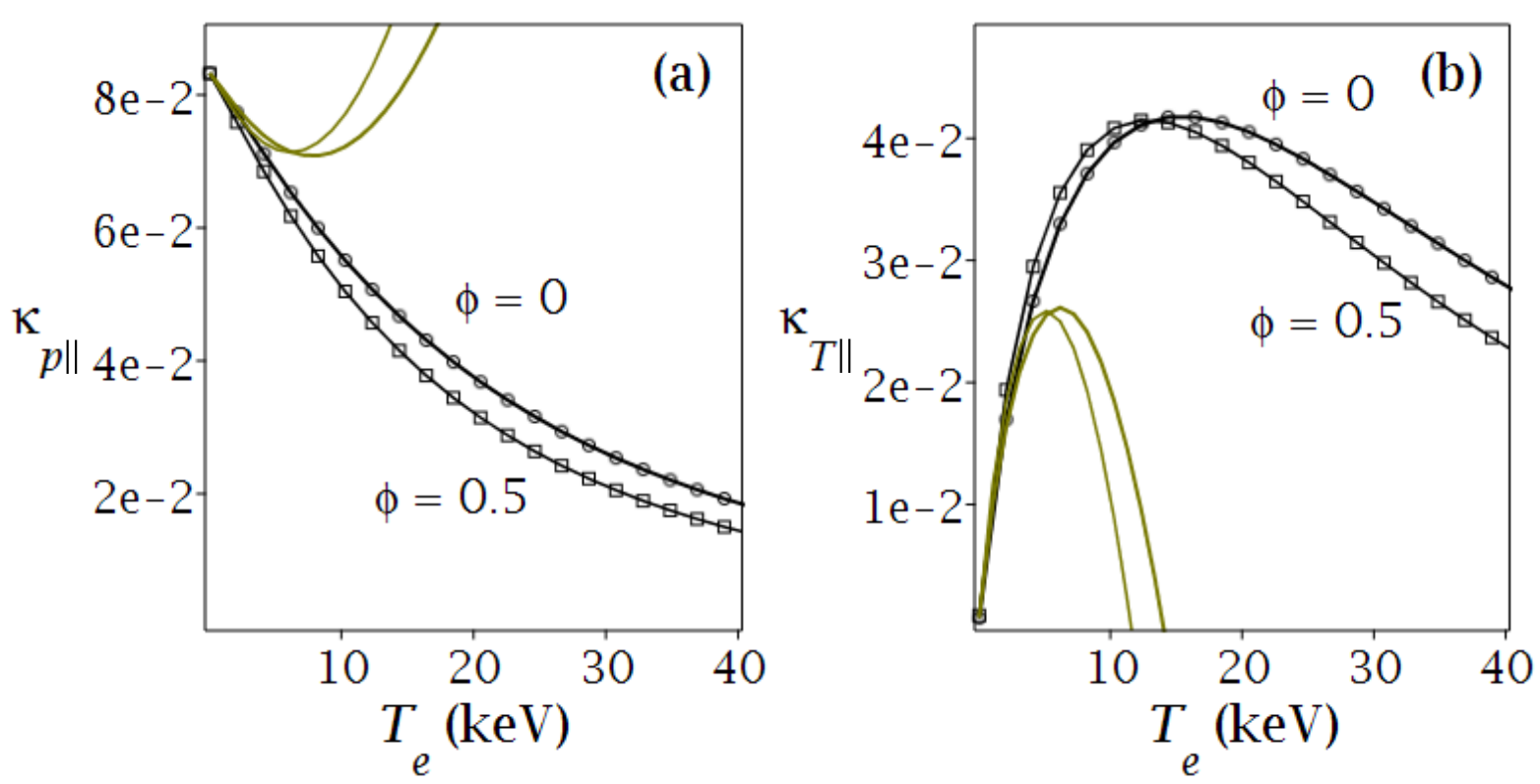

Figure 13. Parallel energy-density diffusion coefficients, $\kappa_{p\parallel}$ (a), and $\kappa_{T\parallel}$ (b), normalized as in Fig. 7, plotted vs. for $\phi = 0$ and 0.5 and the corresponding values of $\mu$, $\langle Z_i \rangle$, and $\langle Z_i^2 \rangle$, at $n = 5\times10^{23}$ cm$^{-3}$ and $10^{24}$ cm$^{-3}$, respectively. Lines with symbols represent exact values; thin and thick lines correspond to the second-order approximation in the small-$\varepsilon$ expansions for the two cases.

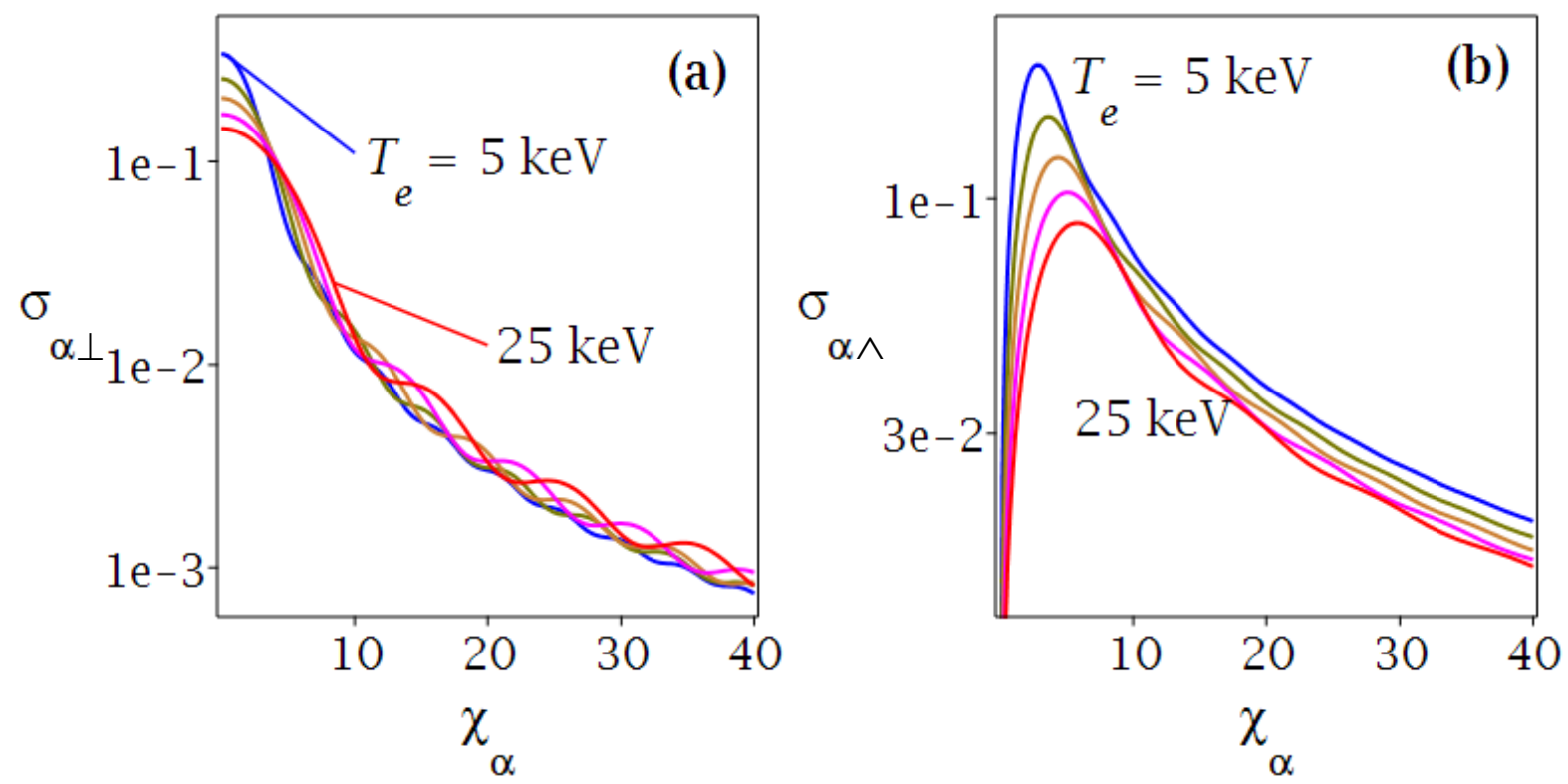

Figure 14. Perpendicular (a) and cross-gradient (b) normalized conductivity, $\sigma_{\alpha\perp}$ and $\sigma_{\alpha\wedge}$, plotted vs. $\chi_\alpha$ for a DT plasma with $\phi = 0$, $n = 5\times10^{23}$ cm$^{-3}$, and electron temperature $T_e = 5$, 10, 15, 20, and 25 keV.

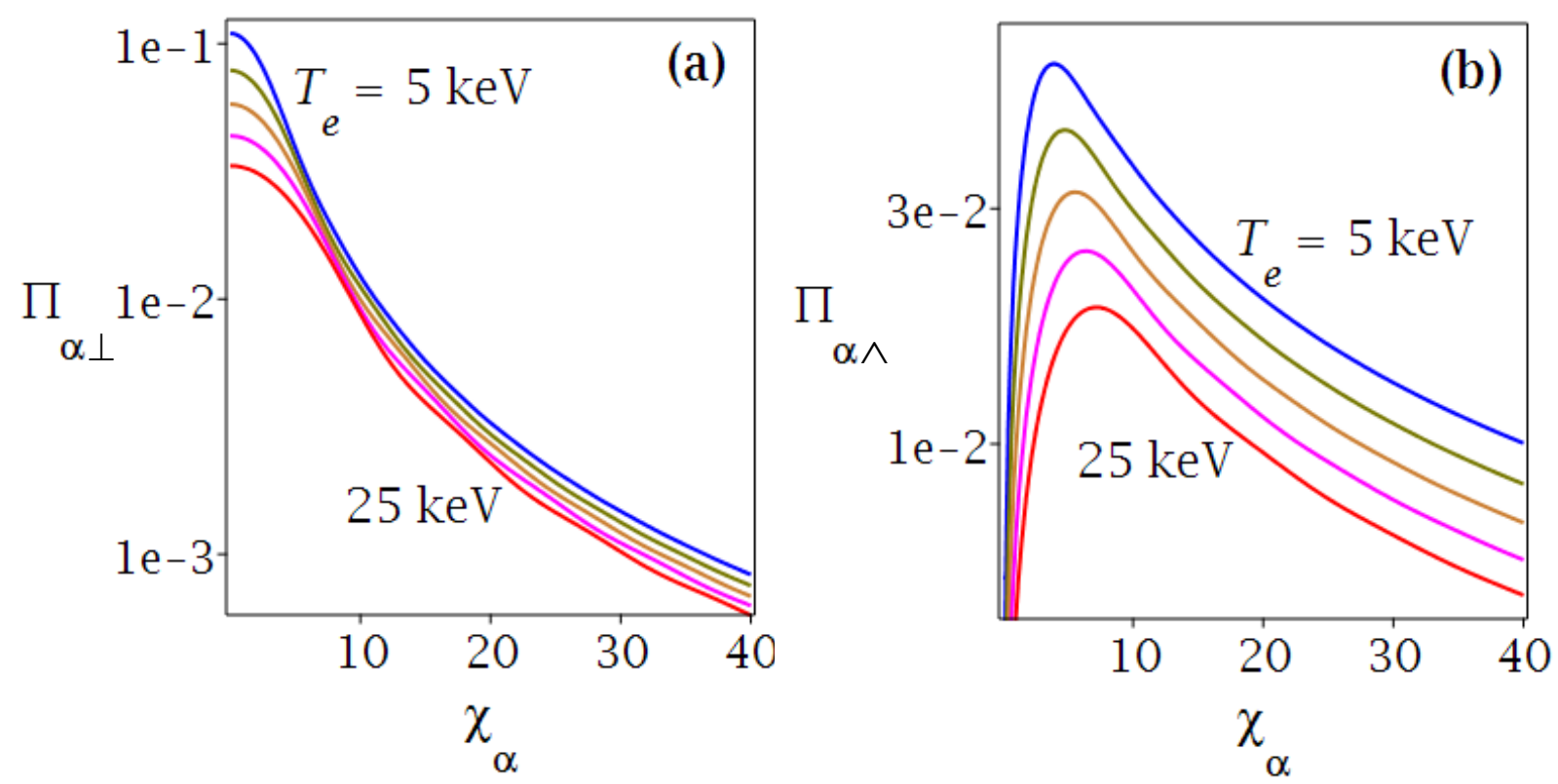

Figure 15. Perpendicular (a) and cross-gradient (b) normalized Peltier coefficients, $\Pi_{\alpha\perp}$ and $\Pi_{\alpha\wedge}$, plotted vs. $\chi_\alpha$ for the same conditions as in Fig. 14.

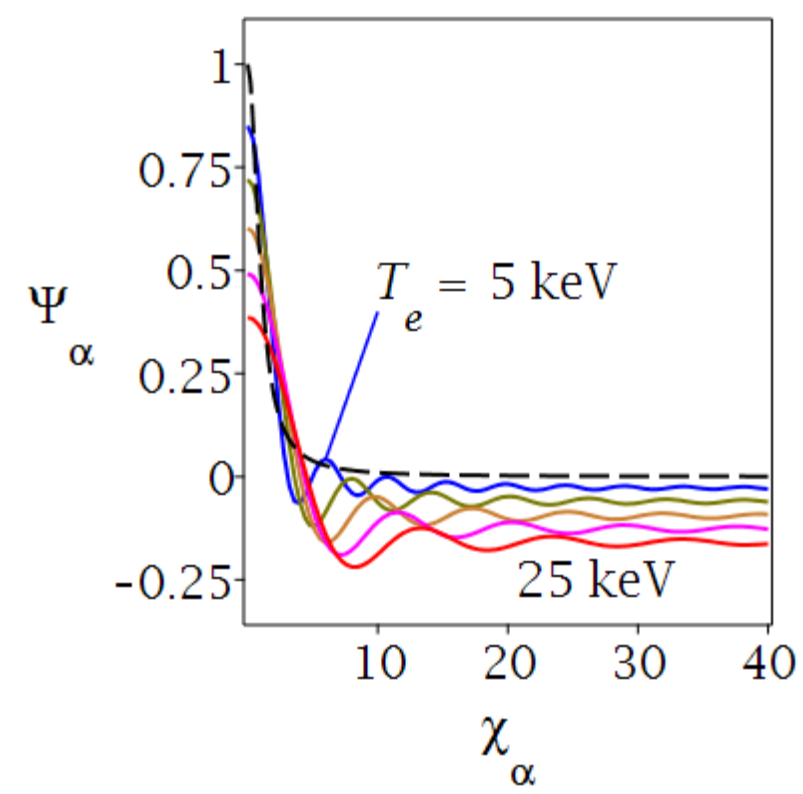

Figure 16. The force balance factor $\Psi_\alpha$ plotted vs. $\chi_\alpha$ for a DT plasma with $\phi = 0$, $n = 5\times10^{23}$ cm$^{-3}$, and electron temperature $T_e = 5$, 10, 15, 20, and 25 keV.

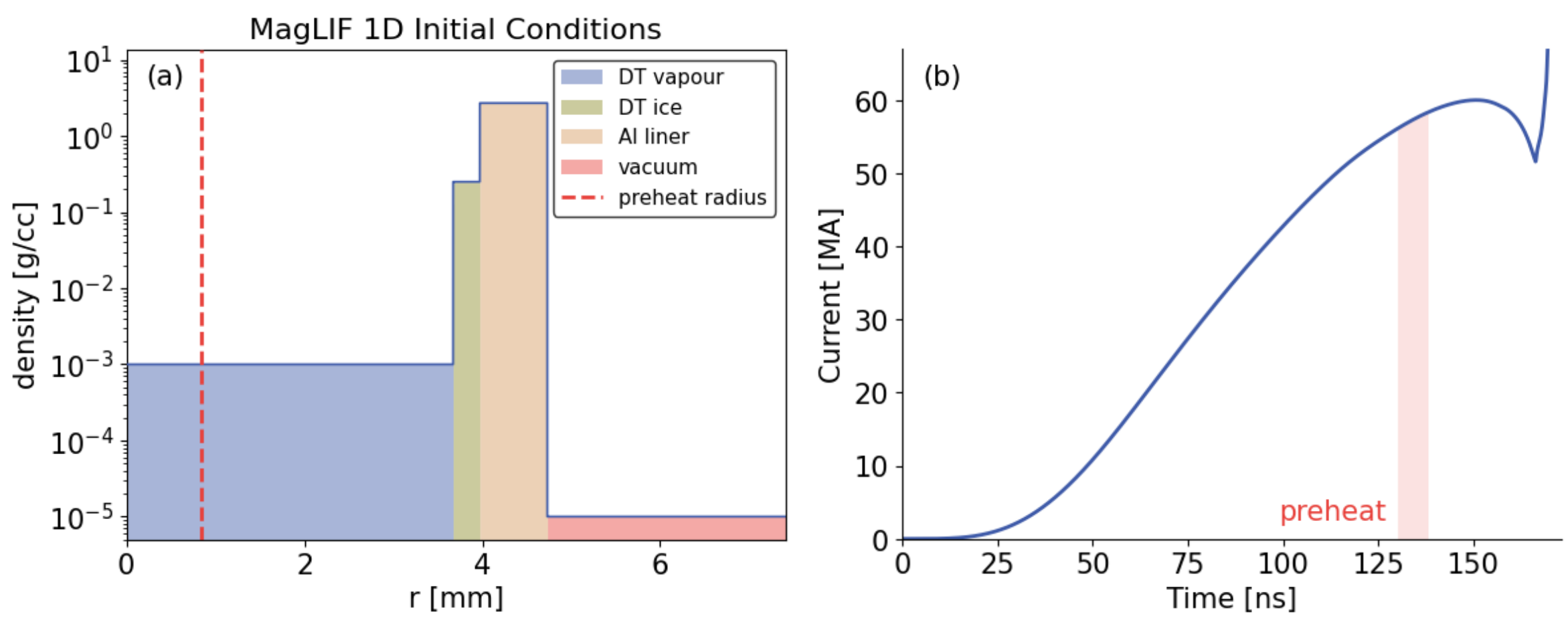

Figure 17. (a) Initial density profile of ice-layered MagLIF target. The dotted red line delimits the central region in which the preheat energy is deposited. (b) Current delivered to the load target as a function of time. The red-shaded area indicates the period during which preheat is active.

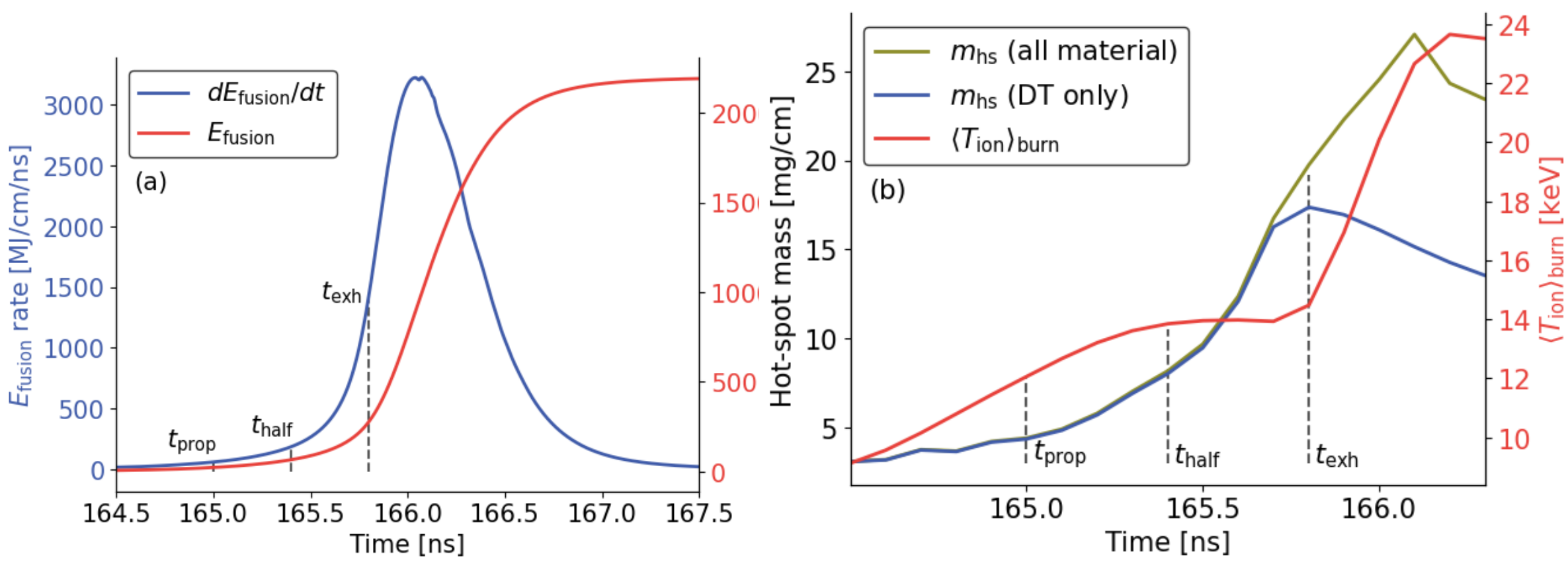

<u>Figure 18.</u> (a) Burn history and total fusion energy production plotted as functions of time. (b) Temporal evolution of the hot spot mass per unit length, fuel mass per unit length in the hot spot, and the burn-averaged ion temperature. Vertical dashed lines indicate the onset of burn propagation ($t_{\mathrm{prop}}$), the time at which burn has propagated through approximately half of the DT ice mass ($t_{\mathrm{half}}$), and the time at which burn propagation has exhausted all the DT ice mass ($t_{\mathrm{exh}}$).

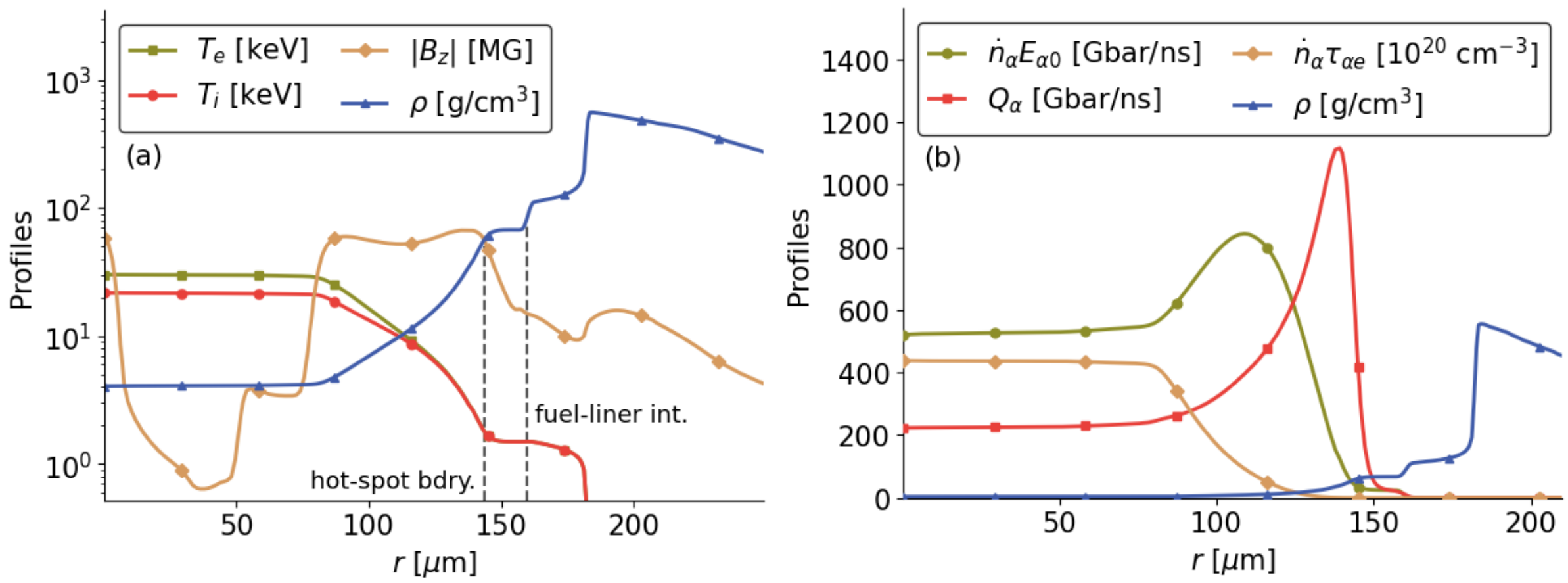

Figure 19. Profiles at $t = t_{\text{half}}$ : (a) Electron temperature ($T_e$), ion temperature ($T_i$), axial magnetic field ($B_z$), and plasma density ($\rho$). The vertical dashed lines show the hot spot boundary and the interface between the fuel and the liner (defined as the location where the aluminum species concentration reaches 0.5). (b) Alpha energy production rate ($\dot{n}_\alpha E_{\alpha 0}$), alpha energy deposition rate ($Q_\alpha$), characteristic α-particle number density ($\dot{n}_\alpha \tau_{\alpha e}$), and plasma density.

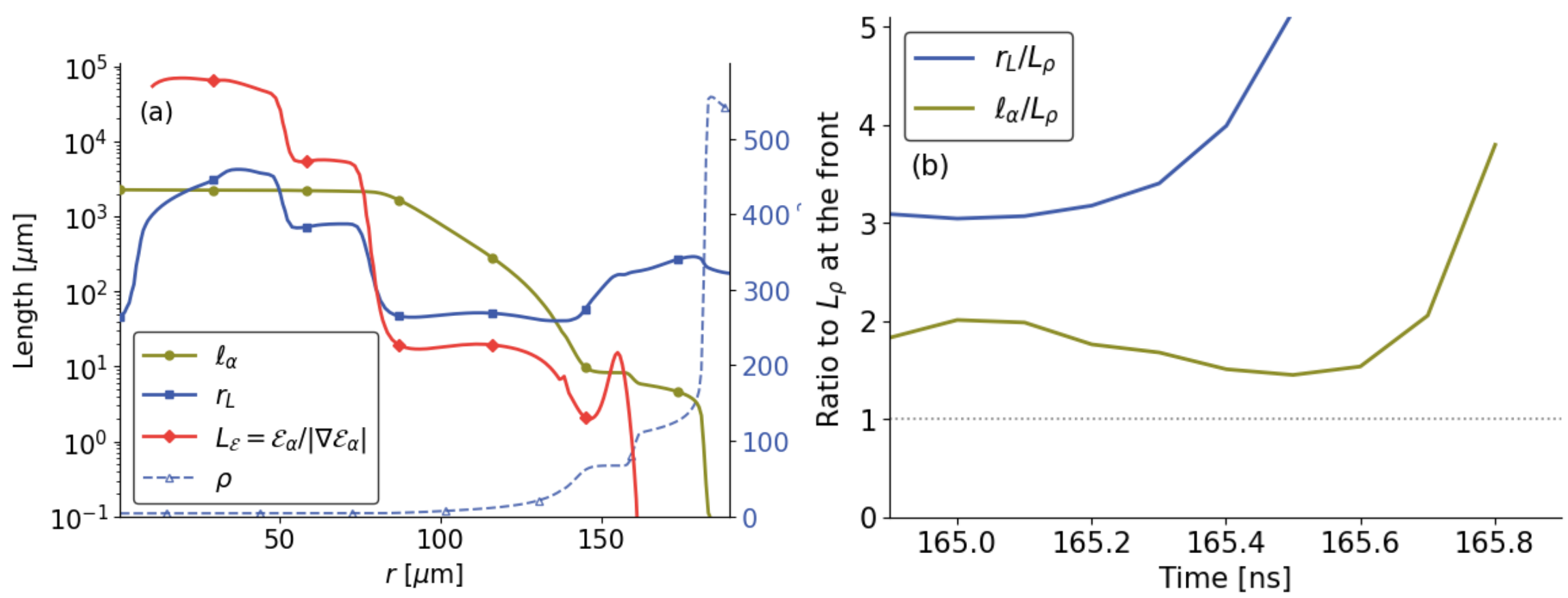

Figure 20. (a) Profiles of alpha deceleration path length $l_\alpha$ [defined in Eq. (30)], alpha-particle gyroradius $r_L$, and α-energy gradient scale length $L_\mathcal{E}$ at $t = t_{\text{half}}$. Plasma density is depicted in dashed lines for reference. (b) Temporal evolution of the ratios of α-particle gyroradius and deceleration path to the density-gradient scale length $L_\rho$ at the ablation front, defined as the location in the fuel where $L_\rho$ reaches its minimum.

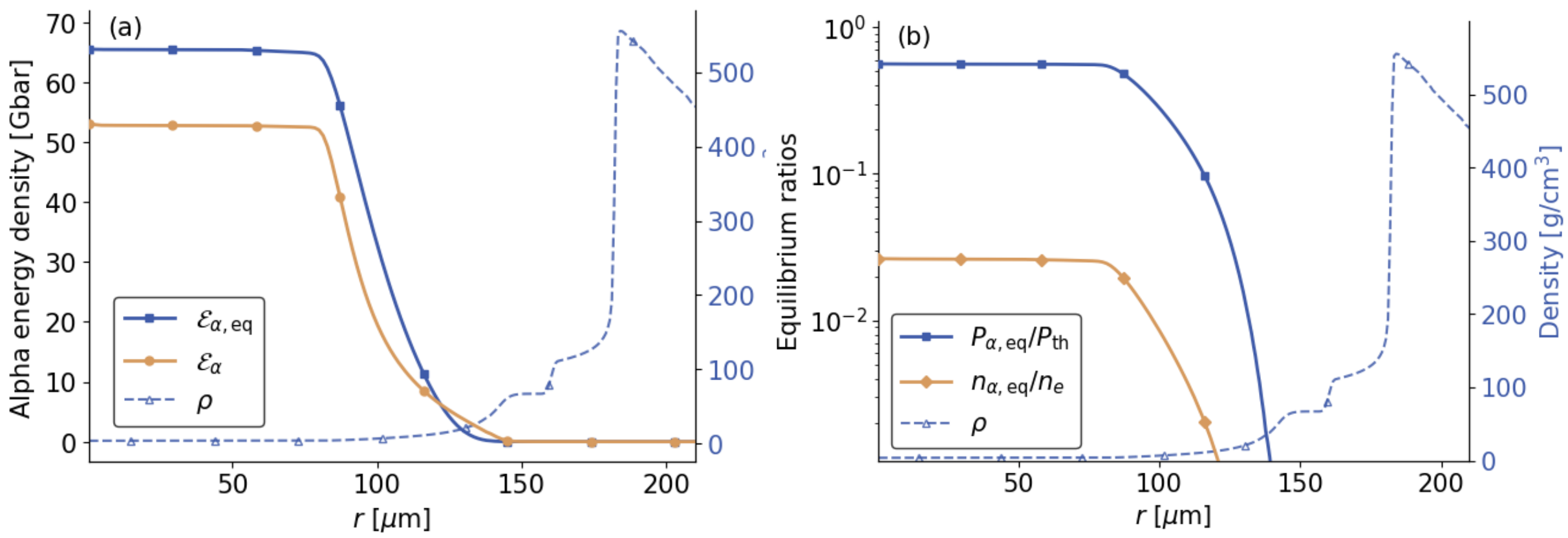

Figure 21. Profiles at $t = t_{\mathrm{half}}$ : (a) Alpha-energy density $\mathcal{E}_{\alpha}$ and its equilibrium value, $\mathcal{E}_{\alpha,\mathrm{eq}}$ given by the leading order of Eq. (65). (b) Ratios of the equilibrium α-pressure $P_{\alpha,\mathrm{eq}}$ [Eq. (66)] and α-particle number density $n_{\alpha,\mathrm{eq}}$ [leading term of Eq. (63)] to the thermal pressure of the plasma and the electron number density, respectively.

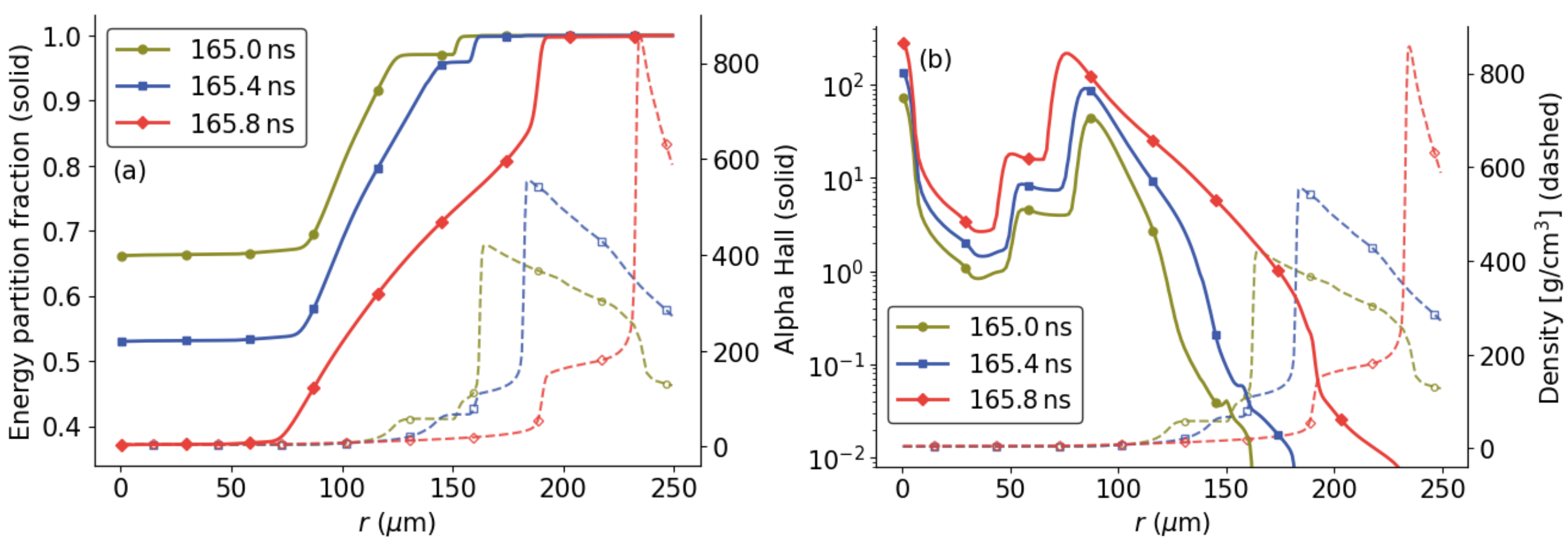

Figure 22. (a) Energy partition fraction to plasma electrons, $F_e$ [given by Eq. (34)], during burn propagation. (b) Alpha Hall parameter, $\chi_\alpha$, during burn propagation. Plasma density $\rho$ is depicted in dashed lines for reference.

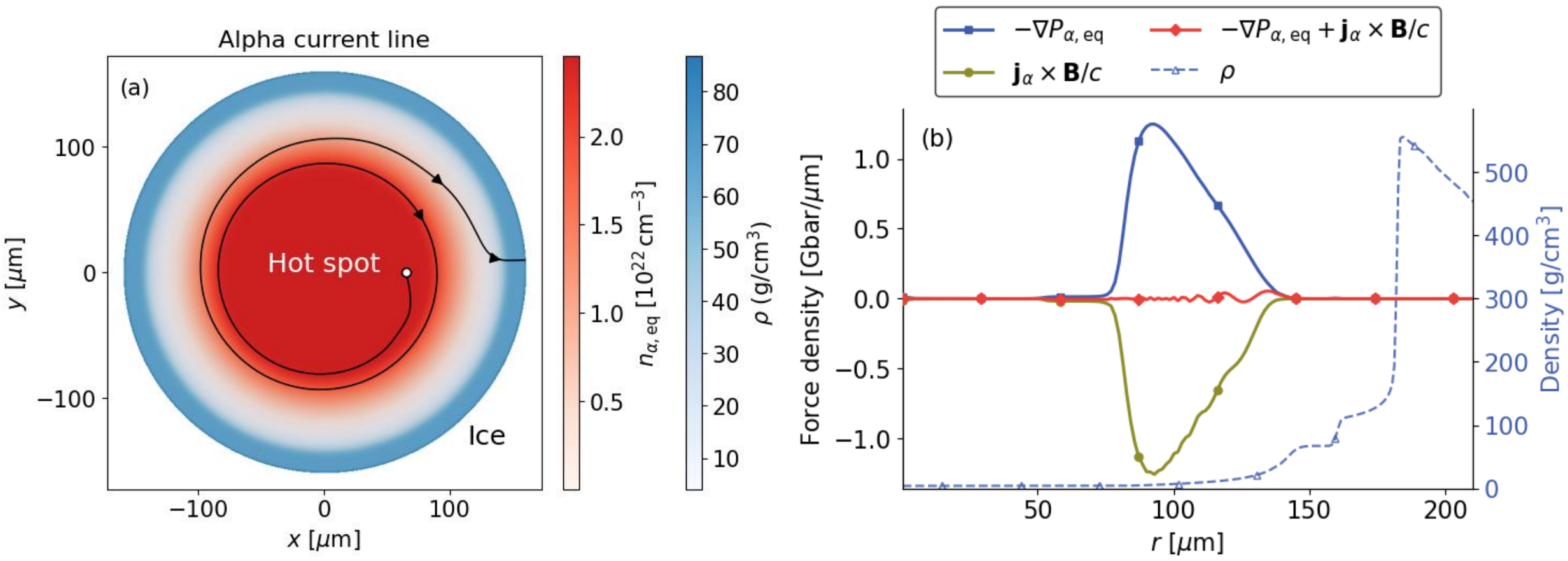

Figure 23. (a) Alpha current line in the $(r,\theta)$ plane over a composite background that combines the azimuthally symmetric maps of equilibrium α-particle number density $n_{\alpha,\mathrm{eq}}$ (inner region) and background plasma density $\rho$ (outer region). (b) Profiles of α-forces and plasma density at $t = t_{\mathrm{half}}$.